\documentclass[12pt]{article}
\usepackage[margin=1in]{geometry}
\usepackage{lineno}
\usepackage[colorlinks=true,citecolor=blue,pdfcreator={},pdfproducer={},linkcolor=blue]{hyperref}
\modulolinenumbers[1]
\usepackage{amsmath}
\usepackage{amsfonts}
\usepackage{bm}
\usepackage{graphicx}
\allowdisplaybreaks
\usepackage{amsbsy}
\usepackage{amssymb}
\usepackage[ruled]{algorithm2e}
\usepackage{natbib}

\usepackage{chngcntr}

\usepackage{subfig}

\usepackage[ruled]{algorithm2e}

\usepackage{indentfirst}

\newcommand*\patchAmsMathEnvironmentForLineno[1]{%
\expandafter\let\csname old#1\expandafter\endcsname\csname #1\endcsname
\expandafter\let\csname oldend#1\expandafter\endcsname\csname end#1\endcsname
\renewenvironment{#1}%
{\linenomath\csname old#1\endcsname}%
{\csname oldend#1\endcsname\endlinenomath}}%
\newcommand*\patchBothAmsMathEnvironmentsForLineno[1]{%
\patchAmsMathEnvironmentForLineno{#1}%
\patchAmsMathEnvironmentForLineno{#1*}}%
\AtBeginDocument{%
\patchBothAmsMathEnvironmentsForLineno{equation}%
\patchBothAmsMathEnvironmentsForLineno{align}%
\patchBothAmsMathEnvironmentsForLineno{flalign}%
\patchBothAmsMathEnvironmentsForLineno{alignat}%
\patchBothAmsMathEnvironmentsForLineno{gather}%
\patchBothAmsMathEnvironmentsForLineno{multline}%
}

\usepackage[nomarkers]{endfloat}

\graphicspath{{./data/}}

\usepackage{xspace} 

\DeclareMathOperator*{\argmin}{arg\,min}

\newcommand{\bfd}{\mathbf{d}}
\newcommand{\bfu}{\mathbf{u}}
\newcommand{\bfm}{\mathbf{m}}
\newcommand{\bfw}{\mathbf{w}}
\newcommand{\bfh}{\mathbf{h}}
\newcommand{\bfs}{\mathbf{s}}

\usepackage{authblk}
\usepackage{times}
\usepackage{mathptmx}

\title{\textbf{Acoustic- and elastic-waveform inversion with total 
generalized p-variation regularization}}
\author[,1]{Kai Gao\thanks{Corresponding Author; kaigao87@gmail.com (K.~Gao); ljh@lanl.gov (L.~Huang)}}
\author[1]{Lianjie Huang}
\affil[1]{Los Alamos National Laboratory, Geophysics Group, MS~D452, Los Aamos, NM 
87545, USA}

\date{}
\begin{document}
\maketitle

\begin{abstract}

Geophysical models usually contain both sharp interfaces and smooth 
   variations, and it is difficult to accurately account for both of 
   these two types of medium parameter variations using conventional 
   full-waveform inversion methods. In addition, sparse geometry, noisy 
   data and source encoding usually lead to strong inversion artifacts.
   We develop a novel full-waveform inversion method for acoustic and 
   elastic waves using a total generalized p-variation regularization 
   scheme to address these challenging problems. We decompose the 
   full-waveform inversion into two subproblems and solve these two 
   minimization subproblems using an alternating-direction minimization 
   strategy. One important advantage of the total generalized p-variation 
   regularization scheme is that it can simultaneously reconstruct sharp 
   interfaces and smooth background variations of geophysical parameters.  
   Such capability can also effectively suppress the noises in 
   source-encoded inversion and sparse-data inversion, or inversion of 
   noisy data. We demonstrate the advantages of our new full-waveform 
   inversion algorithm using a checkerboard model, a modified elastic 
   SEG/EAGE overthrust model, and a land field seismic dataset. Our 
   results of synthetic and field seismic data demonstrate  that our new 
   method reconstructs both smooth background variations and sharp 
   interfaces of subsurface geophysical properties accurately, reduces 
   the inversion artifacts caused by source encoding, noisy data or 
   insufficient data coverage effectively, and provides a useful tool for 
   accurate and reliable inversion of field seismic data.

\end{abstract}

\section{Introduction}

Full-waveform inversion (FWI) attempts to reconstruct subsurface medium 
properties by iteratively minimizing the difference between synthetic and 
observed data \cite[]{Tarantola_1984,Tarantola_1986,Mora_1987,Mora_1988}.  
As self-explained by its name, FWI uses full wavefield information, 
including both the amplitude and the traveltime of seismic signals, to 
invert for subsurface medium properties. 

Theoretically, FWI should be the most accurate inversion method for 
subsurface model building. However, numerous studies have shown that FWI 
is a highly nonlinear, ill-posed inverse problem 
\cite[e.g.,][]{Luo_Schuster_1991,Virieux_Operto_2009}. The misfit 
function of FWI may have numerous local minima. This characteristics of 
FWI usually leads to unsuccessful inversions and unreliable results when  
the initial model used in FWI is far away from the true model. This 
problem is often true in practical applications, particularly when the 
subsurface geology is complicated and/or the acquired seismic data are 
sparse and noisy. 

Given the difficulties in applying the conventional FWI to practical 
problems, numerous studies have been conducted to seek a practically 
applicable and reliable inversion scheme, and continuous efforts is one 
of the most active research areas in the geophysical community.
These studies are mostly in three categories.

The first category is employing a more convex function as the misfit 
function of FWI, rather than using the traditional $\ell_2$-norm waveform 
difference. The misfit function in the conventional FWI relies on the 
absolute waveform matching between synthetic data and observed 
data, and cycle skipping issue can easily 
occur if there is a significant phase and/or amplitude difference.  
Studies of seeking more convex function include the wave-equation 
traveltime based misfit function 
\cite[]{Luo_Schuster_1991,Luo_etal_2016}, the envelope-based misfit 
function \cite[]{Wu_etal_2014,Chi_etal_2014}, the 
instantaneous-phase-based misfit function 
\cite[]{Bozdag_etal_2011,Jiao_etal_2015}, the correlation-based misfit 
function 
\cite[]{vanLeeuwen_Mulder_2010,Luo_Sava_2011,Chi_etal_2015,Choi_Alkhalifah_2016}, 
the deconvolution-based misfit function 
\cite[]{Luo_Sava_2011,Warner_Guasch_2016}, the dynamic-time-warping-based 
misfit function \cite[]{Ma_Hale_2013}, the Huber-norm misfit function 
\cite[]{Guitton_Symes_2003,Ha_etal_2009}, and the optimal transport 
approach 
\cite[]{Metivier_etal_2016b,Metivier_etal_2016a,Yang_etal_2017,Yang_etal_2018}, 
etc.  These misfit functions are generally more convex with respect to 
model perturbations, and therefore less prone to the cycle skipping 
issue. 

The second category is conducting FWI in a Gauss-Newton or quasi-Newton 
minimization framework. Conventional FWIs are essentially based on the 
first-order perturbation theory. The second-order terms (or equivalently 
the Hessian) can be important in FWI to obtain accurate and 
high-resolution model estimations 
\cite[]{Pratt_etal_1998,Tang_etal_2010,Fichtner_Trampert_2011}. The 
Hessian can also be important in multi-parameter inversion 
\cite[]{Pan_etal_2016}. A direct computation of the inverse Hessian can 
be prohibitively expensive using current computational architecture 
\cite[]{Fichtner_Trampert_2011}, and therefore several approximation 
methods were developed, including the limited-memory BFGS (L-BFGS) scheme 
\cite[e.g.,][]{Nocedal_Wright_2006}, the truncated Newton method 
\cite[]{Metivier_etal_2013,Metivier_etal_2017}, the quasi-Newton method 
with projected Hessian \cite[]{Ma_Hale_2013}, combined Newton and 
conjugate gradient schemes \cite[]{Epan_etal_2008}, and 
pseudo-Gauss-Newton scheme \cite[]{Pan_etal_2015}, etc. Several 
preconditioners were also developed to approximated the Hessian 
\cite[e.g.,][]{Zhang_etal_2012b}. 

The third category is employing regularization to accelerate the 
convergence. Applying regularization to geophysical inverse problems has 
a fairly long history \cite[]{Zhdanov_2002}, and the Tikhonov 
regularization is probably the most frequently used regularization scheme 
\cite[]{Tikhonov_etal_1995,Asna_etal_2013}. 

The Tikhonov regularization tends to produce smooth models. In image 
analysis and processing, the total variation (TV) method 
\cite[]{Rudin_etal_1992} was developed to promote sharp interfaces in the 
image. The method has been applied in geophysical inverse problems to 
promote the sharp interfaces of layers \cite[]{Anagaw_2011}.  \cite{Guitton_2012} designed a blocky regularization scheme for FWI. He employed a $\ell_1$-norm and a Cauchy function to enforce blockiness of the model. The blocky regularization is essentially edge-promoting: it can 
enforce sharp interfaces even at locations where interfaces probably do 
not exist, a feature in accordance with minimizing the total variation of 
a model. Therefore, the inversion results tend to be piecewise constant.  
This feature is also an inherent limitation of the first-order TV regularization. 

The first-order TV regularization can be less efficient when the data is noisy or the model is complex. \cite{Lin_Huang_2014} developed a novel regularized FWI 
based on a modified TV (MTV) regularization scheme. They decomposed FWI 
into two interlacing inversion problems: one is the conventional FWI with 
a Tikhonov regularization term, and the other is an image denoising 
problem using the first-order TV method. The second problem is solved 
efficiently with the split-Bregman iteration, a technique that is proven 
to be especially suitable to produce clean and accurate TV denoising 
results \cite[]{Goldstein_Osher_2009}. The model parameter and the 
auxiliary model parameter are updated in an alternating-direction 
approach, resulting in an efficient first-order TV regularized FWI.  
However, this regularization scheme is based on the first-order TV.  
Although it avoids spike noises and provides more reliable inversion 
results, it still cannot avoid the inherent disadvantage of the 
first-order TV, i.e., the staircase artifacts in inversion results.

\cite{Esser_etal_2016} developed an asymmetric TV regularized FWI. The 
asymmetric TV, or hinge-loss constraint TV as they called, penalizes the 
model discontinuities only in the vertical direction. They show that this 
asymmetric TV regularized FWI can facilitate the automatic delineation of 
high-contrast medium parameter anomalies such as salt bodies, as well as depth structures.

The Tikhonov and TV regularizations are not the only regularization 
schemes used in full-waveform inversion. \cite{Guitton_etal_2012} 
developed a preconditioned FWI using the geological information derived 
from the migration image. The preconditioner is solved via estimating the 
local dip information from the migration image followed by a directional 
Laplacian filter \cite[]{Hale_2007}. \cite{Lewis_etal_2014} developed 
a similar approach based on the anisotropic diffusion filtering, where 
the structural tensor is estimated from a migration image. In their 
approach, the gradient is preconditioned based on the geological 
structures derived from structural images, thus becomes geologically 
meaningful. In the case of sparse sources and therefore insufficient 
subsurface model coverage, such preconditioning can facilitate faster 
convergence towards meaningful inversion results. \cite{Xue_etal_2017} 
developed a similar regularization scheme using sparsity promotion in the 
seislet domain. 

The total generalized variation 
\cite[]{Bredies_2010,Knoll_etal_2011,Zhang_etal_2016} is a technique that 
incorporates the higher-order total variations in image reconstruction.  
It is usually applied to various problems in its second-order form with 
a $\ell_1$-norm framework. The total generalized variation can 
reconstruct both the sharp interfaces and smooth variations of an image, 
leading to fewer artifacts compared with the first-order total-variation 
approach. 

In medical imaging, the $\ell_p$-norm ($0<p<1$) compressive-sensing 
\cite[]{Chartrand_2009,Chartrand_2013} is surprisingly effective to 
reconstruct images using extremely sparse measurements compared with the  
$\ell_1$-norm compressive sensing techniques 
\cite[]{Candes_etal_2006,Donoho_2006,Candes_Wakin_2008,Goldstein_Osher_2009}.  

We develop a novel full-waveform inversion method for acoustic and 
elastic waves using a total generalized p-variation regularization scheme 
(TGPV-FWI). We combine the advantages of $\ell_p$-norm compressive 
sensing techniques and the second-order total variation to reconstruct 
both sharp interfaces and smooth background variations of geophysical 
parameters. We formulate our TGPV-FWI in an alternating-direction 
minimization framework. We decompose the TGPV-FWI into two interlacing 
minimization problems. The first minimization problem is a conventional 
FWI problem with the Tikhonov regularization, whereas the second 
minimization problem is a second-order compressive sensing model 
denoising problem. Once the second subproblem is solved correctly, we 
only need to add the difference between the FWI model in the last 
inversion and its ``denoising'' result from the second minimization 
problem to the gradient in the current inversion step. This strategy 
forces the inversion to evolve towards the TGPV model that is less prone 
to artifacts when seismic data are sparse, noisy or when using dynamic 
source encoding. With this alternating minimization framework, the FWI 
eventually converges to a more accurate and reliable result.

Our paper is organized as follows. In the Methodology section, we present 
the formulation of our new TGPV-FWI. We present the detailed algorithms 
for solving the second subproblem and our new TGPV-FWI in two appendices.  
In the Numerical Results section, we give three numerical examples, 
including two synthetic data examples and one field data example, to 
verify the advantages of our new TGPV-FWI over FWI with the conventional 
regularization schemes including the Tikhonov and TV regularizations. We 
give our findings in the Conclusions section. 

\section{Methodology}

We formulate our new full-waveform inversion with the total generalized p-variation regularization (TGPV-FWI) as the following minimization problem:
\begin{equation}\label{eq:fwilp}
\bfm^* = \argmin\limits_{\bfm} \left\{\frac{1}{2} \|\mathbf{d}-f(\bfm)\|_2^2 + \lambda \mathcal{T}_p (\bfm) \right\},
\end{equation}
where $\mathbf{d}$ is an observed dataset, $f(\bfm)$ is a synthetic 
dataset, $\bfm$ is the medium parameter model to be inverted, and 
$\mathcal{T}_p(\bfm)$ is the TGPV regularization term. 

Unlike the conventional regularization schemes such as the Tikhonov and 
TV regularizations that are defined using the $\ell_2$-norm and 
$\ell_1$-norm, respectively, the TGPV
regularization term $\mathcal{T}_p (\bfm)$ is defined through a $\ell_p$-norm minimization problem:
\begin{equation}\label{eq:reg_tgpv}
\mathcal{T}_p(\bfm) = \argmin\limits_{\bfw} \left\{\alpha_0 \|\nabla \bfm - \bfw \|_p^p + \alpha_1 \|\varepsilon(\bfw)\|_p^p\right\},  \qquad (0<p<1)
\end{equation}
where $\bfw=(\bfw_x,\bfw_y)$ is an auxiliary vector variable. The $\ell_p$-norm (or more precisely, the $\ell_p$ quasi-norm) with $0<p<1$ of a number $\mathbf{x} \in \mathbb{R}^n$ is defined as
\begin{equation}
\|\mathbf{x}\|_p = \left(\sum_{i=1}^{n} |x_i|^p\right)^{1/p}. 
\end{equation}

In the 2D case, the gradient of model $\bfm$ is
\begin{equation}
\nabla \bfm = \begin{bmatrix}
\nabla_x \bfm \\
\nabla_y \bfm 
\end{bmatrix},
\end{equation}
and the symmetric gradient $\varepsilon(\cdot)$ reads
\begin{equation}
\varepsilon(\bfw) = \begin{bmatrix}
\nabla_x \bfw_x & \frac{1}{2}(\nabla_x \bfw_y + \nabla_y \bfw_x) \\
\frac{1}{2}(\nabla_x \bfw_y + \nabla_y \bfw_x) & \nabla_y \bfw_y 
\end{bmatrix}.
\end{equation}


The minimization problem defining the TGPV regularization term is based 
on the total generalized variation in image processing 
\cite[]{Bredies_2010,Knoll_etal_2011,Zhang_etal_2016}. The advantage of 
the total generalized variation is that it penalizes both the first-order 
gradient and high-order gradients of an image or a model, and thus it can 
effectively avoid the staircase effect compared with the first-order 
total variation. The advantage of the $\ell_p$-norm penalty over the 
traditional $\ell_1$-norm penalty is observed from the results in 
compressive-sensing medical imaging \cite[]{Chartrand_2009}, in which it 
can handle even sparse data than the $\ell_1$-norm for accurate image 
reconstruction. The $\ell_p$-norm regularization term leads to 
a nonconvex minimization problem. This introduces difficulties in solving 
the FWI problem. In the context of compressive sensing, 
\cite{Chartrand_2009} shows that there exist techniques to efficiently 
solve the $\ell_p$-norm nonconvex problem. We solve our TGPV-FWI using 
the similar approaches. 


Unlike the conventional TV-regularized FWI, it is impossible to 
solve the minimization problem in eq.~\eqref{eq:fwilp} directly. To solve 
our TGPV-FWI efficiently, we reformulate the minimization in 
eq.~\eqref{eq:fwilp} into an alternating-direction minimization problem 
as by \cite{Lin_Huang_2014}:
\begin{equation} \label{eq:dualvar}
\{\bfm^*,\bfu^*\} =  \argmin\limits_{\bfm,\bfu} \left\{\frac{1}{2} 
\|\mathbf{d}-f(\bfm)\|_2^2 + \lambda_1 \|\bfm-\bfu\|_2^2 + \lambda_2 
\mathcal{T}_p(\bfu)\right\},
\end{equation}
which can be decomposed into two interlacing minimization problems:
\begin{subequations}
\begin{align}
\bfm^{(l+1)} & = \argmin\limits_{\bfm} \left\{\frac{1}{2}\|\mathbf{d}-f(\bfm)\|_2^2 + \lambda_1 \|\bfm-\bfu^{(l)}\|_2^2\right\},  \label{eq:fwi_m} \\
\bfu^{(l+1)} & = \argmin\limits_{\bfu} \left\{\frac{1}{2}\|\bfm^{(l+1)}-\bfu\|_2^2 + \lambda_2 \mathcal{T}_p (\bfu)\right\}, \label{eq:fwi_u}
\end{align}
\end{subequations}
where $l$ is the iteration number in the FWI. That is, the medium 
parameter $\bfm$ and the auxiliary variable $\bfu$ are updated 
alternatively in the inversion procedure. The auxiliary variable serves 
as the prior information in the TGPV-FWI, and it is also updated through 
iterations. 

The first minimization problem in eq.~\eqref{eq:fwi_m} is a conventional 
FWI problem with a zeroth-order Tikhonov regularization term. Numerous methods 
can be adopted to solve this nonlinear minimization problem, such as the 
conjugate-gradient (CG) method and the limited-memory BFGS (L-BFGS) 
method \cite[]{Nocedal_Wright_2006}, etc. Misfit functions other than the 
simple $\ell_2$-norm-squared waveform difference in eq.~\eqref{eq:fwilp}, as 
discussed in the Introduction, can also be applied to the first 
subproblem. We adopt the split-Bregman iteration methodology \cite[]{Goldstein_Osher_2009} to solve the second minimization problem. 

We show detailed algorithms for solving the second minimization problem in the 2D and 3D cases in Appendices A and B, respectively. 

Our TGPV-FWI contains several parameters to be adjusted. We describe how 
to select these parameters in the following.

The positive parameter $\lambda_1$ controls the ``strength'' of the regularization term in the first minimization problem, and a larger $\lambda_1$ leads to a stronger TGPV regularization. In applications, we use the following simple rule to determine $\lambda_1$: 
\begin{equation}
\lambda_1 (\mathbf{m}) = \gamma \frac{\|\nabla \chi(\mathbf{m})\|_2}{\|\mathbf{m}-\mathbf{u}\|_2}, 
\label{eq:regl}
\end{equation}
where $\gamma$ is a scaling factor that can be tuned during iterations, 
and $\nabla \chi(\mathbf{m})$ represents the gradient of the misfit 
function w.r.t. some model parameter at certain iteration. $\|\cdot\|_2$ 
represents the $\ell_2$-norm  of a quantity. For different applications, the 
value of $\gamma$ can vary. Normally, a value between 0.05 to 0.5 should 
be suitable for most applications, and larger values of $\gamma$ are not 
encouraged. In our numerical tests, we determine different $\lambda_1$s 
for different model parameters (e.g., $V_p$ or $V_s$ in elastic-waveform 
inversion) in each iteration using eq.~\eqref{eq:regl}. We find 
$\gamma=0.1$ can serve as a suitable value.

The method to determine the regularization parameter $\lambda_1$ is 
slightly different from that in such as \cite{Lin_Huang_2014}. The 
general principle, however, is in common: the regularization term and the 
data misfit term should be in decent balance to avoid excessive or 
insufficient regularization. 

There are six parameters in the second minimization problem in 
eq.~\eqref{eq:fwi2sub}: $\alpha_0$, $\alpha_1$, $\eta_0$, $\eta_1$, $\mu$ 
and the norm $p$. The positive parameter $\lambda_2$ (or equivalently 
$1/\mu$) controls the ``strength'' of denoising in the second 
minimization problem, and a larger $\lambda_2$ (or equivalently a smaller 
$\mu$) leads to a stronger smoothing. Note that the smoothing is not 
a spatial smoothing in a usual sense, such as that of Gaussian spatial 
filtering; it is a smoothing in the total-variation sense. In extreme 
cases, $\mu \rightarrow 0$ can smooth out all the features of a model, 
while $\mu \rightarrow + \infty$ leaves the model unchanged. The most 
important feature of this smoothing is that it simultaneously preserves 
sharp interfaces and smooth variations of the model in the framework of 
the high-order total variation.

For the other five parameters, our extensive tests show that for most of, 
if not all, practical applications, $p=0.5$, $\alpha_0=1$, $\alpha_1=1$ 
or 2, and $\eta_0=2\mu$ and $\eta_1 = \frac{\alpha_1}{\alpha_0} \eta_0$ 
can solve the second subproblem to produce fairly accurate results.  

Therefore, for the second subproblem, only the parameter $\mu$ should be 
tuned for different FWI problems. This indicates that in our TGPV-FWI, in most cases we only need to tune two parameters, the Tikhonov regularization coefficient $\lambda_1$ and the regularization parameter $\mu$ (or equivalently $\lambda_2$) as described above, leading to an efficient inversion system. 

We summarize our TGPV-FWI algorithm in Appendix C. 

\section{Numerical Results}

In the following, we use Tikhonov-TV to denote the FWI with the Tikhonov regularization, TV-FWI to denote the FWI with the TV regularization, and TGPV-FWI to denote the FWI with our TGPV regularization.

\subsection{Synthetic data example I: Checkerboard model}

Our TGPV-FWI for acoustic and elastic waves can be used for either 
large-scale tomography of the Earth using earthquake data or small-scale 
seismic reflection inversion. For the former case, transmission seismic 
waves are mostly used. To verify the improved inversion accuracy of our 
new TGPV-FWI algorithm using mostly transmission signals, we design 
a checkerboard model with randomly distributed sources within the model.  
The goal of this numerical test is to verify the applicability of our new 
method using mostly transmission signals, rather than using Earthquake 
data.

\begin{figure}
\centering
\subfloat[]{\includegraphics[width=0.485\textwidth]{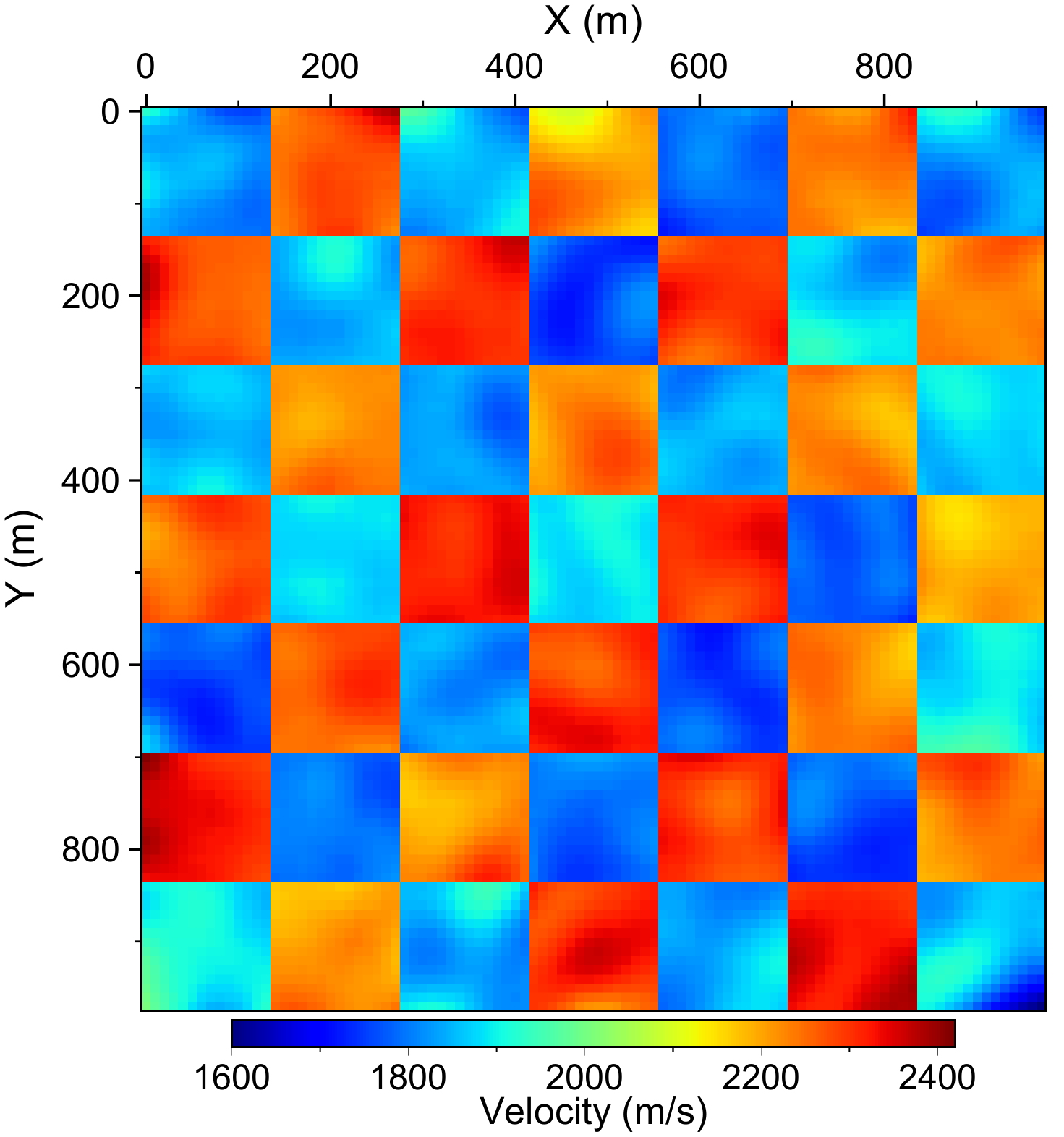}}
\subfloat[]{\includegraphics[width=0.485\textwidth]{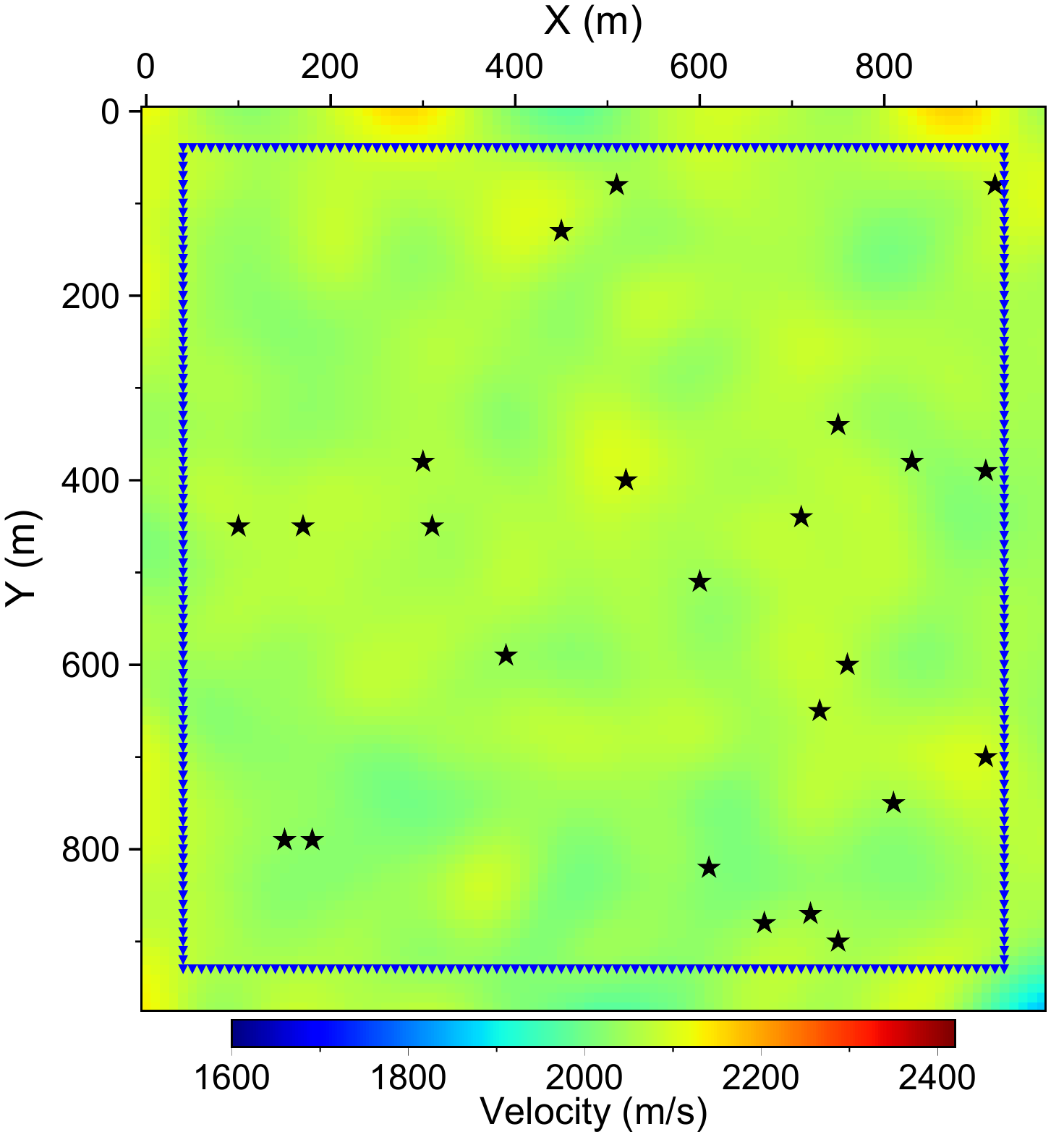}}
\caption{A checkerboard model for FWI tests. (a) The true P-wave velocity 
model and (b) the initial P-wave velocity model for FWI. The black stars 
   in (b) represent 24 randomly-distributed sources, while the blue 
   triangles denote 356 receivers.}
\label{fig:cb}
\end{figure}

The model is defined in a region of 970~m$\times$970~m with a grid 
interval of 10~m in both directions.  This checkerboard model is composed 
of a smoothly-varying background and checkerboard velocity perturbations, 
as shown in Fig.~\ref{fig:cb}a. The velocity perturbation is 
approximately 10 percent throughout the model, and the perturbation can 
be as high as approximately 20 percent in some regions. We use a smooth 
velocity model without the large-contrast velocity perturbations as 
displayed in Fig.~\ref{fig:cb}b as the initial velocity model for FWI.

We place a total of 24 sources randomly distributed throughout the model 
(black stars in Fig.~\ref{fig:cb}b) and a total of 356 receivers near the 
boundaries of the model (blue triangles in Fig.~\ref{fig:cb}b). 

The source wavelet is a Ricker wavelet with a center frequency of 20~Hz. 
We employ the L-BFGS inversion framework and terminate the inversion 
after 150 iterations. We conduct three tests using the Tikhonov-FWI, TV-FWI and TGPV-FWI.

\begin{figure}
\centering
\subfloat[]{\includegraphics[width=0.485\textwidth]{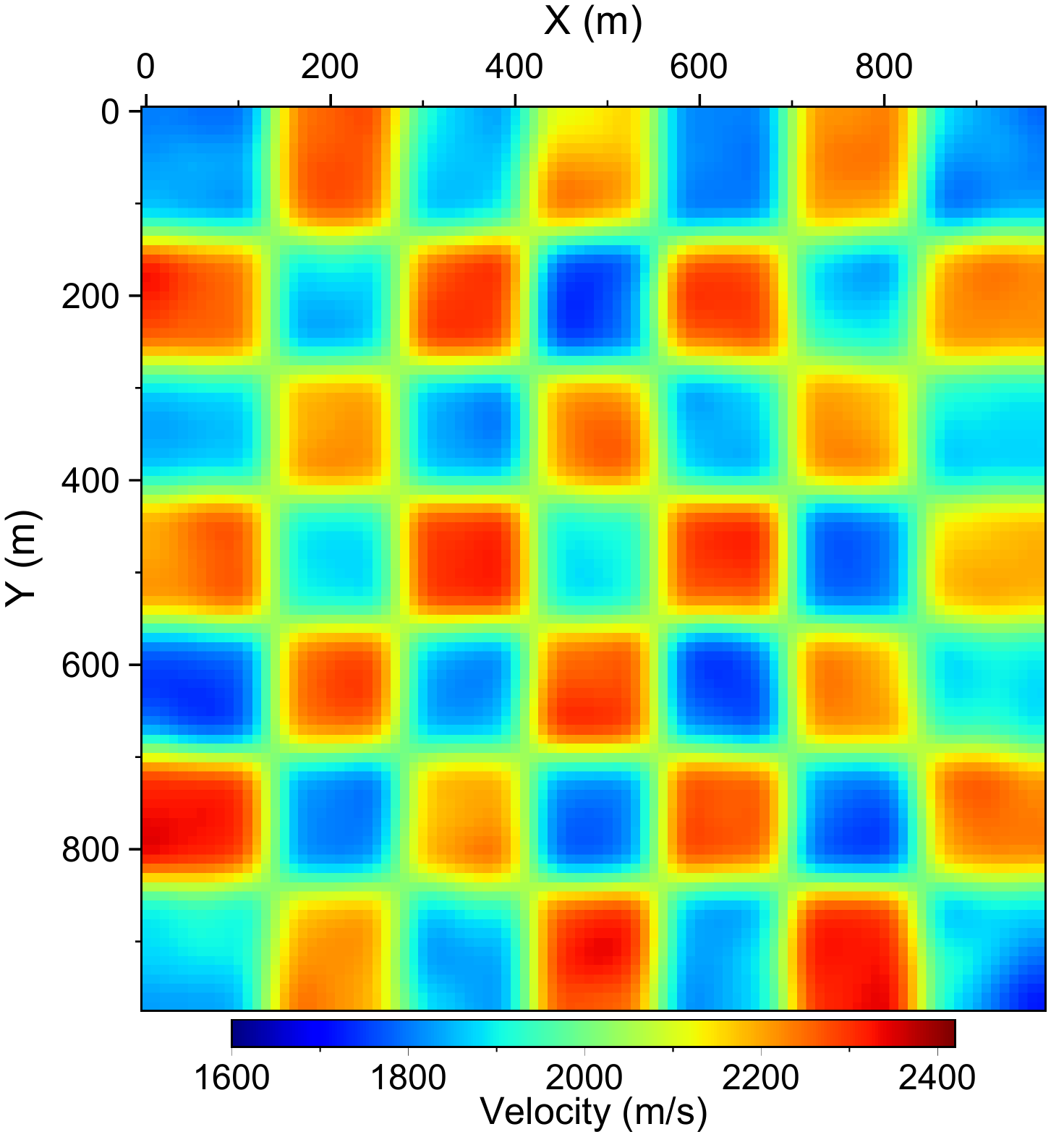}}
\subfloat[]{\includegraphics[width=0.485\textwidth]{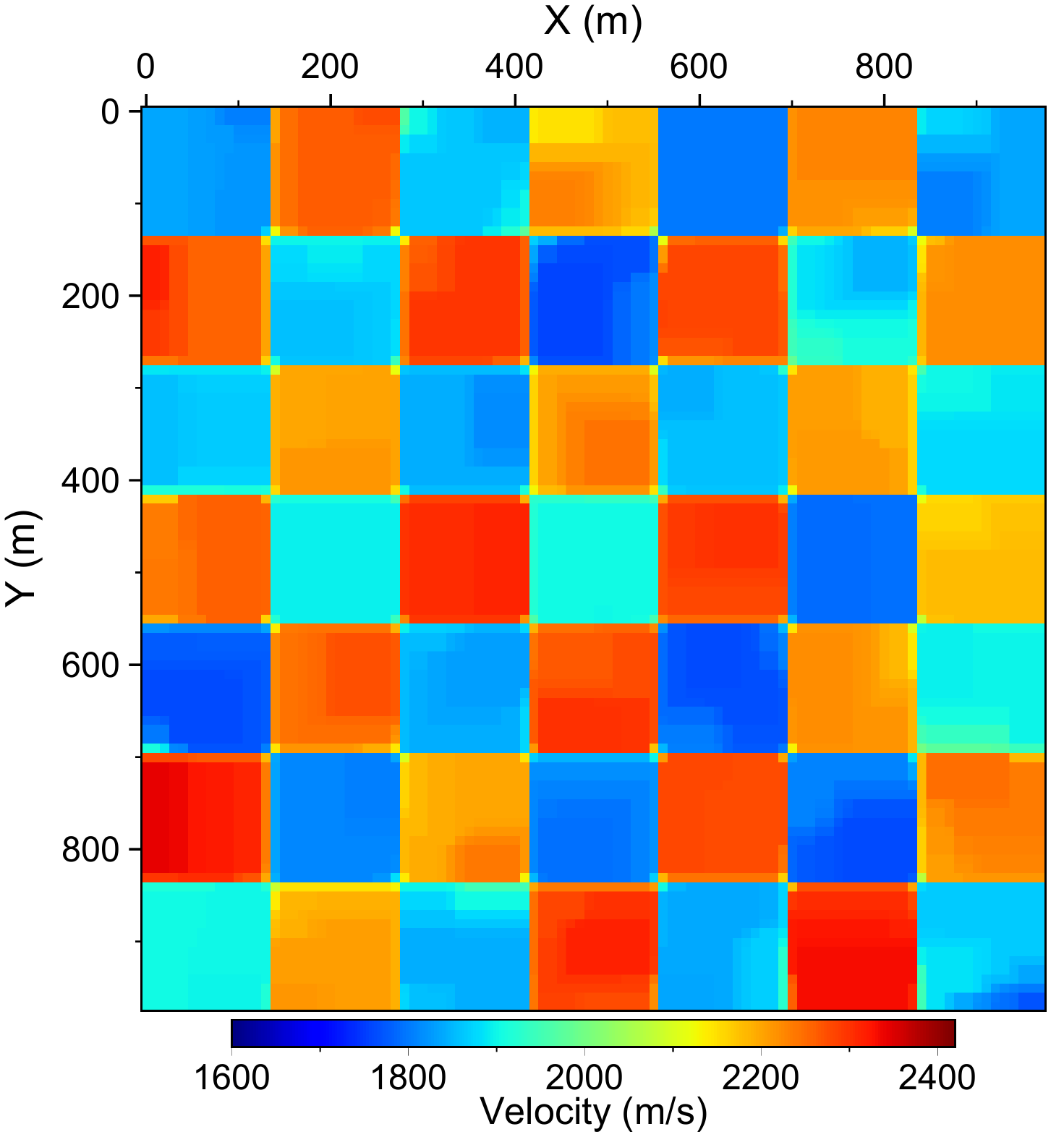}} \\
\subfloat[]{\includegraphics[width=0.485\textwidth]{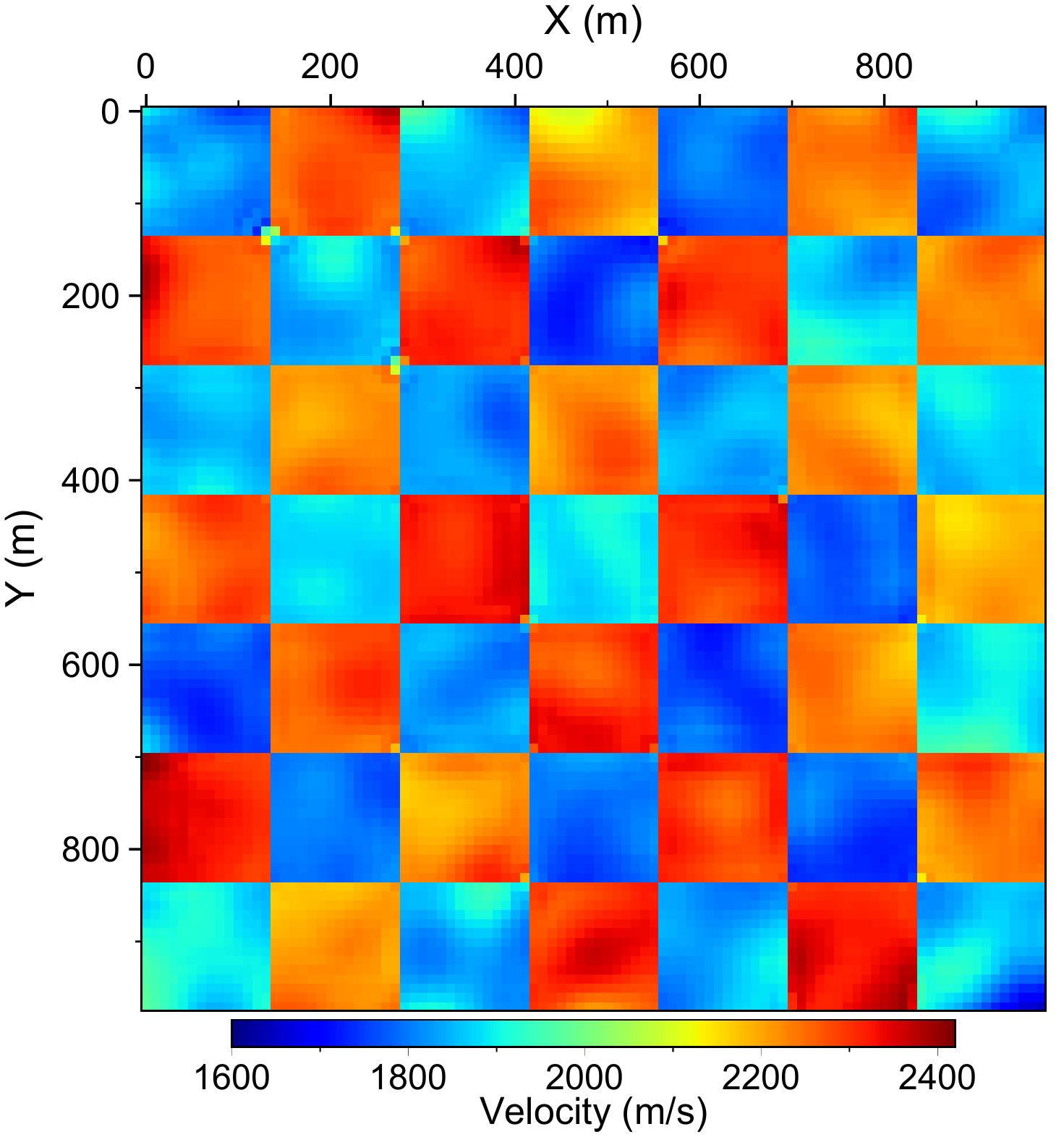}}
\caption{Inverted checkerboard velocity models after 150 iterations using 
(a) Tikhonov-FWI (b) TV-FWI and (c) TGPV-FWI.}
\label{fig:cb_invt}
\end{figure}

Figs.~\ref{fig:cb_invt}a, b and c show the FWI inversion results obtained 
using the Tikhonov-FWI, TV-FWI and TGPV-FWI, respectively. The Tikhonov-FWI 
produces a smooth checkerboard inversion result with missing sharp 
interfaces of checkerboard blocks. The TV-FWI 
preserves the sharp interfaces of the checkerboard blocks, yet within the 
blocks, the inversion result shows clear the piecewise-constant feature, 
or more commonly, the staircase effect. Our TGPV-FWI accurately 
reconstructs not only the sharp block interfaces, but also  the smooth 
variations inside the blocks.

\begin{figure}
\centering
\subfloat[]{\includegraphics[width=0.33\textwidth]{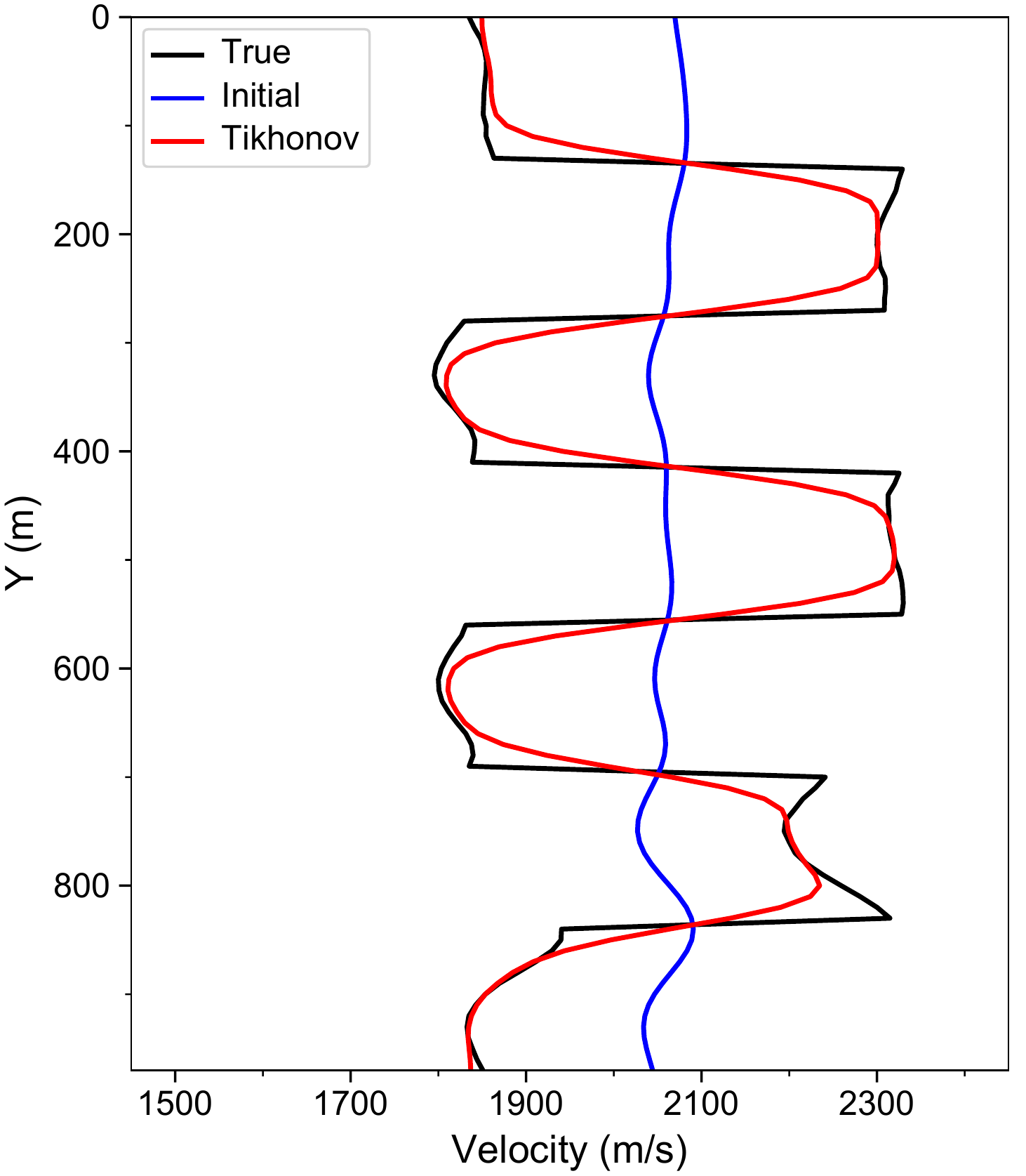}}
\subfloat[]{\includegraphics[width=0.33\textwidth]{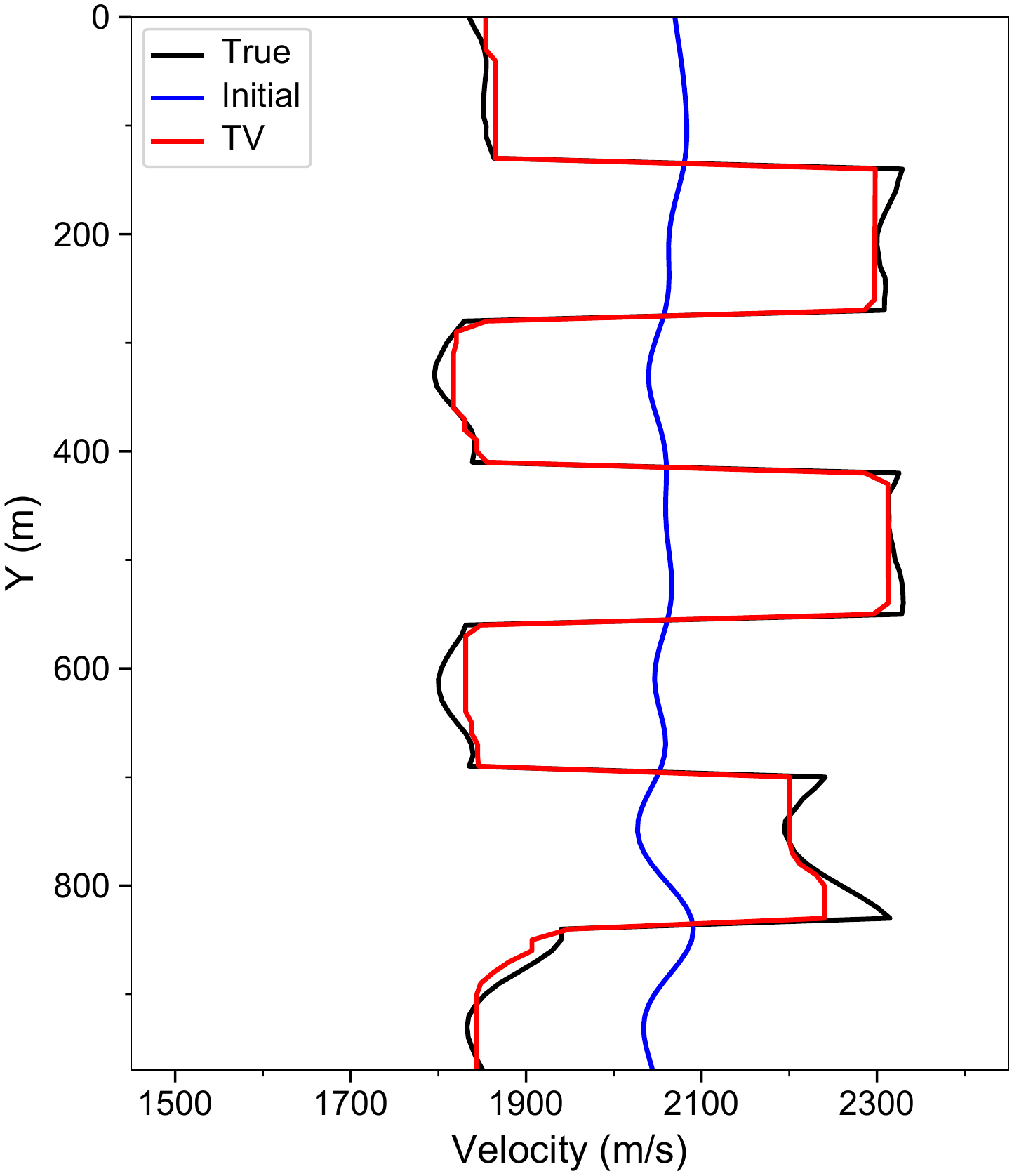}}
\subfloat[]{\includegraphics[width=0.33\textwidth]{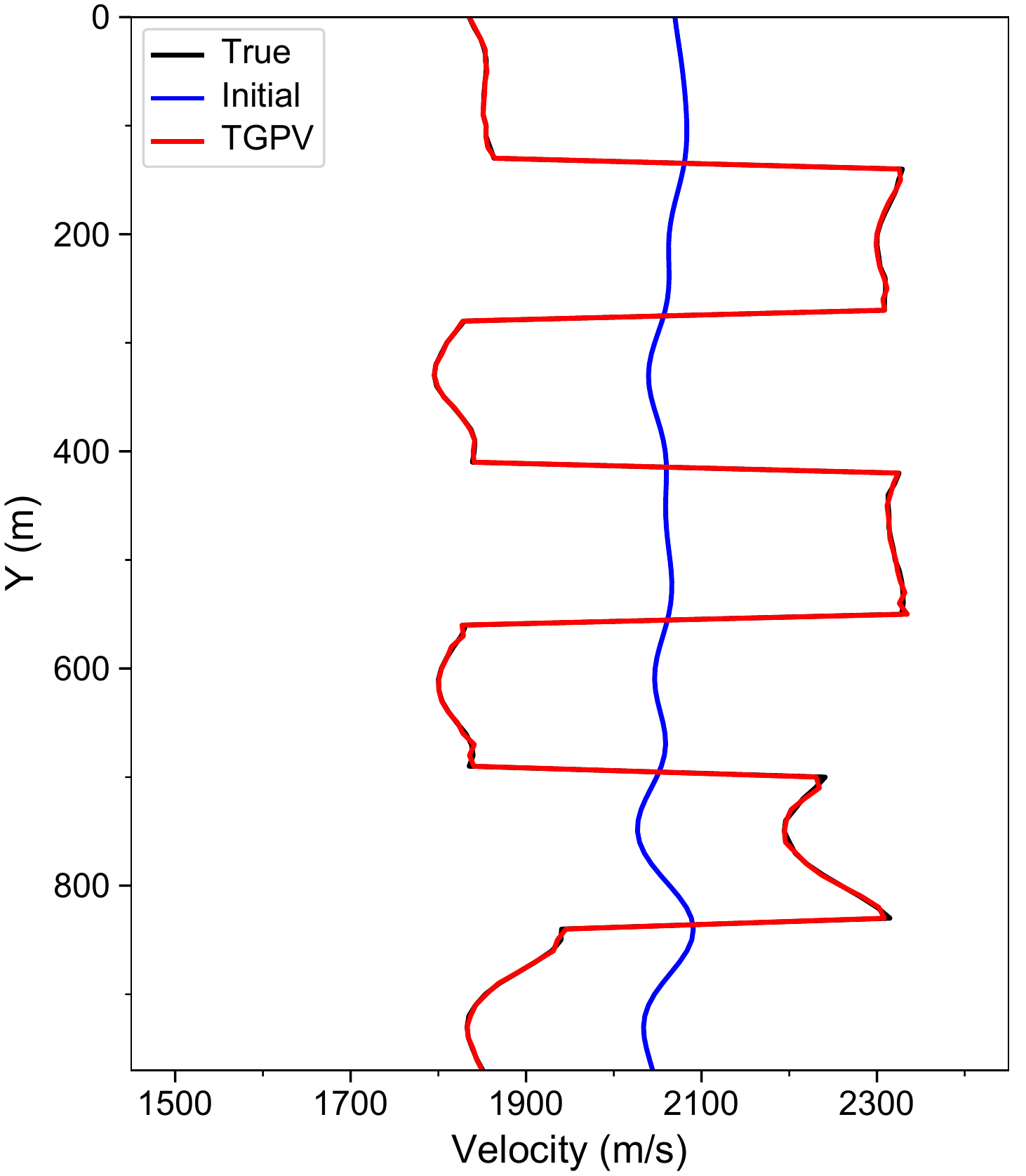}}
\caption{Velocity profiles at $X=500$~m selected for the quantitative 
   comparison among the true, the initial and the inverted model using 
   (a) Tikhonov-FWI (b) TV-FWI and (c) TGPV-FWI.}
\label{fig:cb_profile}
\end{figure}

To quantitatively compare the inversion results in 
Fig.~\ref{fig:cb_invt}, we plot velocity profiles at the position of $X=500$~m from the three inversion results, and show them in 
Figs.~\ref{fig:cb_profile}a, b and c for Figs.~\ref{fig:cb_invt}a, b and 
c, respectively. This comparison further demonstrates the superior 
capability of our TGPV-FWI in reconstructing both the sharp interfaces 
and smooth variations in the checkerboard model. 

\begin{figure}
\centering
\subfloat[]{\includegraphics[width=0.6\textwidth]{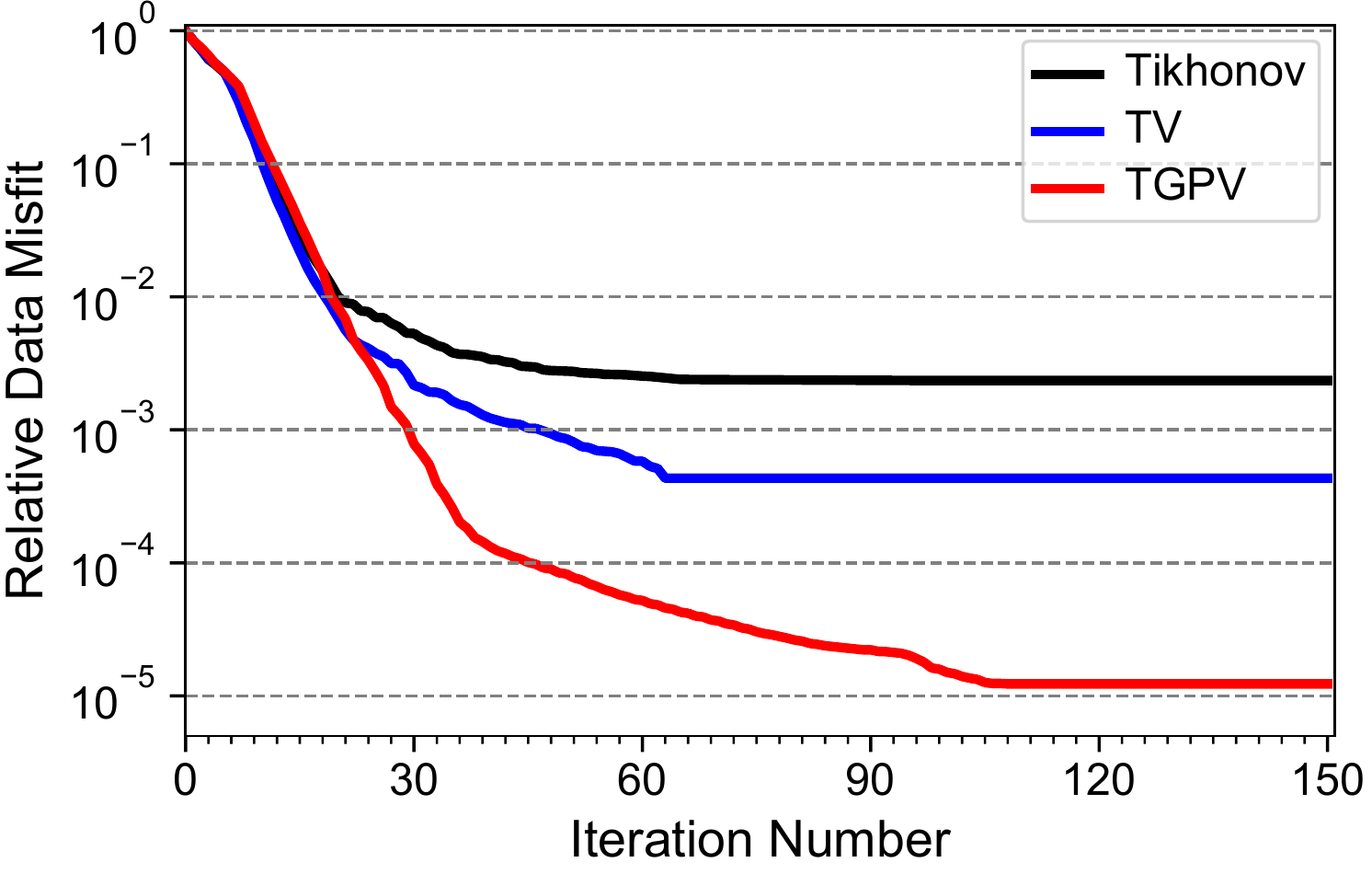}} \\
\subfloat[]{\includegraphics[width=0.6\textwidth]{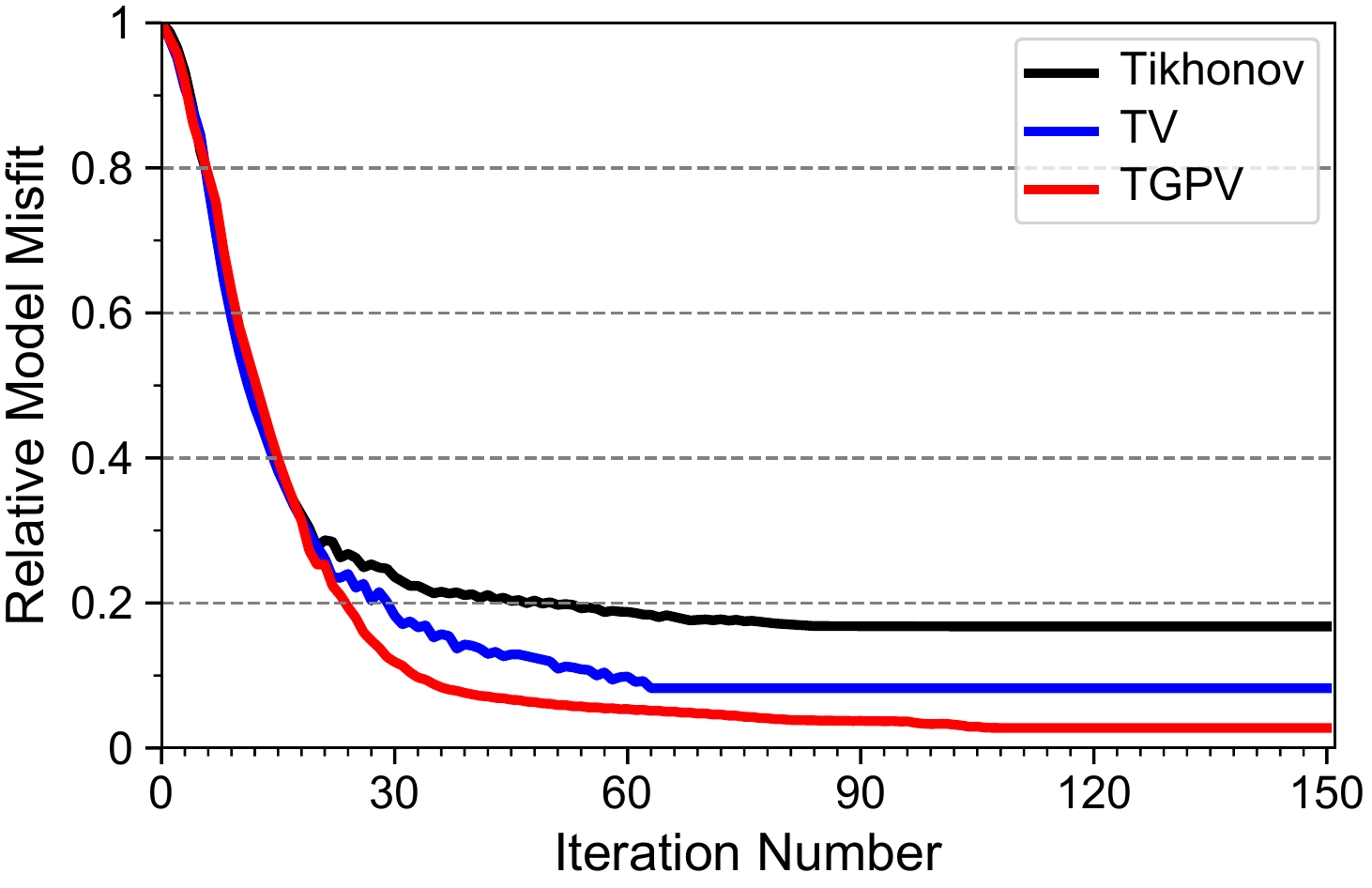}}
\caption{(a) Relative data misfit and (b) relative model misfit over 
   a total of 150 iterations for the Tikhonov-FWI 
   (black), TV-FWI (blue) and the TGPV-FWI 
   (red).}
\label{fig:cb_misfit}
\end{figure}

Fig.~\ref{fig:cb_misfit} shows that the our TGPV-FWI results in smallest  
data and model misfits among the three FWI methods used, demonstrating 
the improved inversion accuracy of our TGPV-FWI.

\subsection{Synthetic data example II: Modified elastic SEG/EAGE 
overthrust model}

We use a modified elastic SEG/EAGE overthrust model to verify the 
efficacy of our TGPV-FWI method for elastic-waveform inversion using 
surface reflection data. In the following numerical tests, we verify our 
TGPV-FWI method for: (1) noise-free data with adequate numbers of  
sources and receivers (a regular source/receiver geometry); (2) 
noise-free, sparse data; (3) noisy data with a regular source/receiver 
geometry; and (4) noise-free data with dynamic source encoding.

\begin{figure}
\centering
\subfloat[]{\includegraphics[width=0.485\textwidth]{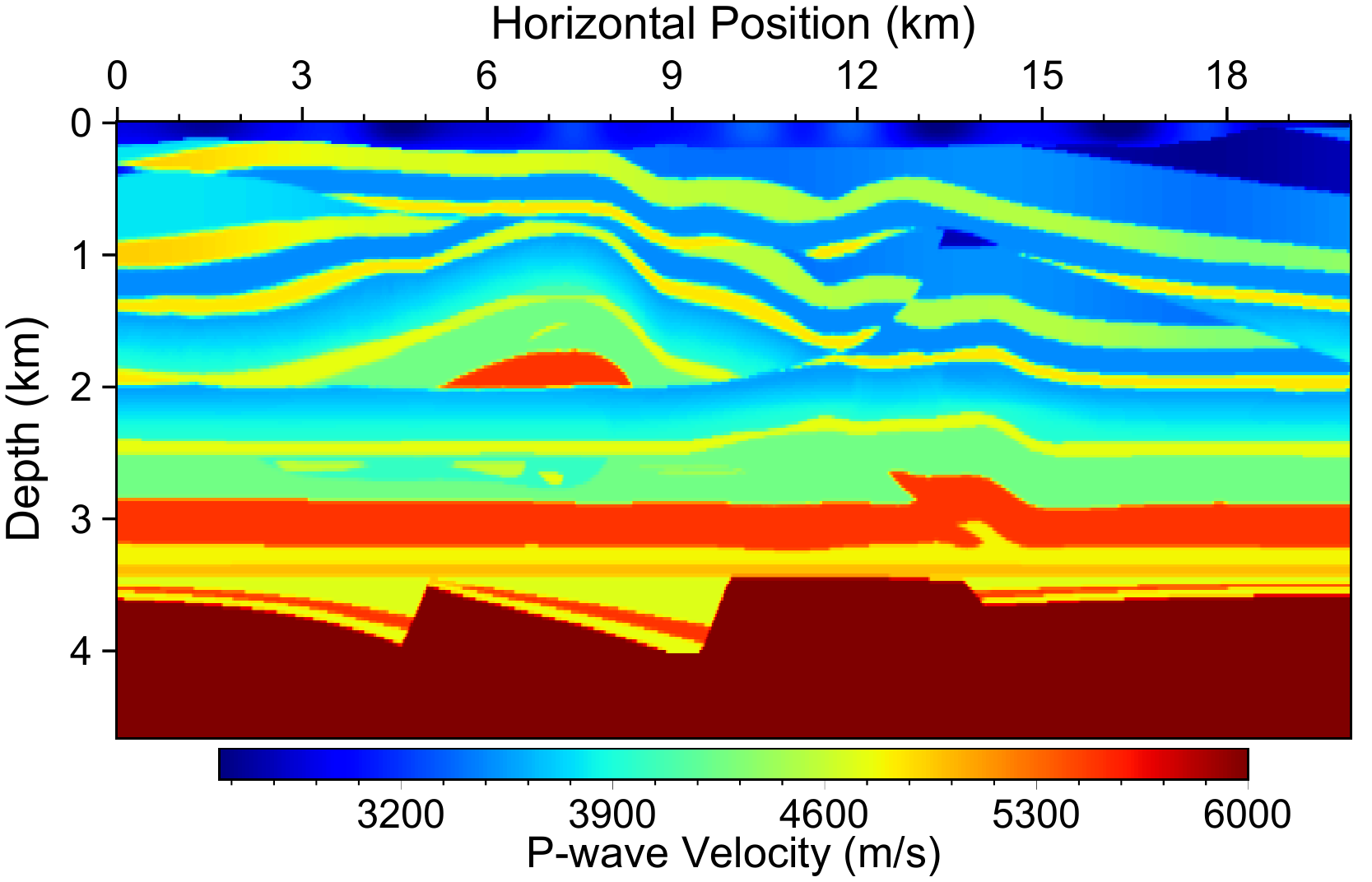}} 
\subfloat[]{\includegraphics[width=0.485\textwidth]{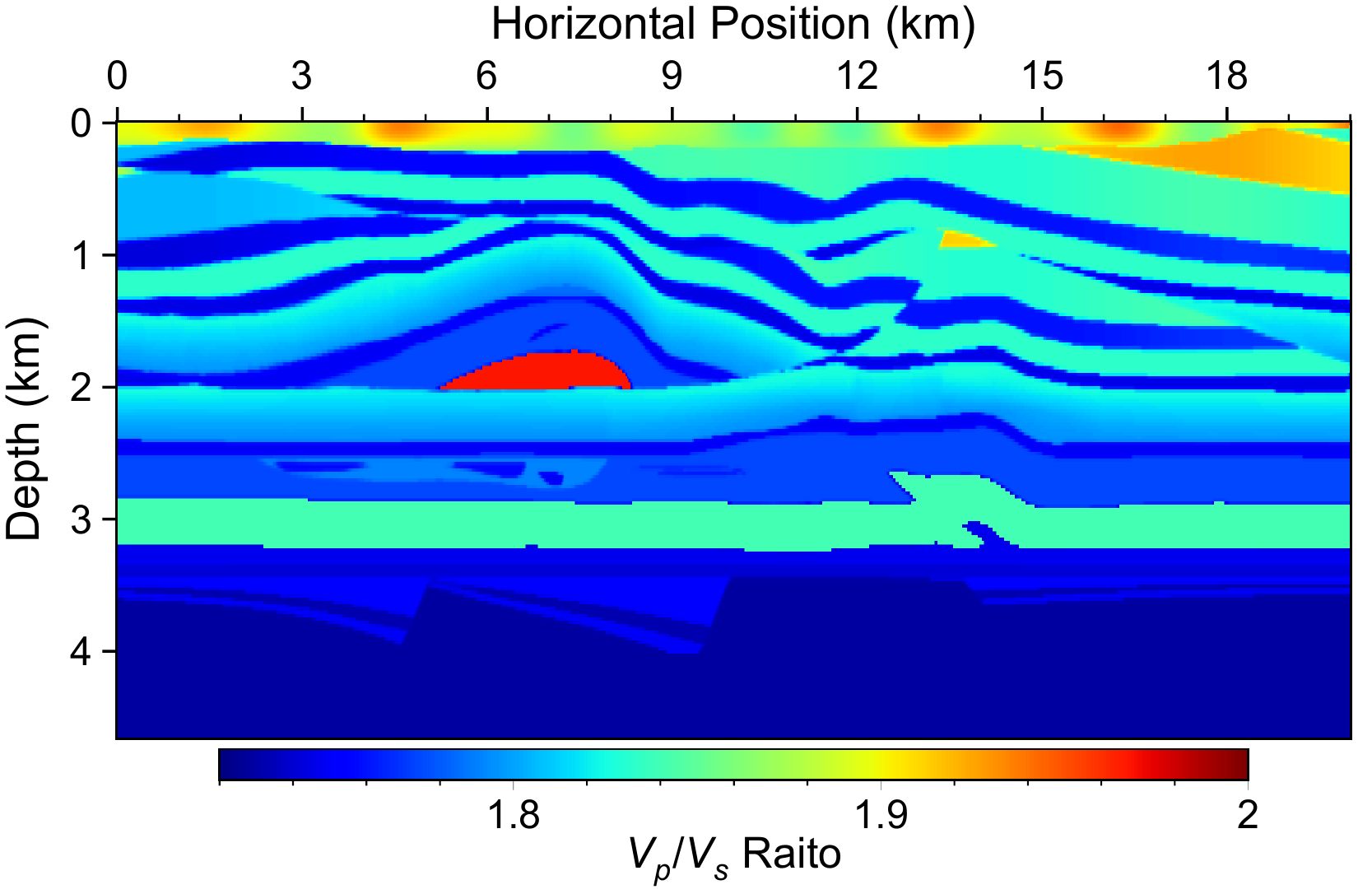}} \\
\subfloat[]{\includegraphics[width=0.485\textwidth]{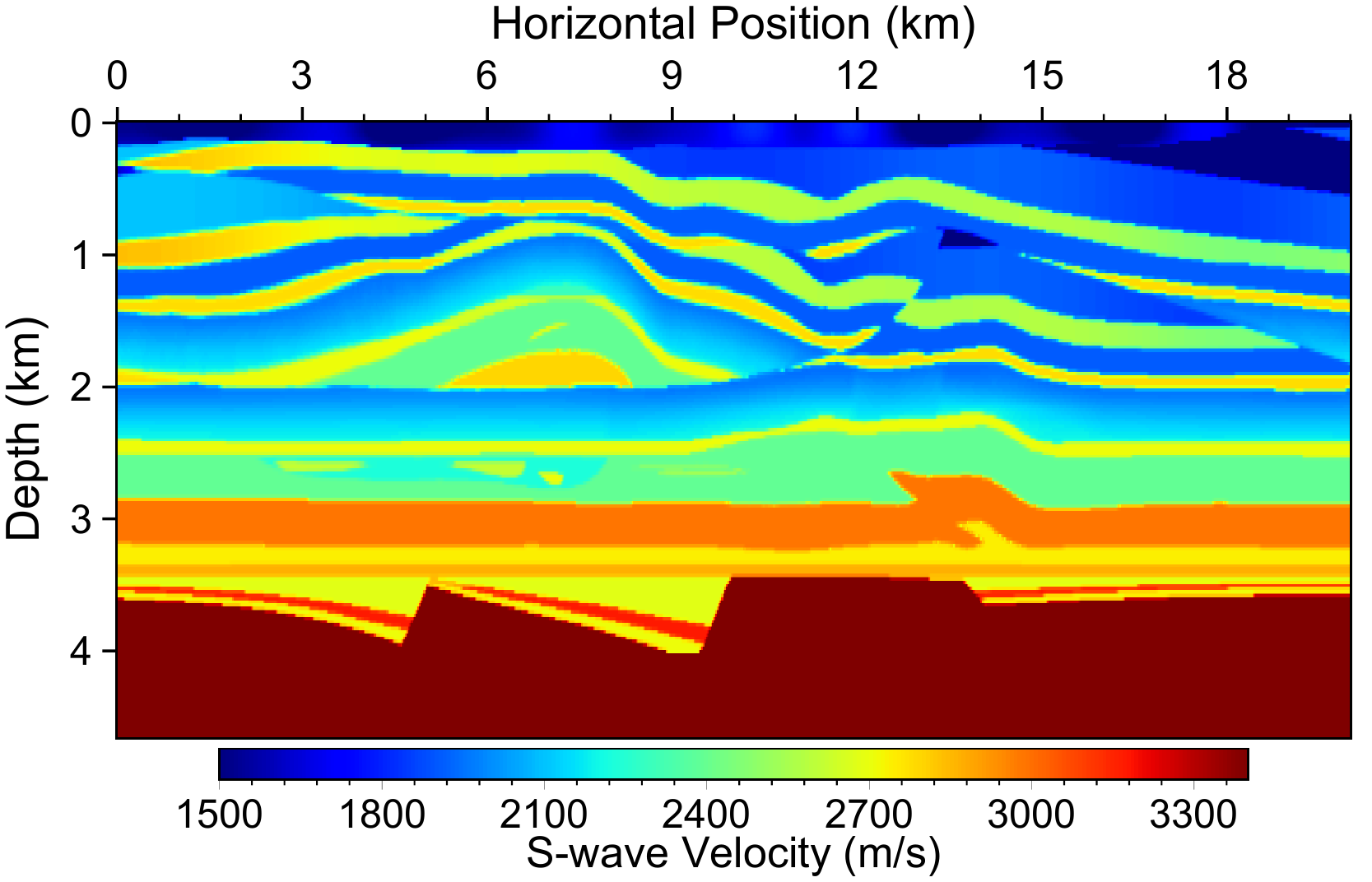}}
\caption{(a) A selected slice from the 3D overthrust P-wave velocity 
model, (b) the spatially varying $V_p/V_s$ ratio, and (c) the resulting 
   S-wave velocity model.}
\label{fig:over_model}
\end{figure}

Fig.~\ref{fig:over_model}a shows a slice of the 3D SEG/EAGE overthrust 
model. The model is composed of 187 grid points in depth and 801 grids 
points in the horizontal direction, with a grid size of 25~m in both 
directions. The initial model (Fig.~\ref{fig:over_model}b) is the 
smoothed version of the true model in Fig.~\ref{fig:over_model}a obtained 
using a strong Gaussian spatial filtering.

The original overthrust model is an acoustic model. For our 
elastic-waveform inversion, we build an S-wave velocity model using 
spatially varying $V_p/V_s$ ratios as shown in 
Fig.~\ref{fig:over_model}b, resulting in an S-wave velocity model in 
Fig.~\ref{fig:over_model}c. Both the P- and S-wave velocity models 
contain sharp interfaces and smoothly varying regions. 

\begin{figure}
\centering
\subfloat[]{\includegraphics[width=0.485\textwidth]{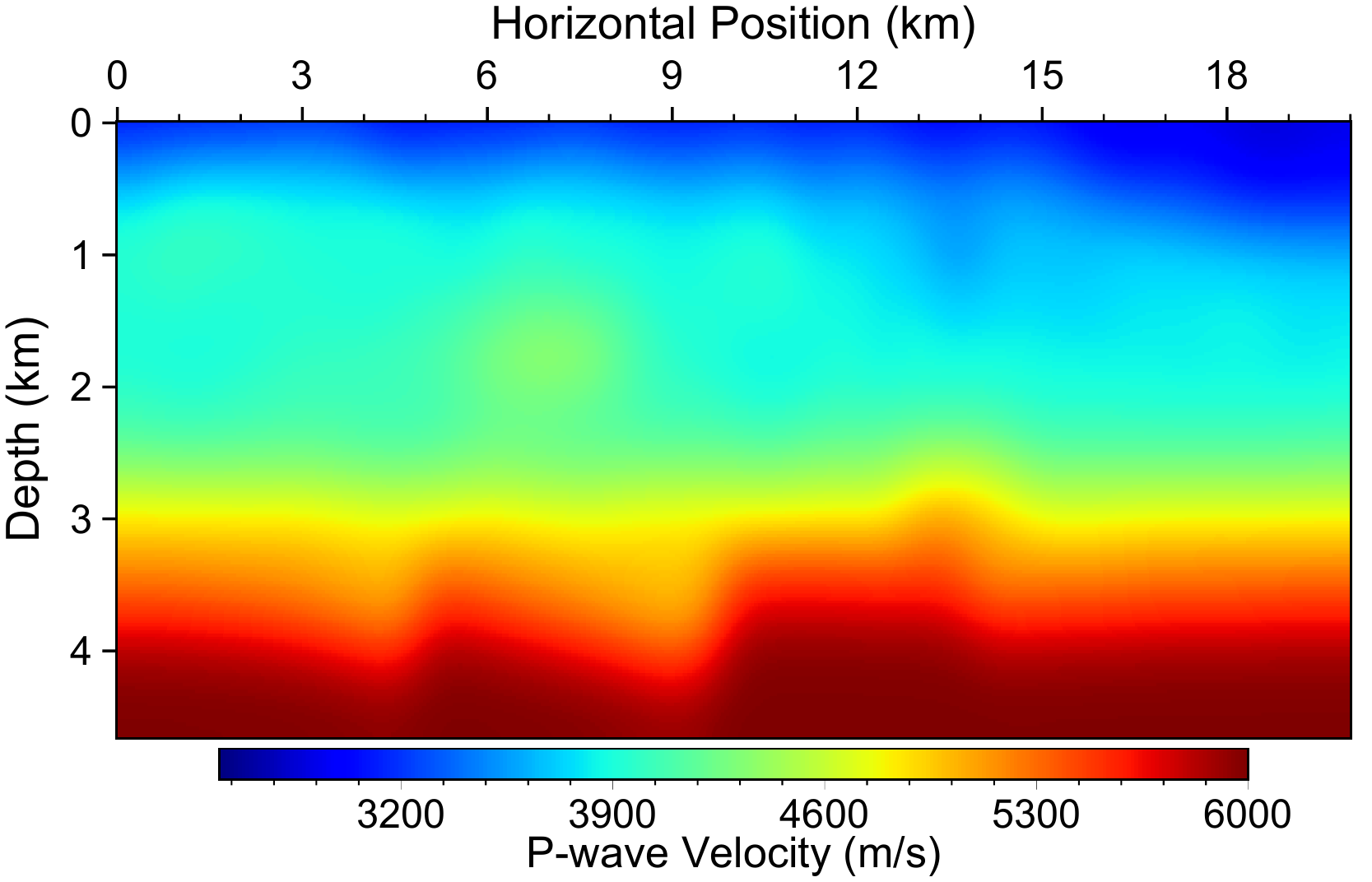}} 
\subfloat[]{\includegraphics[width=0.485\textwidth]{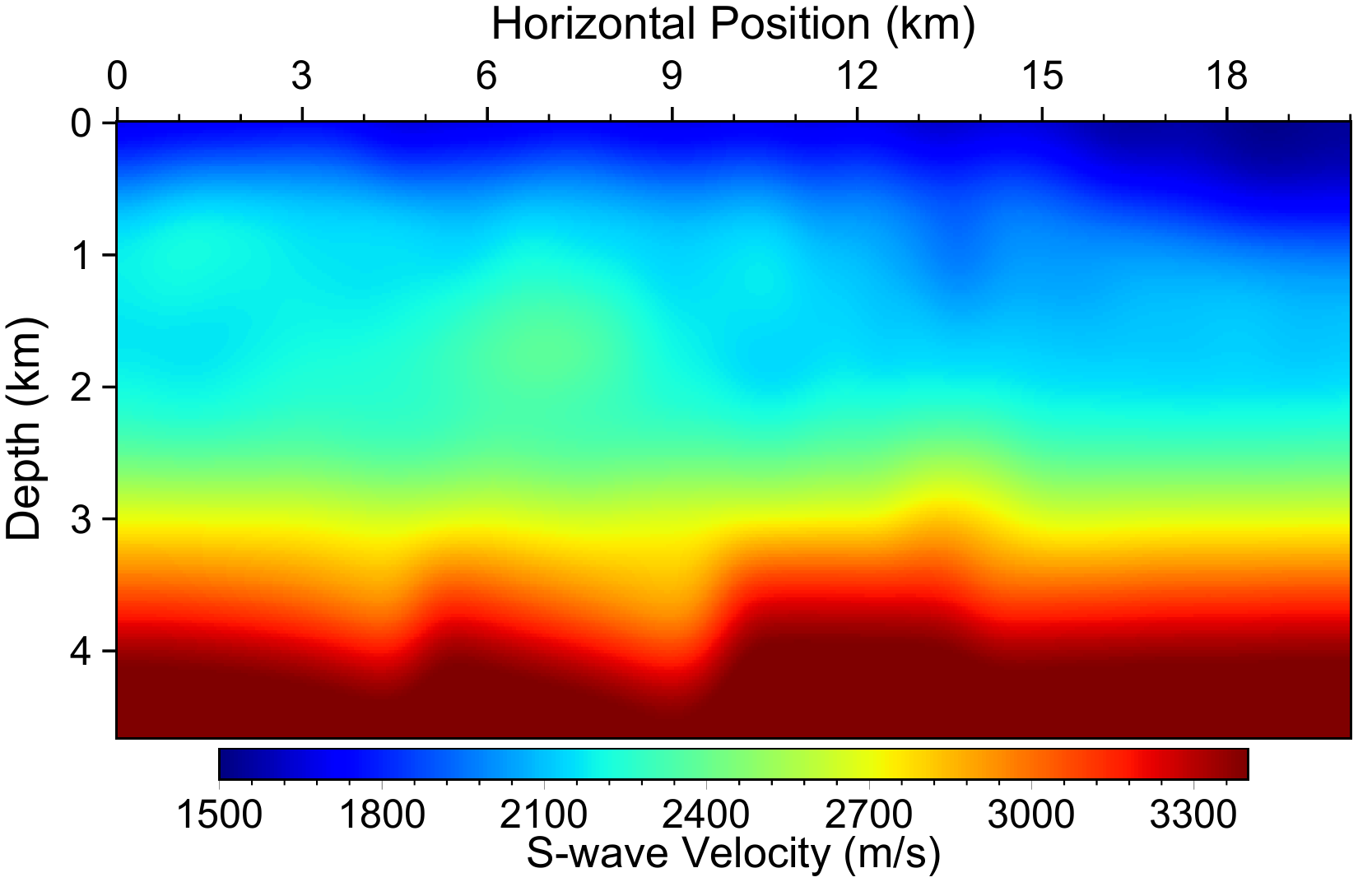}}
   \caption{The initial P-wave (a) and S-wave (b) velocity models for 
   FWI.}
\label{fig:over_model_init}
\end{figure}

\begin{figure}
\centering
\subfloat[]{\includegraphics[width=0.485\textwidth]{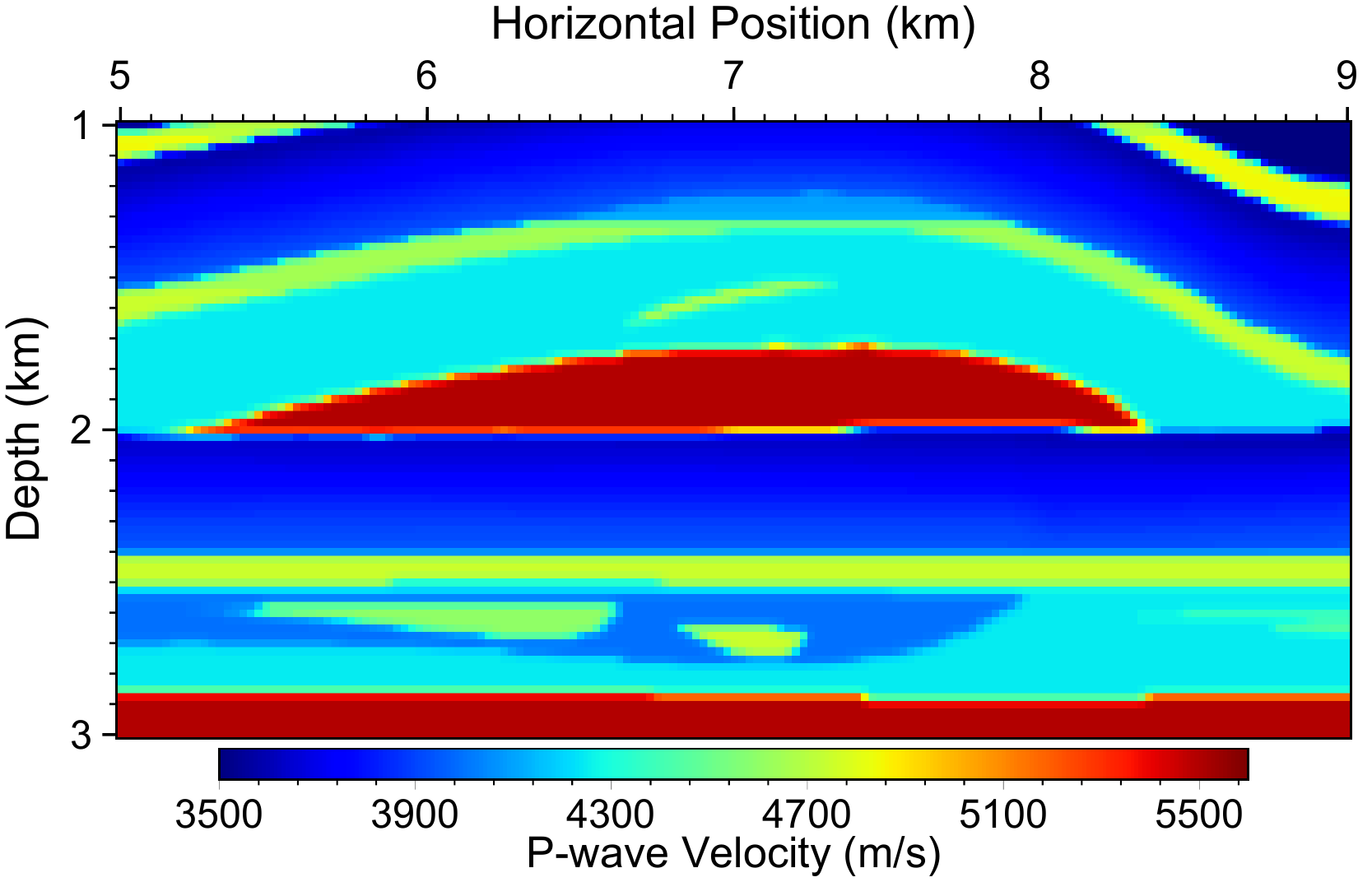}} 
\subfloat[]{\includegraphics[width=0.485\textwidth]{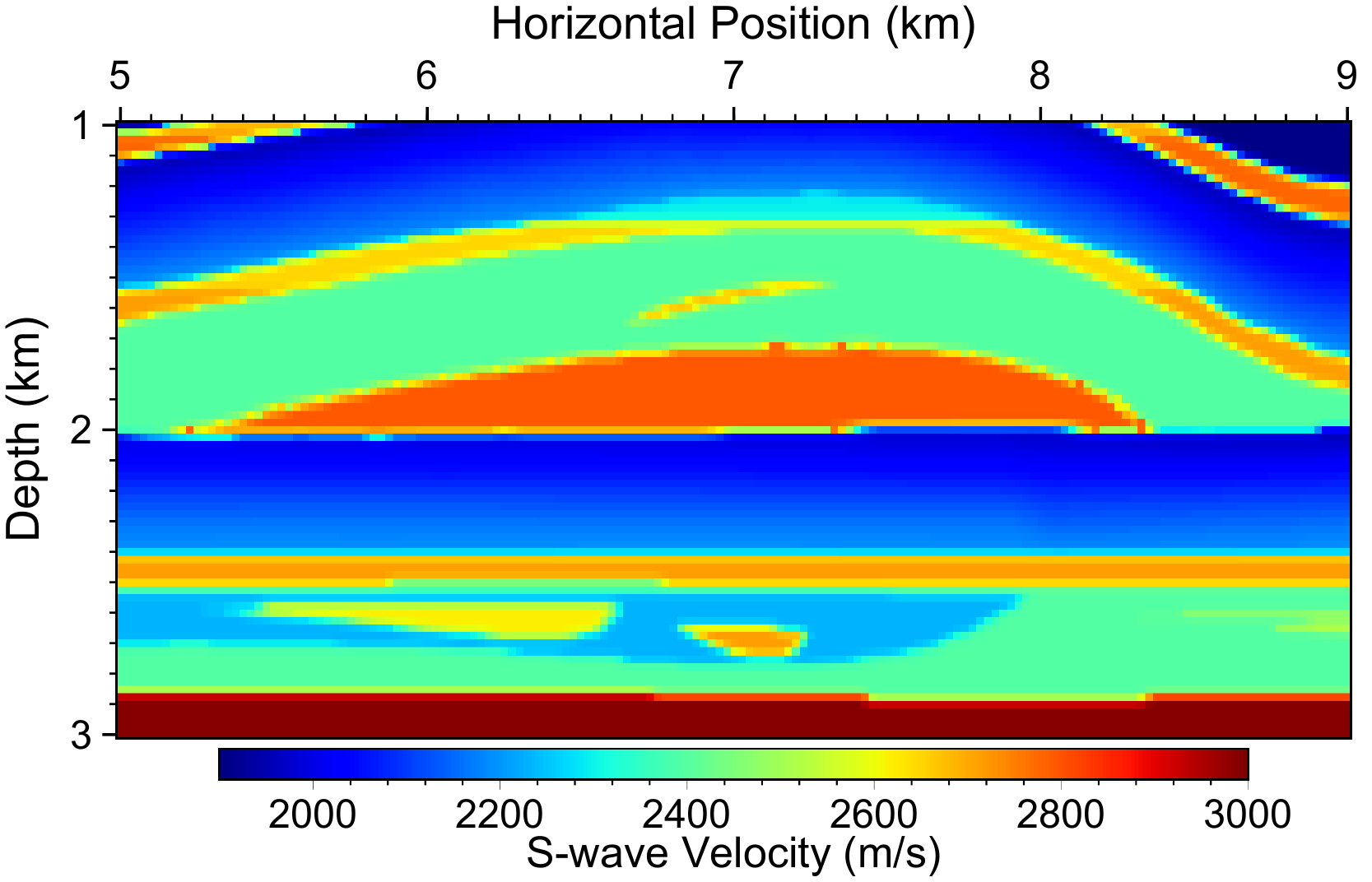}} 
   \caption{Region of interest selected from the models shown in 
   Fig.~\ref{fig:over_model} for comparison of inversion results. (a) The 
   P-wave velocity and (b) the S-wave velocity.}
\label{fig:over_zoom}
\end{figure}

We employ a total of 80 equally-spaced sources and 399 equally-spaced 
receivers at a depth of 50~m in the model. The source interval is 250~m 
and the receiver interval is 50~m. We use a Ricker wavelet with a center 
frequency of 8~Hz as the source wavelet. The source is vertical source 
force acted on the particle velocity wavefield. We employ the CG 
inversion framework for inversion. The initial P- and S-wave velocity 
models are shown in Figs.~\ref{fig:over_model_init}a and b, respectively.  
We also show the region of interest from the true velocity models in 
Figs.~\ref{fig:over_zoom}a and b for comparison of inversion results. We 
compare FWI inversion results obtained using the Tikhonov-FWI, TV-FWI and TGPV-FWI. 

\begin{figure}
\centering
\subfloat[]{\includegraphics[width=0.485\textwidth]{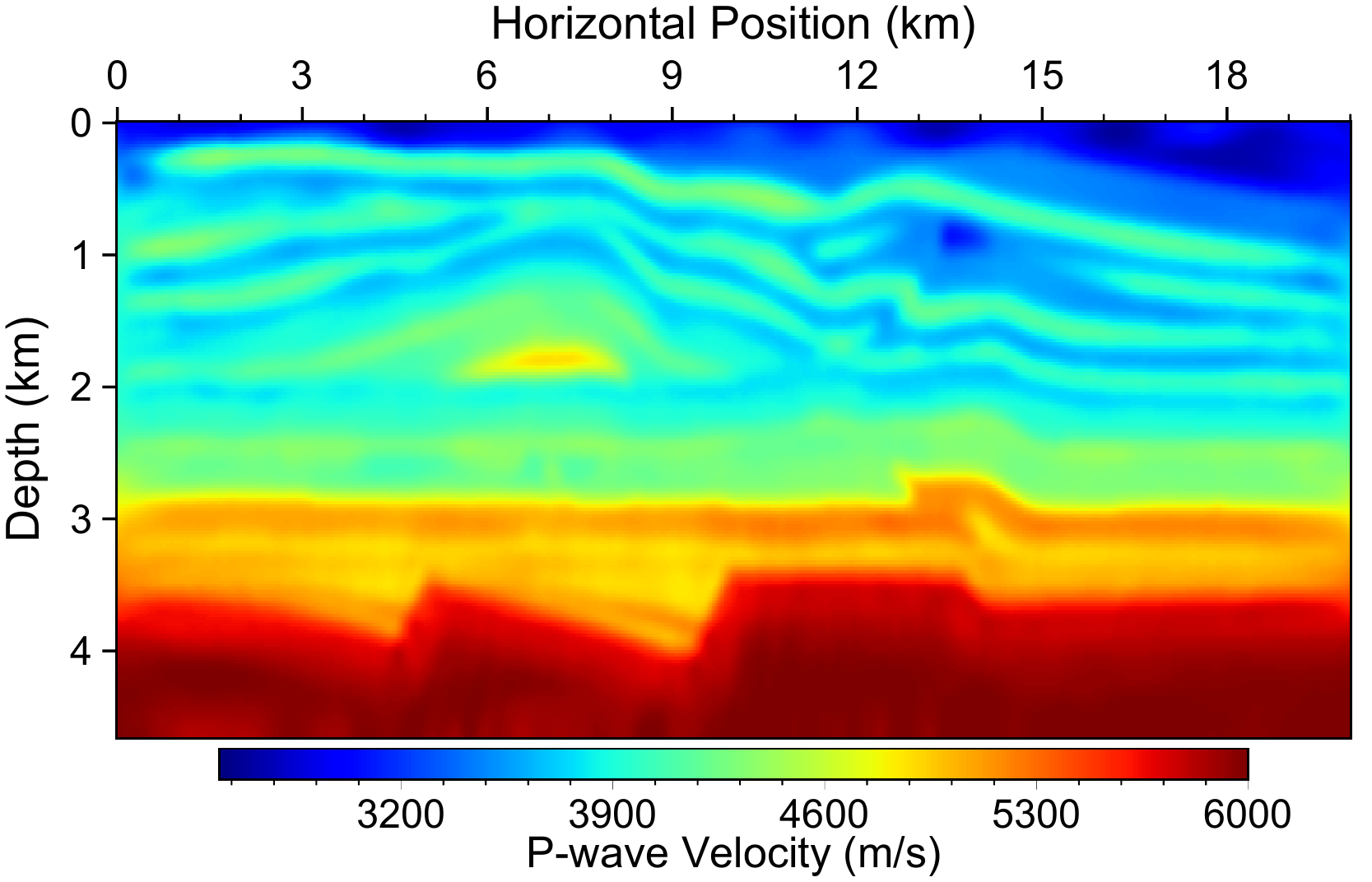}} 
\subfloat[]{\includegraphics[width=0.485\textwidth]{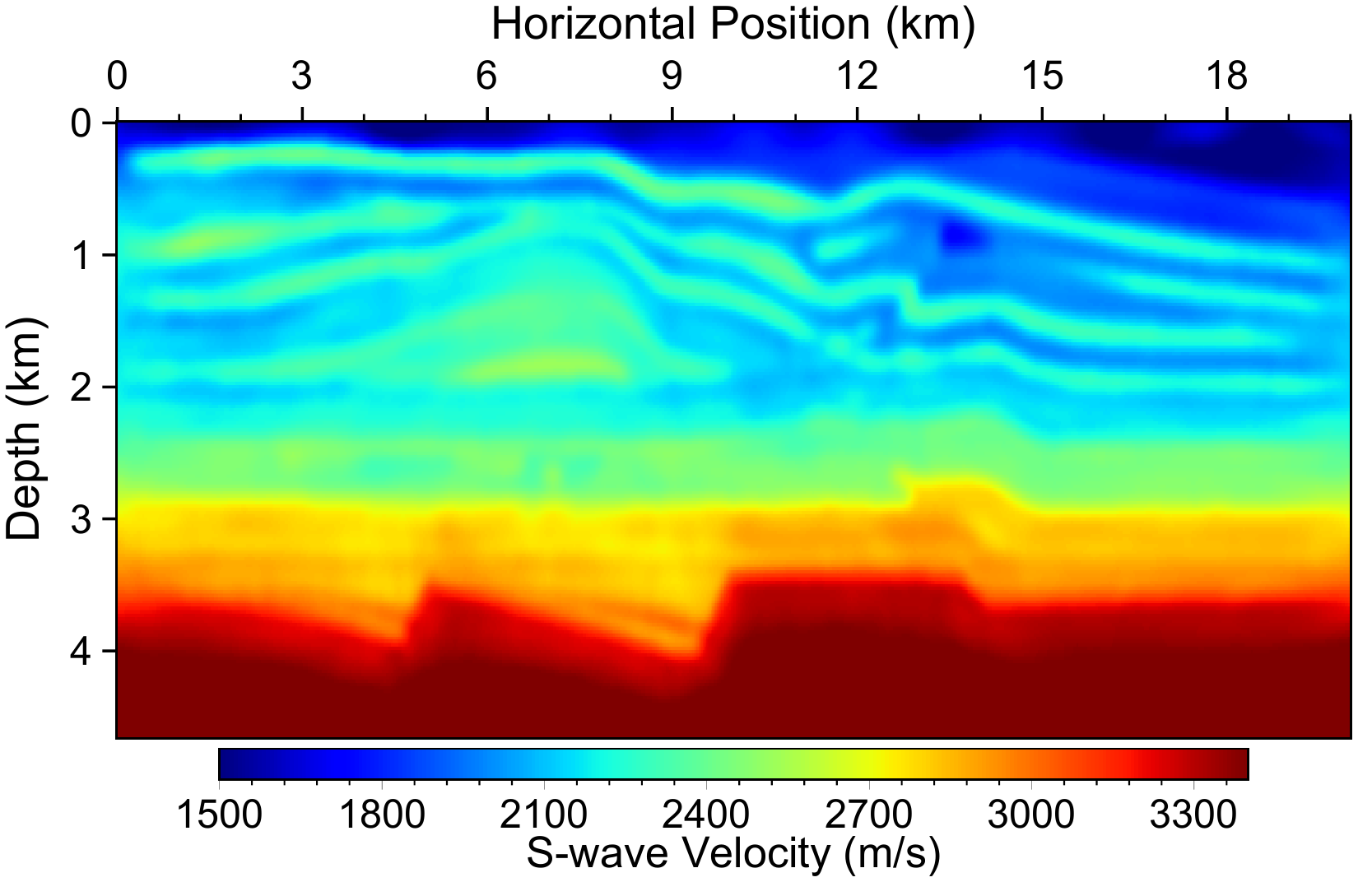}} \\
\subfloat[]{\includegraphics[width=0.485\textwidth]{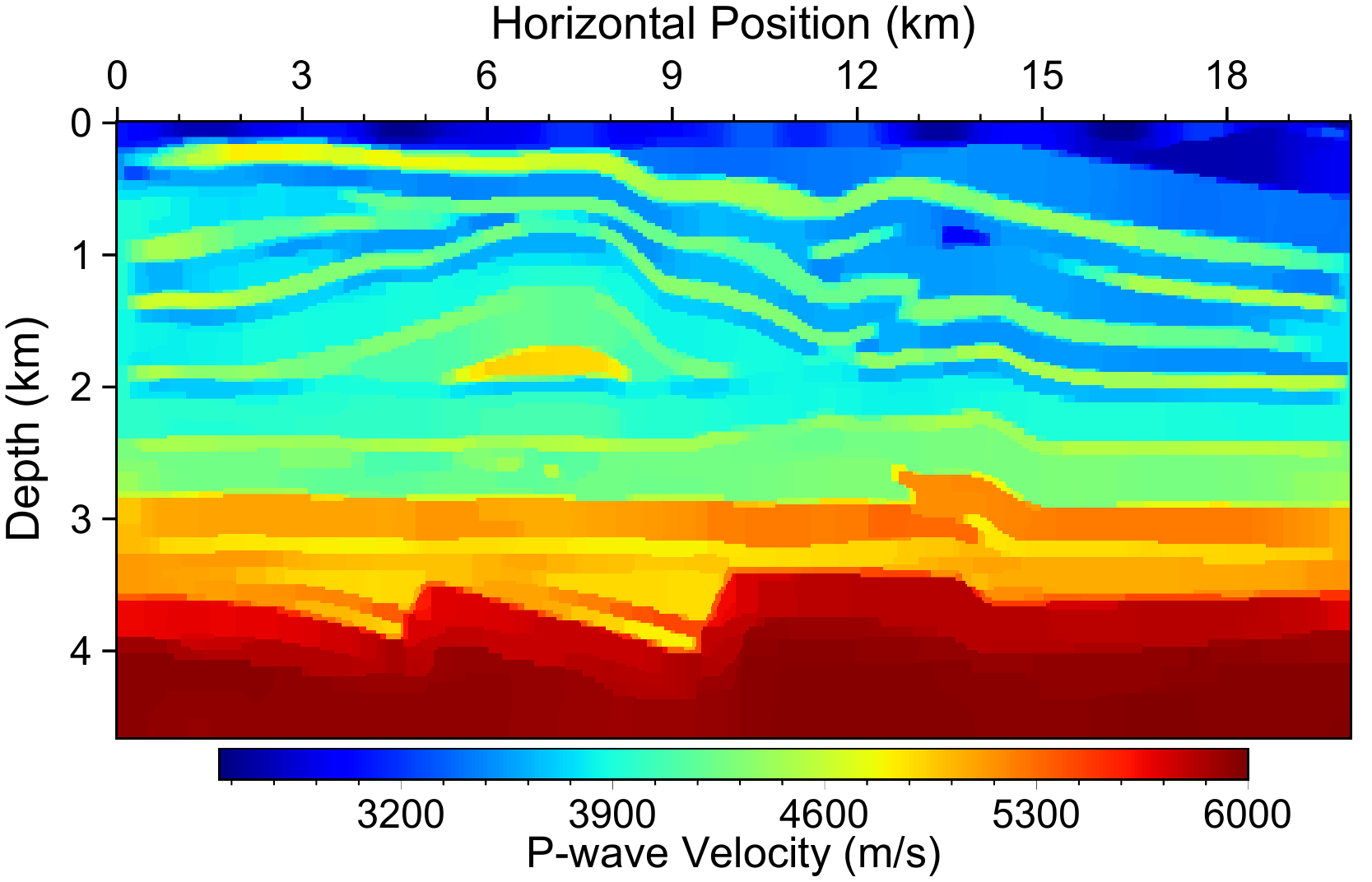}}
\subfloat[]{\includegraphics[width=0.485\textwidth]{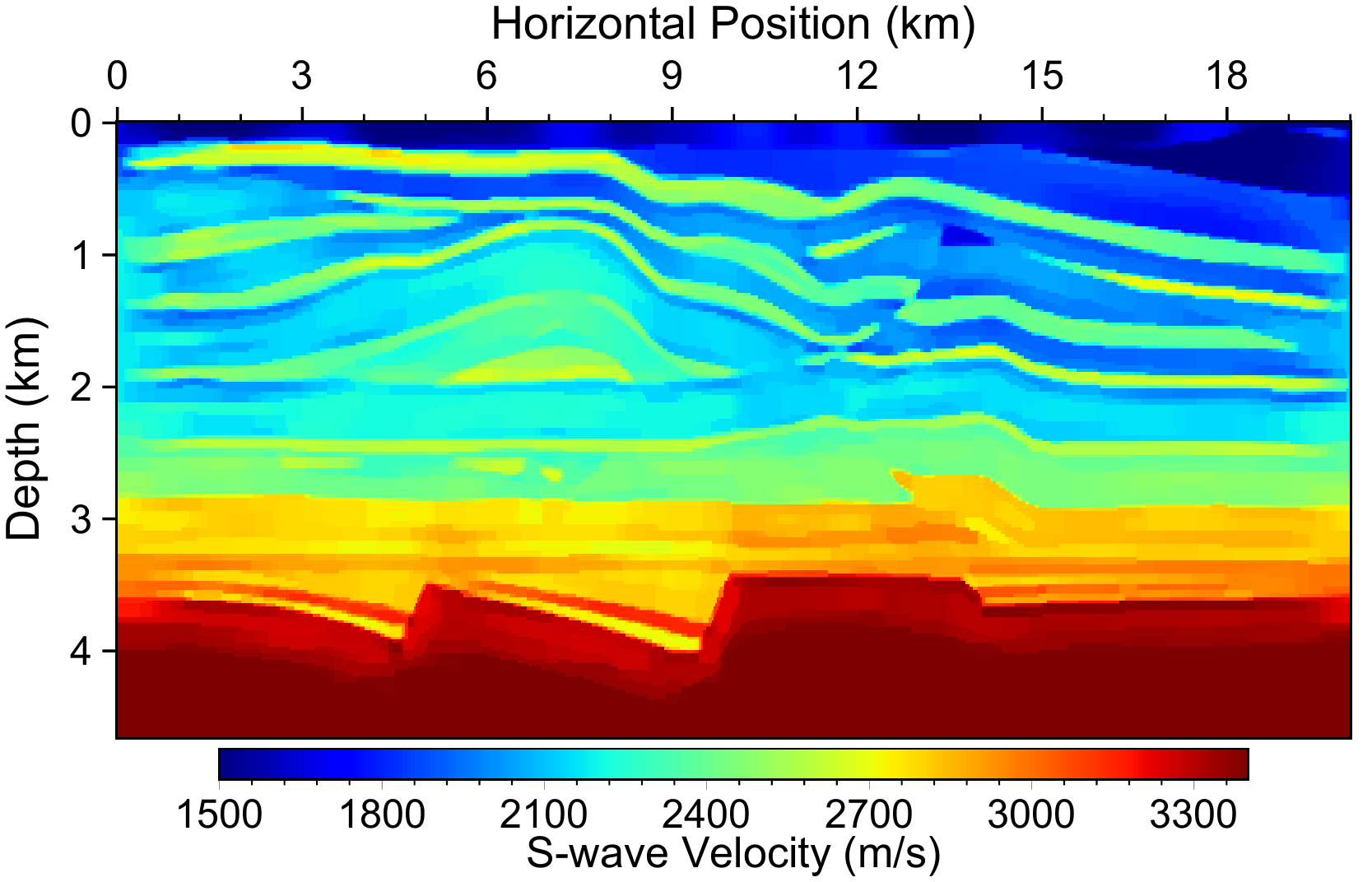}} \\
\subfloat[]{\includegraphics[width=0.485\textwidth]{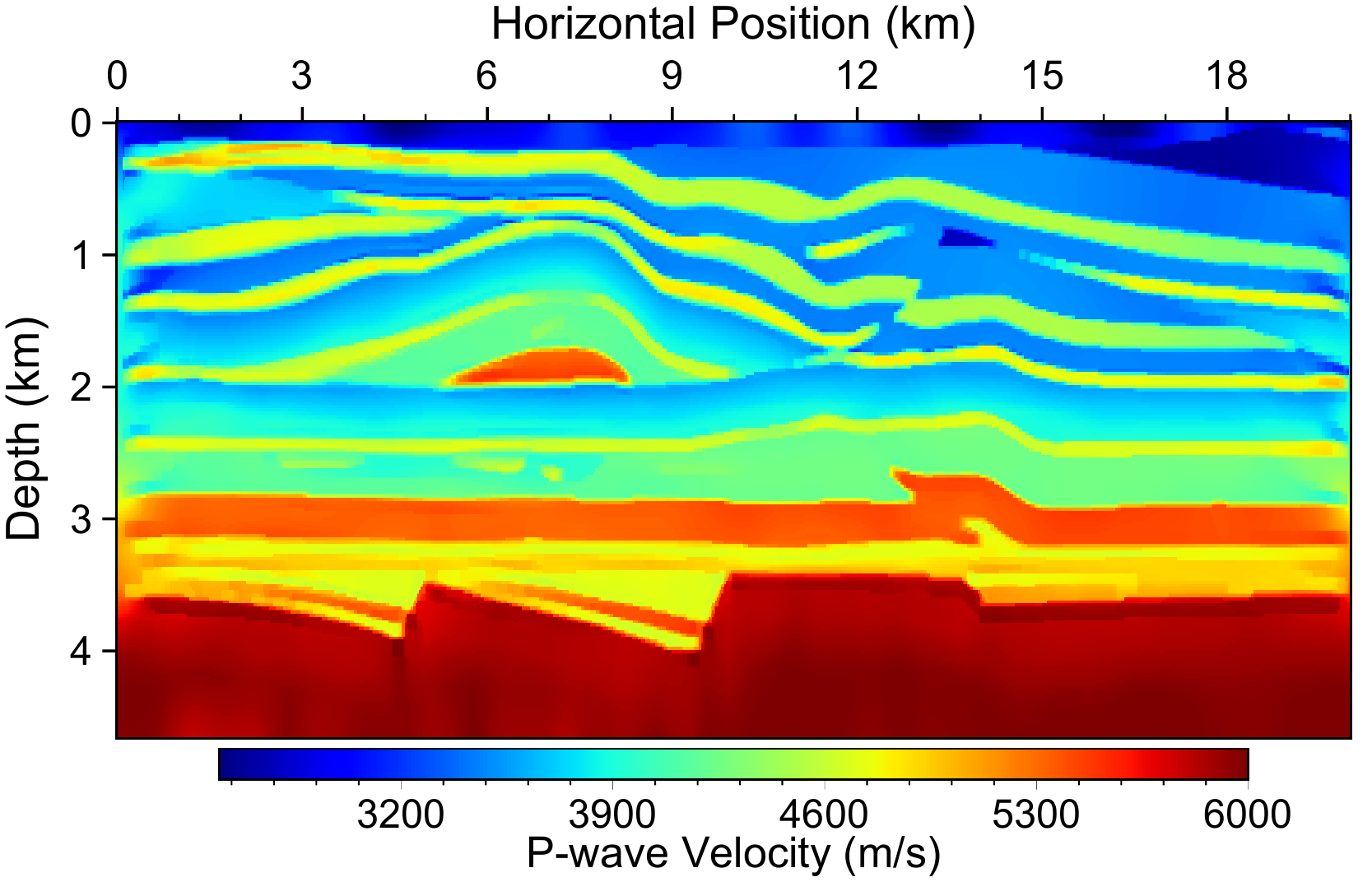}}
\subfloat[]{\includegraphics[width=0.485\textwidth]{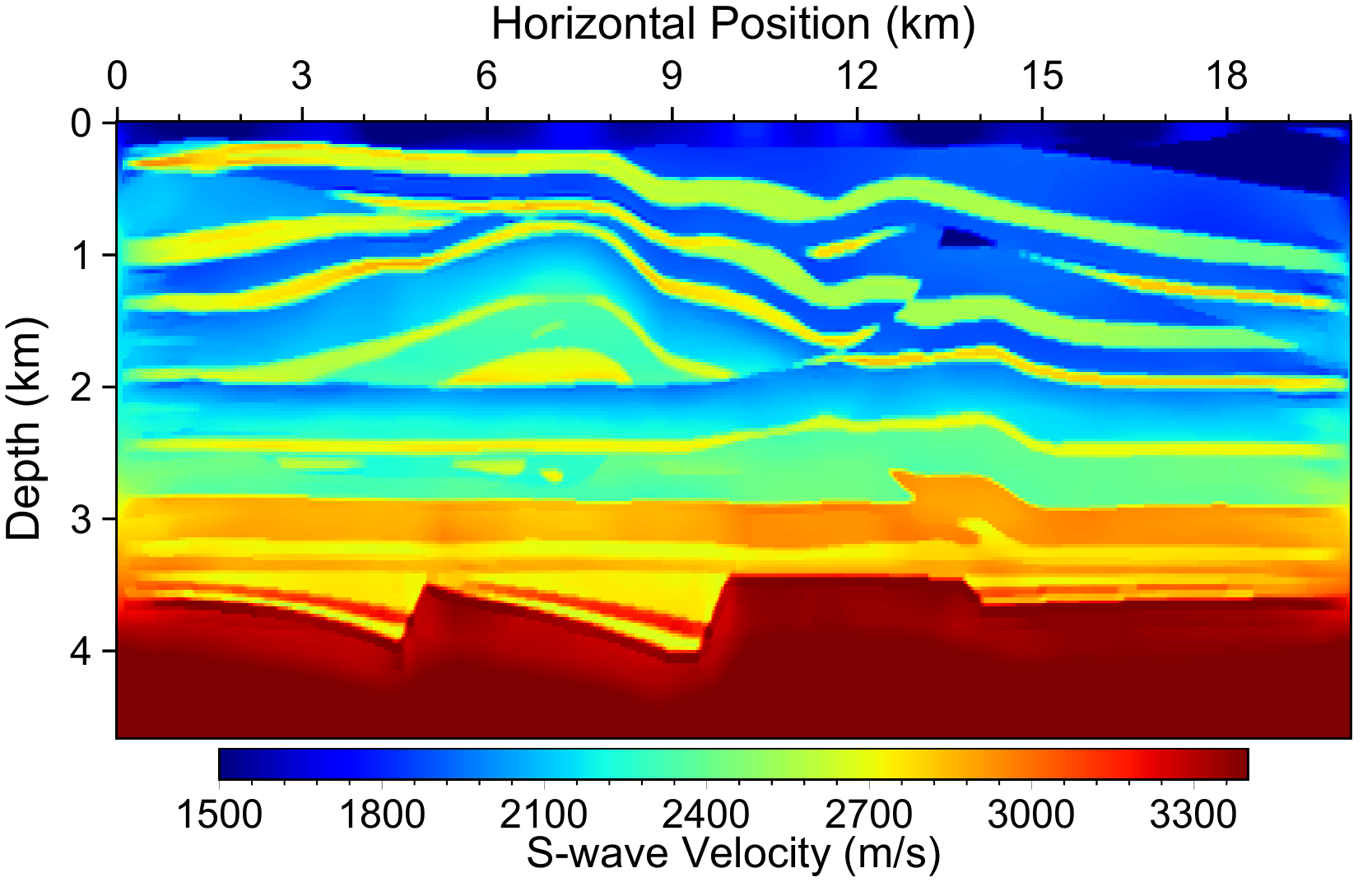}}
\caption{Inverted velocity models using (a) Tikhonov-FWI (b) TV-FWI and (c) TGPV-FWI.  Panels in (a), (c) and (e) are inverted P-wave 
   velocities and those in (b), (d) and (f) are inverted S-wave 
   velocities.}
\label{fig:over_fwi}
\end{figure}

In the first numerical test, we compare the results obtained using FWI 
with three different regularization schemes with all data of the 80 
sources and the 399 receivers. The Tikhonov-FWI 
reconstructs smooth P- and S-wave velocity models as shown in 
Figs.~\ref{fig:over_fwi}a and b, respectively. The TV-FWI improves reconstruction of the interfaces and inversion 
resolution as depicted in Figs.~\ref{fig:over_fwi}c and d. By contrast, 
our TGPV-FWI produces best results among the three 
methods, as displayed in Figs.~\ref{fig:over_fwi}e and f. The method not 
only improves reconstruction of the interfaces, but also inversion 
accuracy.  In addition, Figs.~\ref{fig:over_fwi}e and f contains fewer 
inversion artifacts than the other results in Fig.~\ref{fig:over_fwi}.
Comparing between the inversion results obtained using the TV-FWI and our TGPV-FWI, we find that, the TV-FWI results contain 
staircase artifacts in regions between thin layers where seismic 
velocities vary smoothly in space. By contrast, our TGPV-FWI accurately 
reconstructs these smoothly varying velocities.

\begin{figure}
\centering
\subfloat[]{\includegraphics[width=0.485\textwidth]{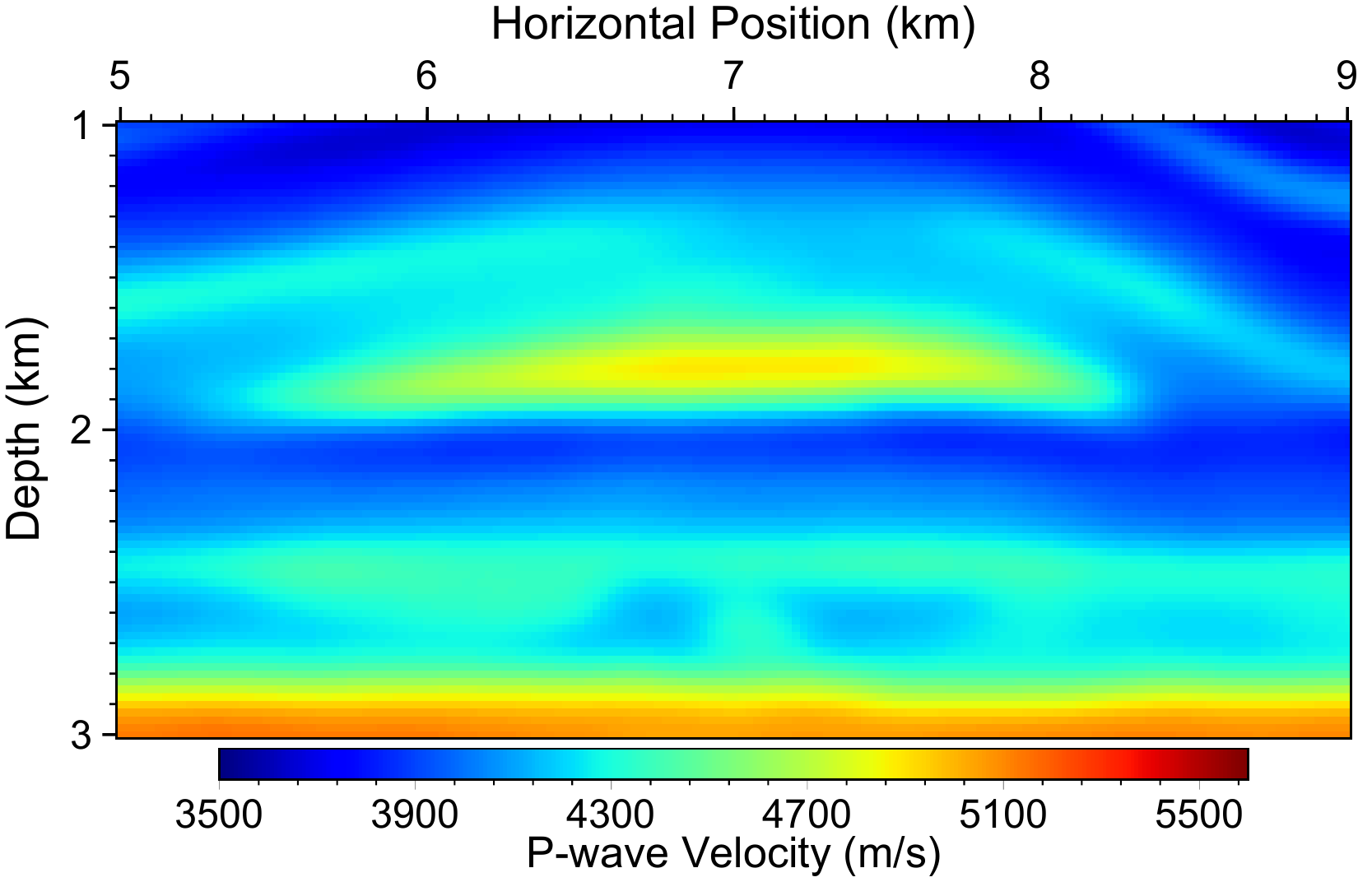}} 
\subfloat[]{\includegraphics[width=0.485\textwidth]{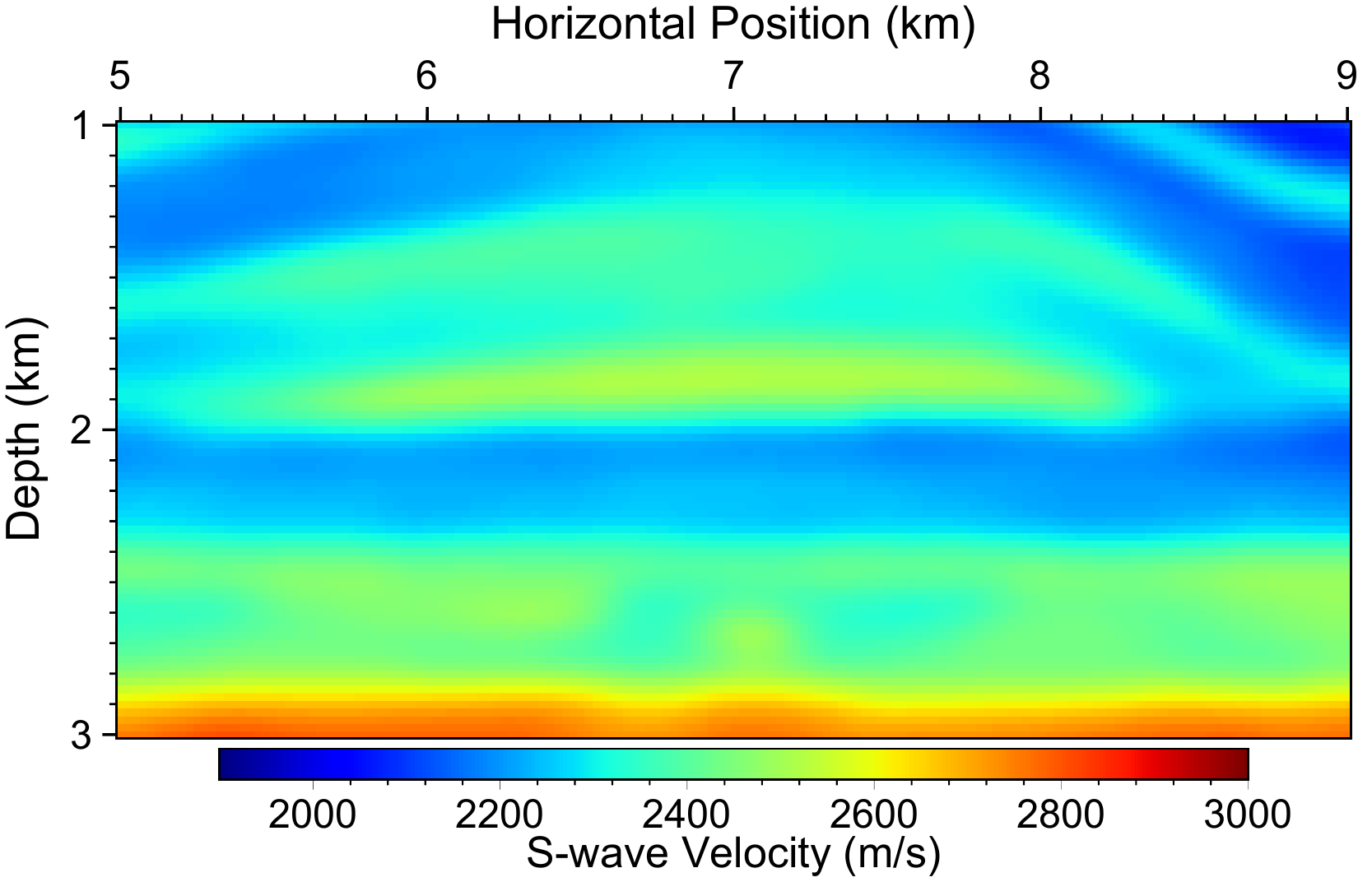}} \\
\subfloat[]{\includegraphics[width=0.485\textwidth]{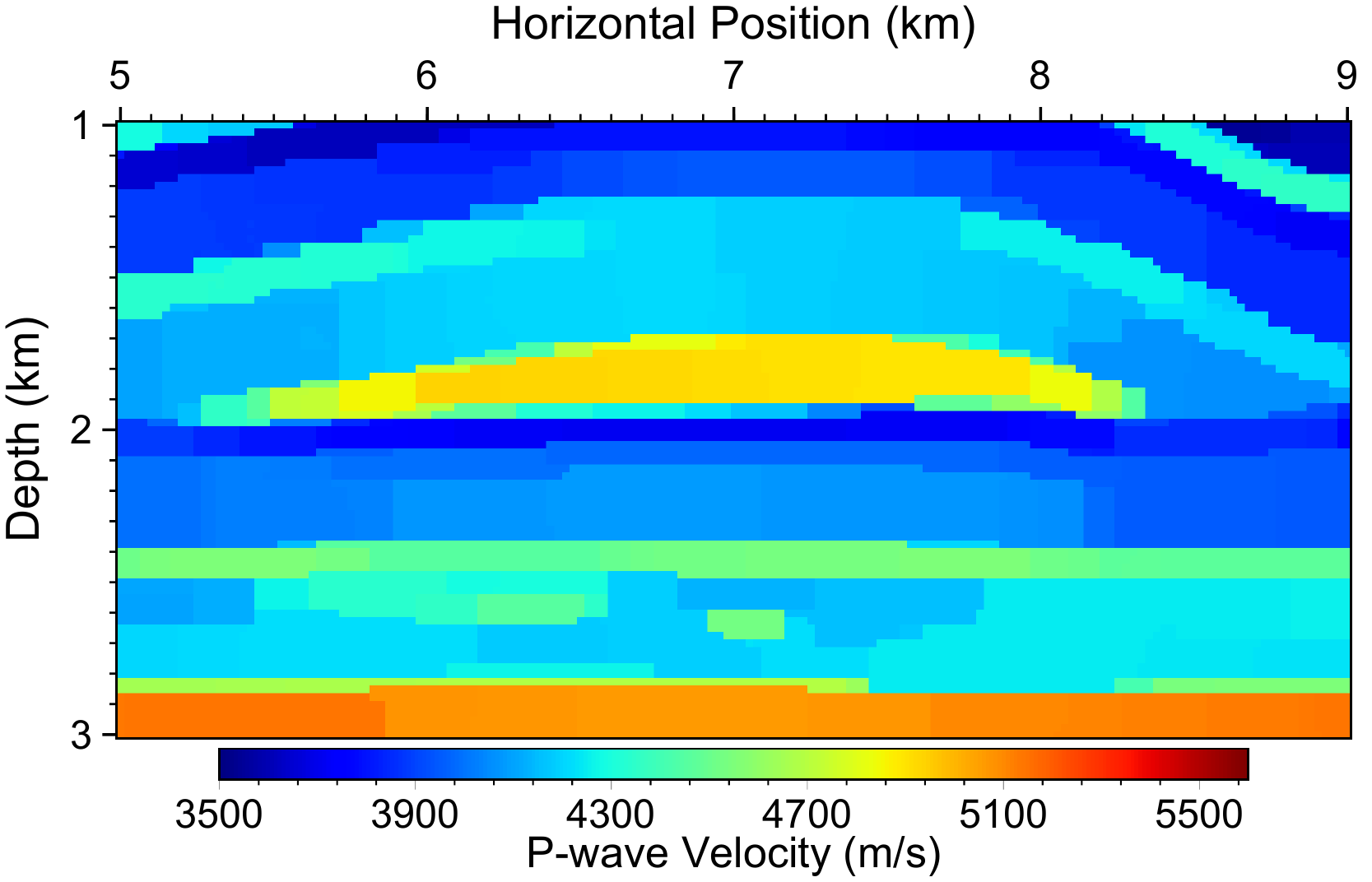}}
\subfloat[]{\includegraphics[width=0.485\textwidth]{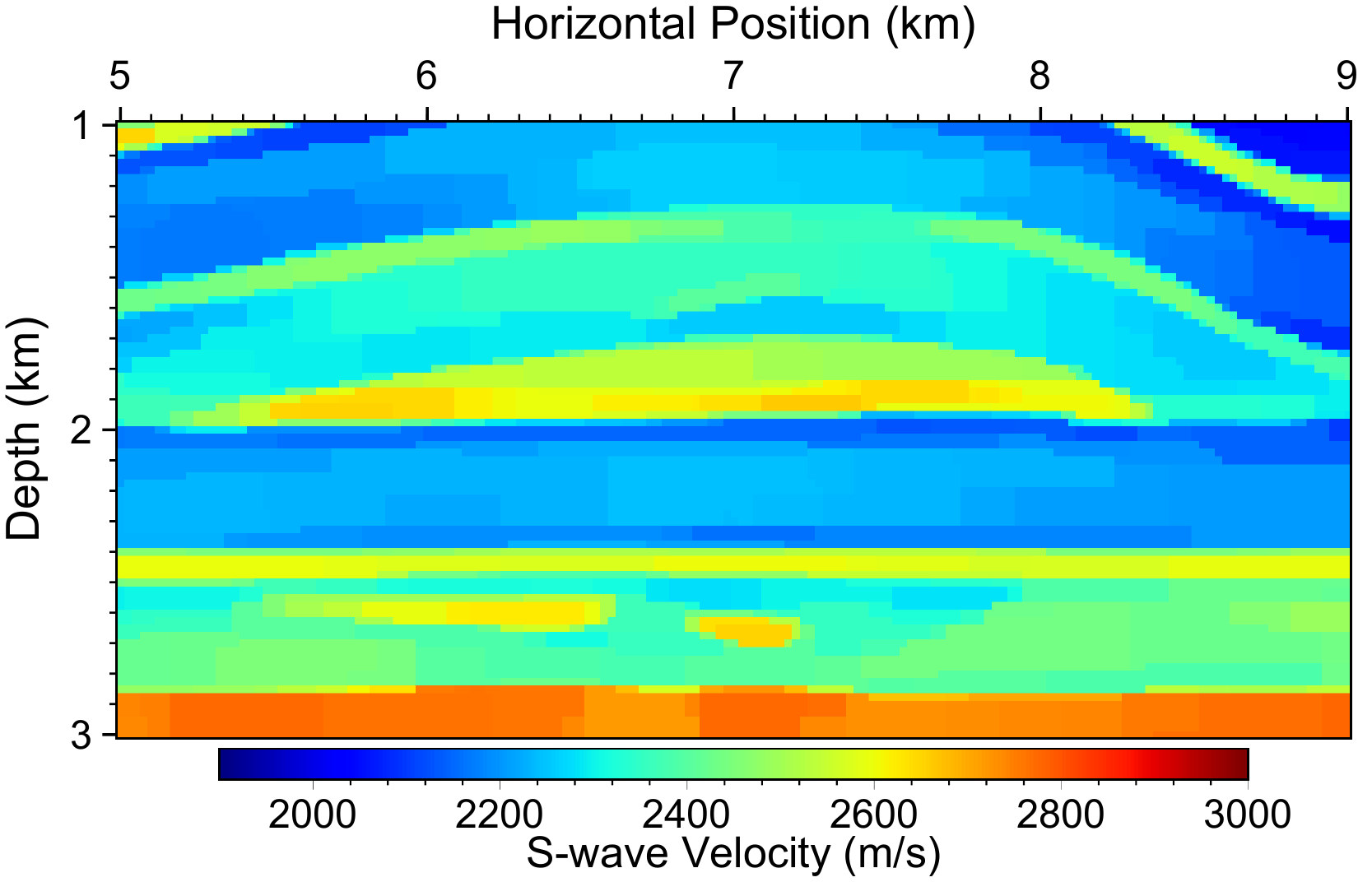}} \\
\subfloat[]{\includegraphics[width=0.485\textwidth]{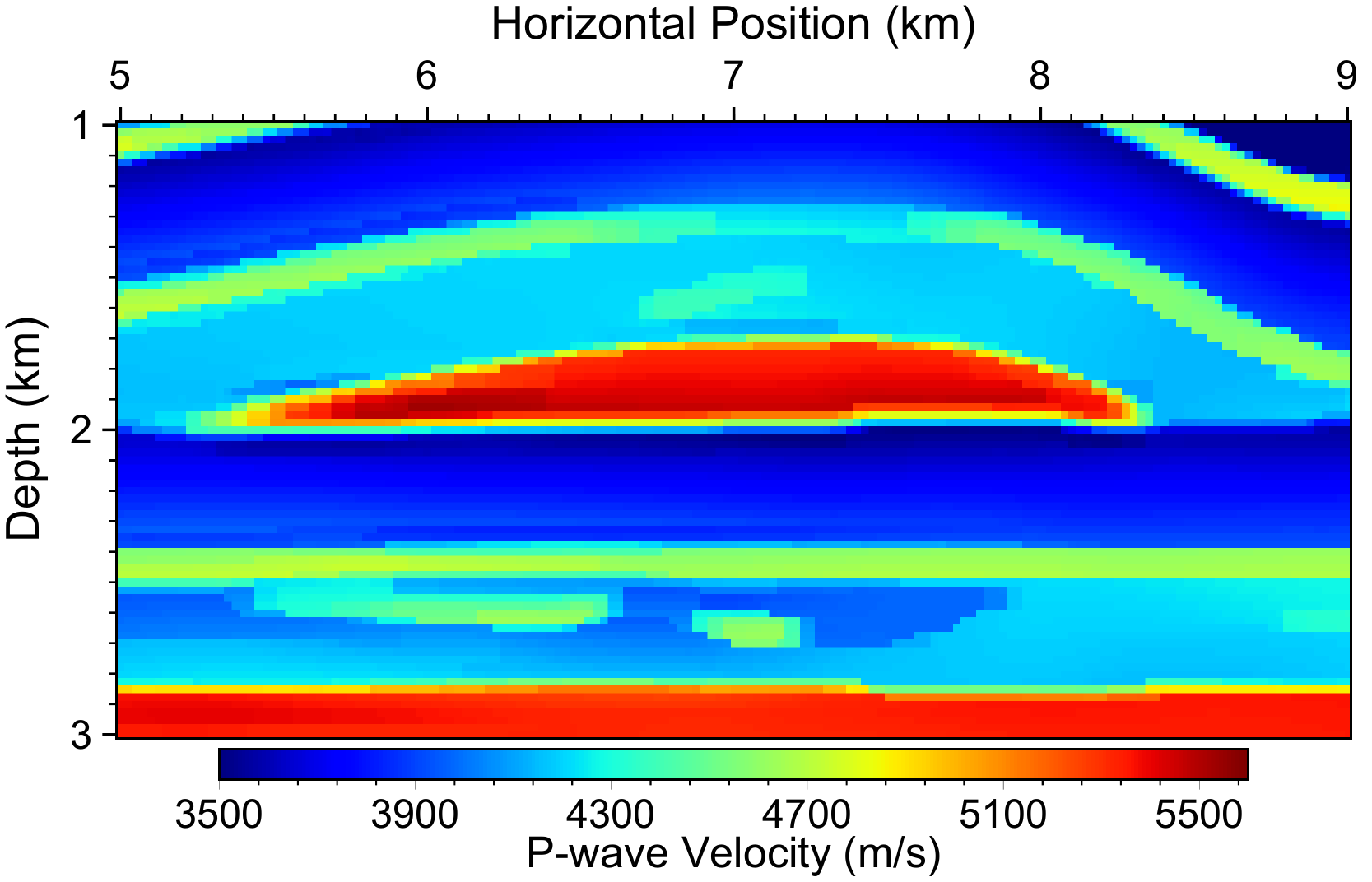}}
\subfloat[]{\includegraphics[width=0.485\textwidth]{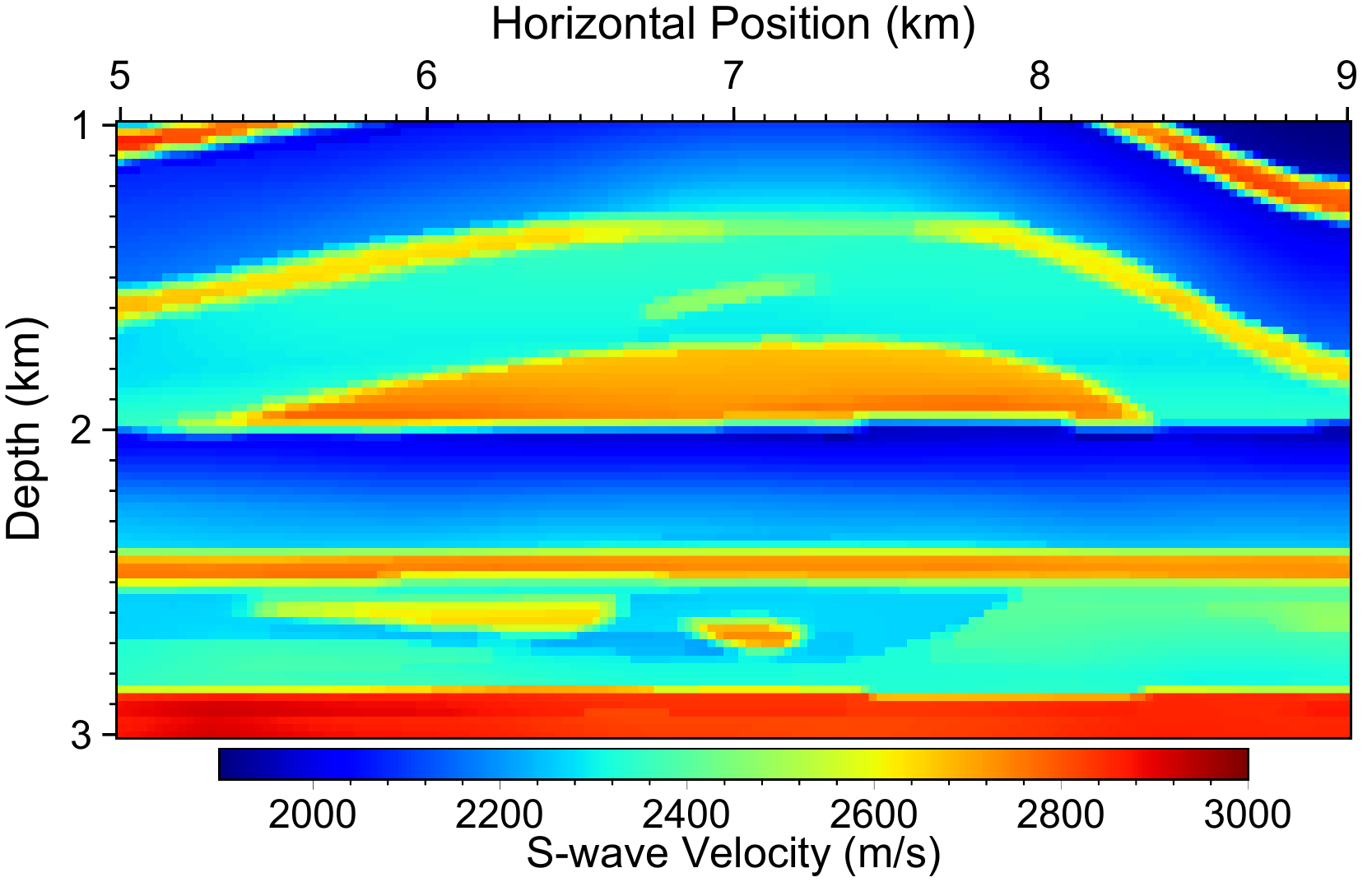}}
\caption{Zoom-in views of the region of interest of inverted velocity 
   models in Fig.~\ref{fig:over_fwi} obtained using (a) Tikhonov-FWI (b) TV-FWI and (c) TGPV-FWI. Panels in (a), (c) and (e) are inverted P-wave 
   velocities and those in (b), (d) and (f) are inverted S-wave 
   velocities.}
\label{fig:over_fwi_zoom}
\end{figure}

The comparison among the inversion results in Fig.~\ref{fig:over_fwi} in 
the region of interest manifests the significant improvements of our 
TGPV-FWI compared with the other two FWI methods, as shown in 
Fig.~\ref{fig:over_fwi_zoom}.

\begin{figure}
\centering
\subfloat[]{\includegraphics[width=0.485\textwidth]{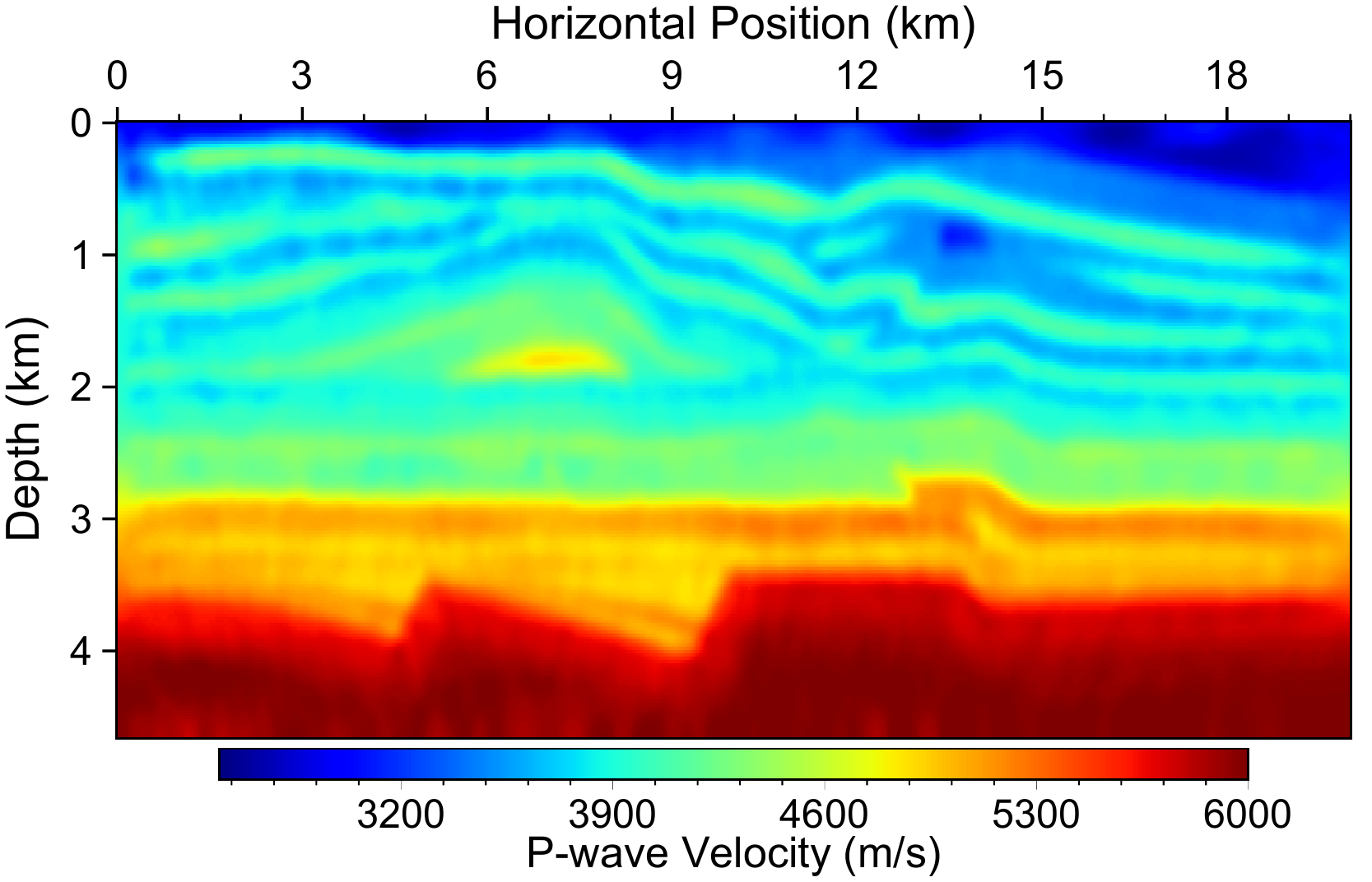}} 
\subfloat[]{\includegraphics[width=0.485\textwidth]{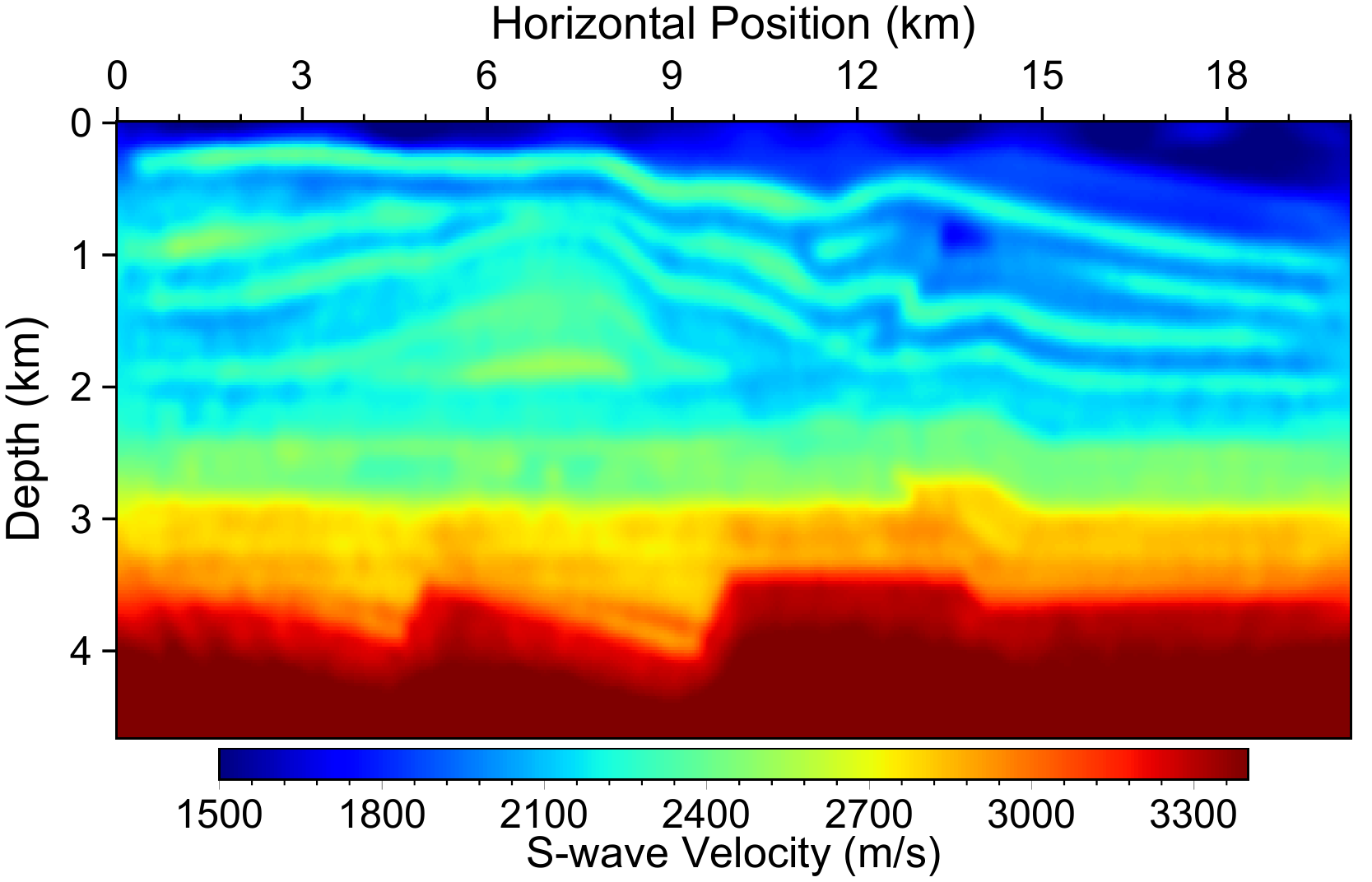}} \\
\subfloat[]{\includegraphics[width=0.485\textwidth]{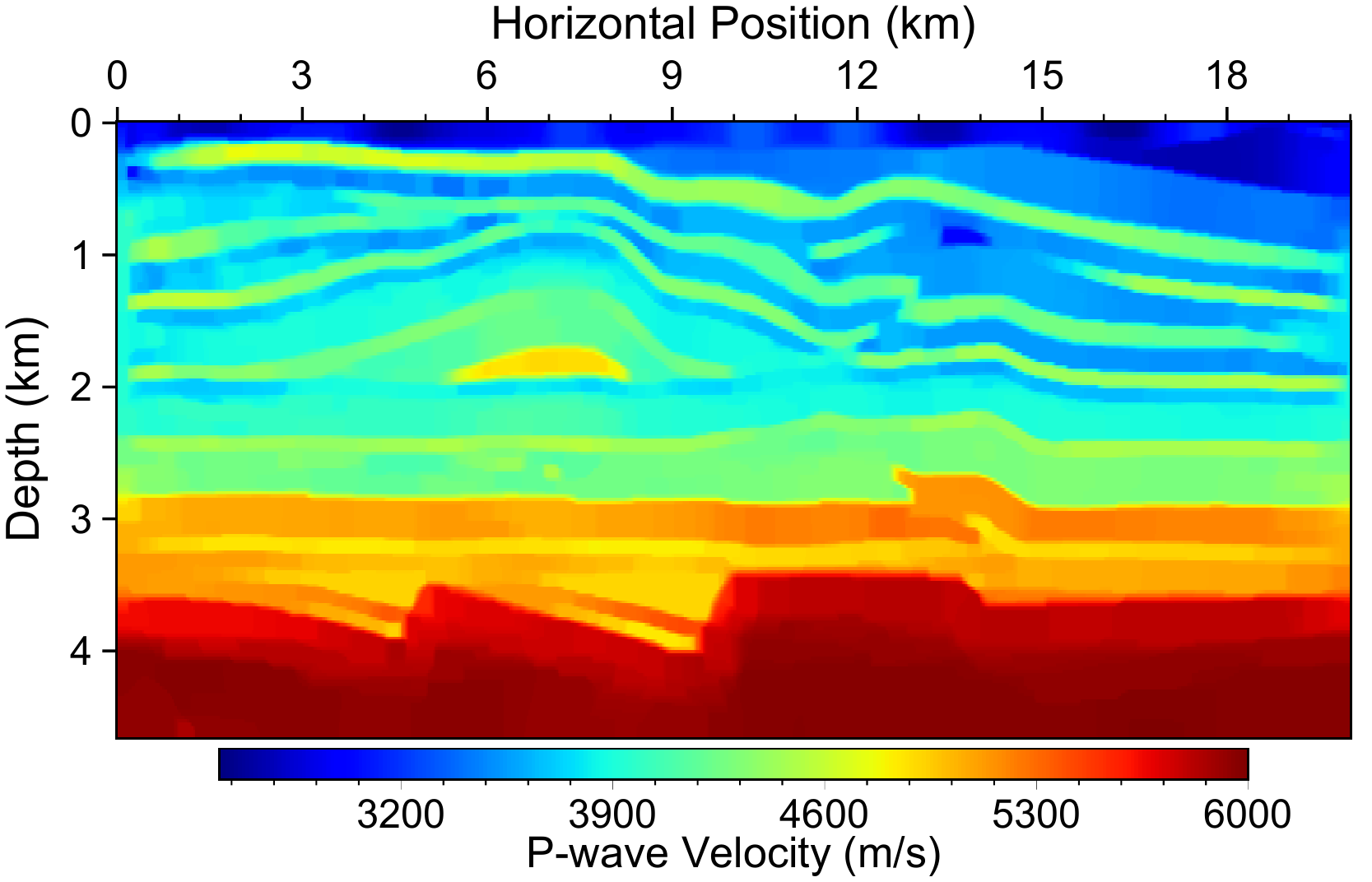}}
\subfloat[]{\includegraphics[width=0.485\textwidth]{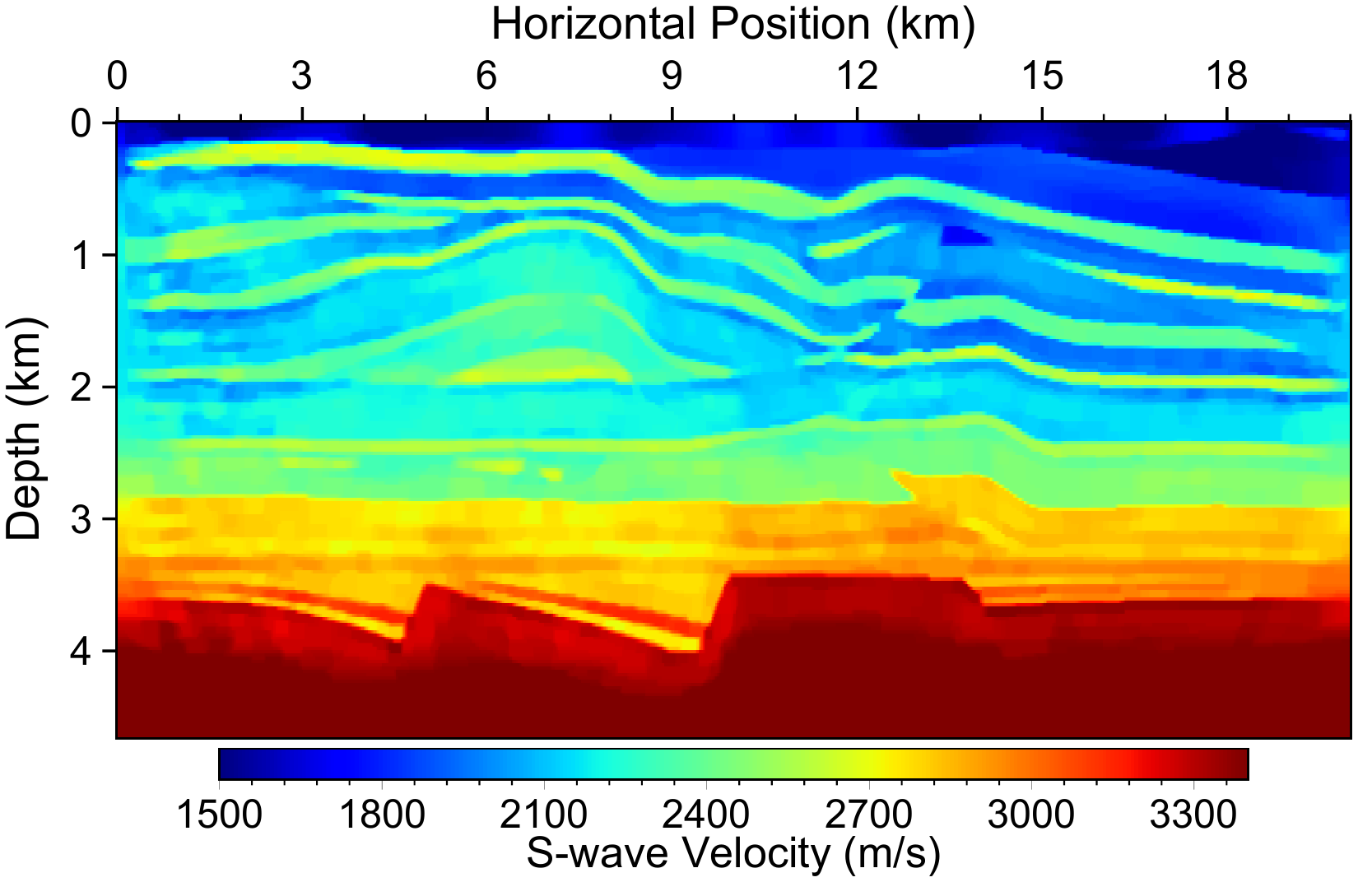}} \\
\subfloat[]{\includegraphics[width=0.485\textwidth]{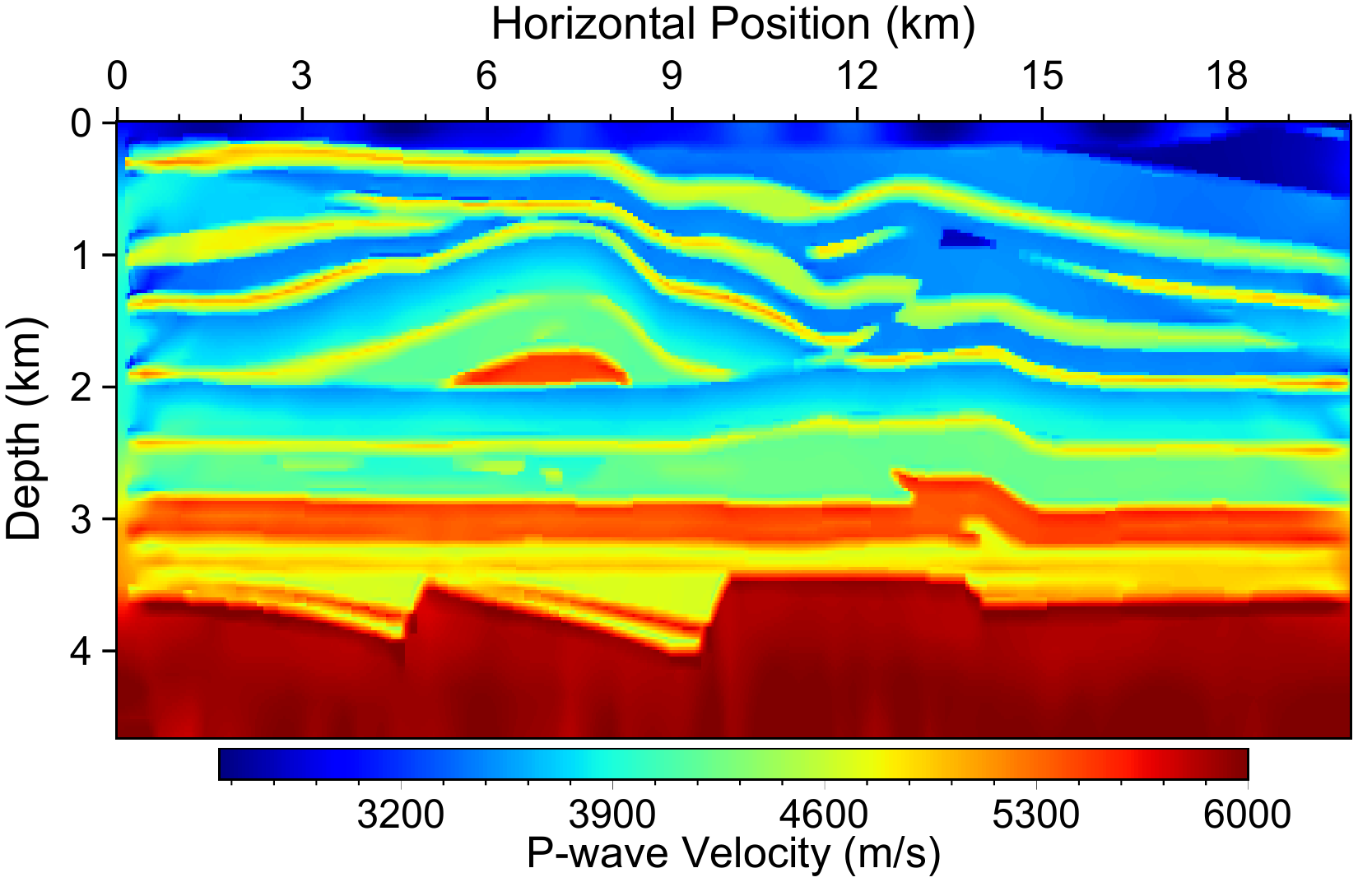}}
\subfloat[]{\includegraphics[width=0.485\textwidth]{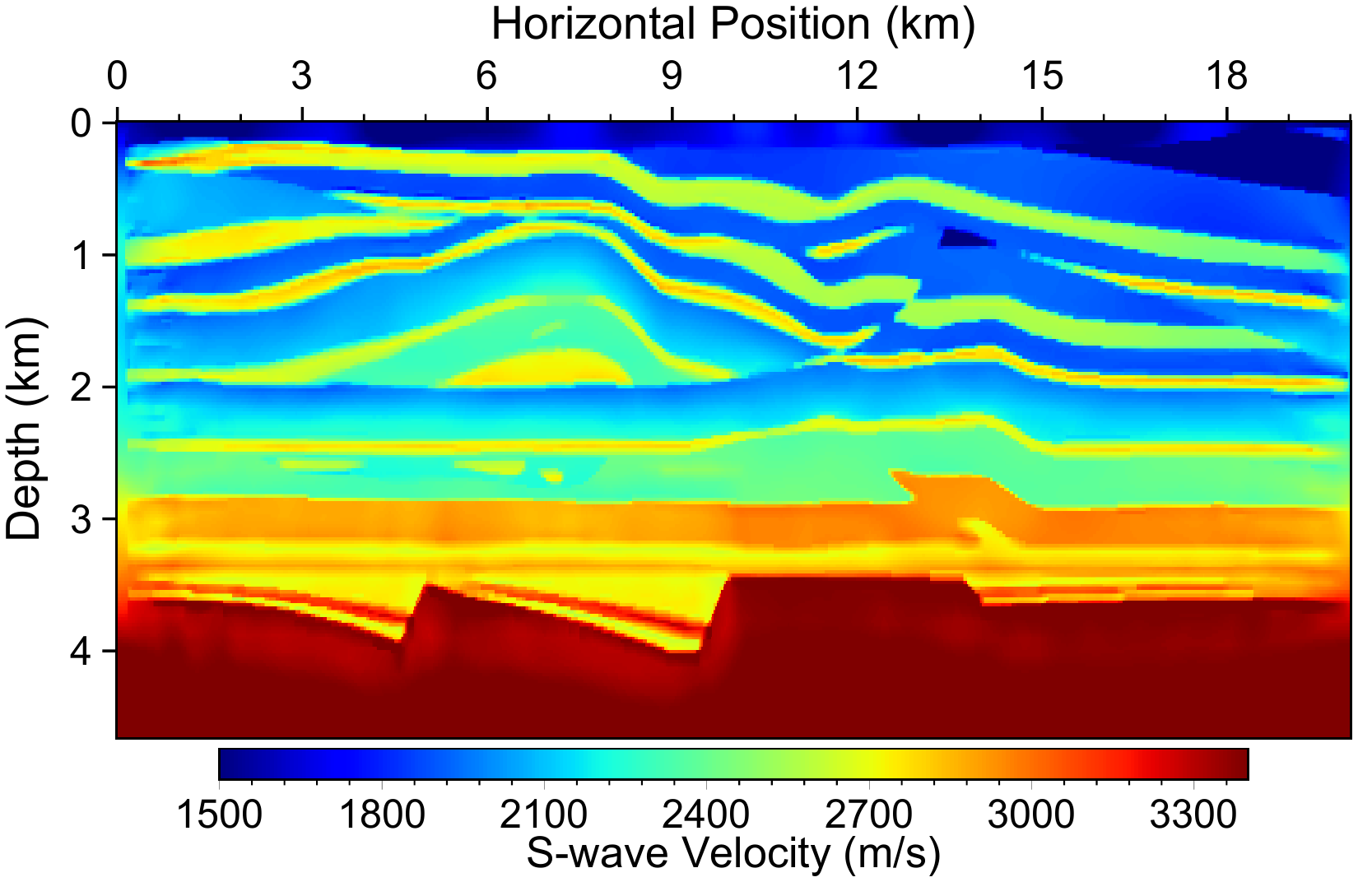}}
\caption{Inverted velocity models using (a) Tikhonov-FWI (b) TV-FWI and (c) TGPV-FWI. Panels in (a), (c) and (e) are inverted P-wave 
   velocities and those (b), (d) and (f) are inverted S-wave velocities.  
   All three inversions use sparse seismic data acquired with inadequate 
   numbers of sources and receivers.}
\label{fig:over_fwi_sparse}
\end{figure}

\begin{figure}
\centering
\subfloat[]{\includegraphics[width=0.485\textwidth]{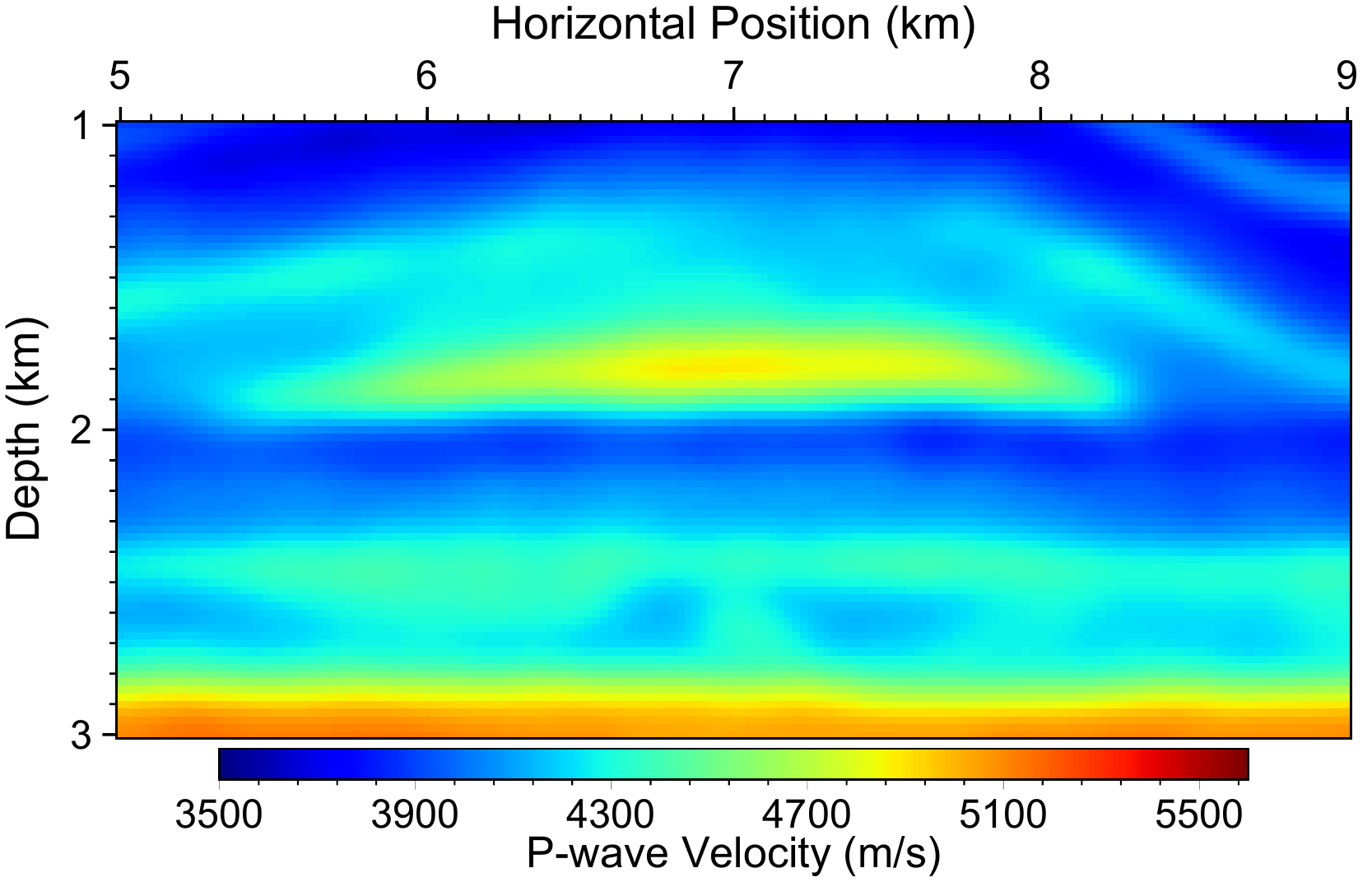}} 
\subfloat[]{\includegraphics[width=0.485\textwidth]{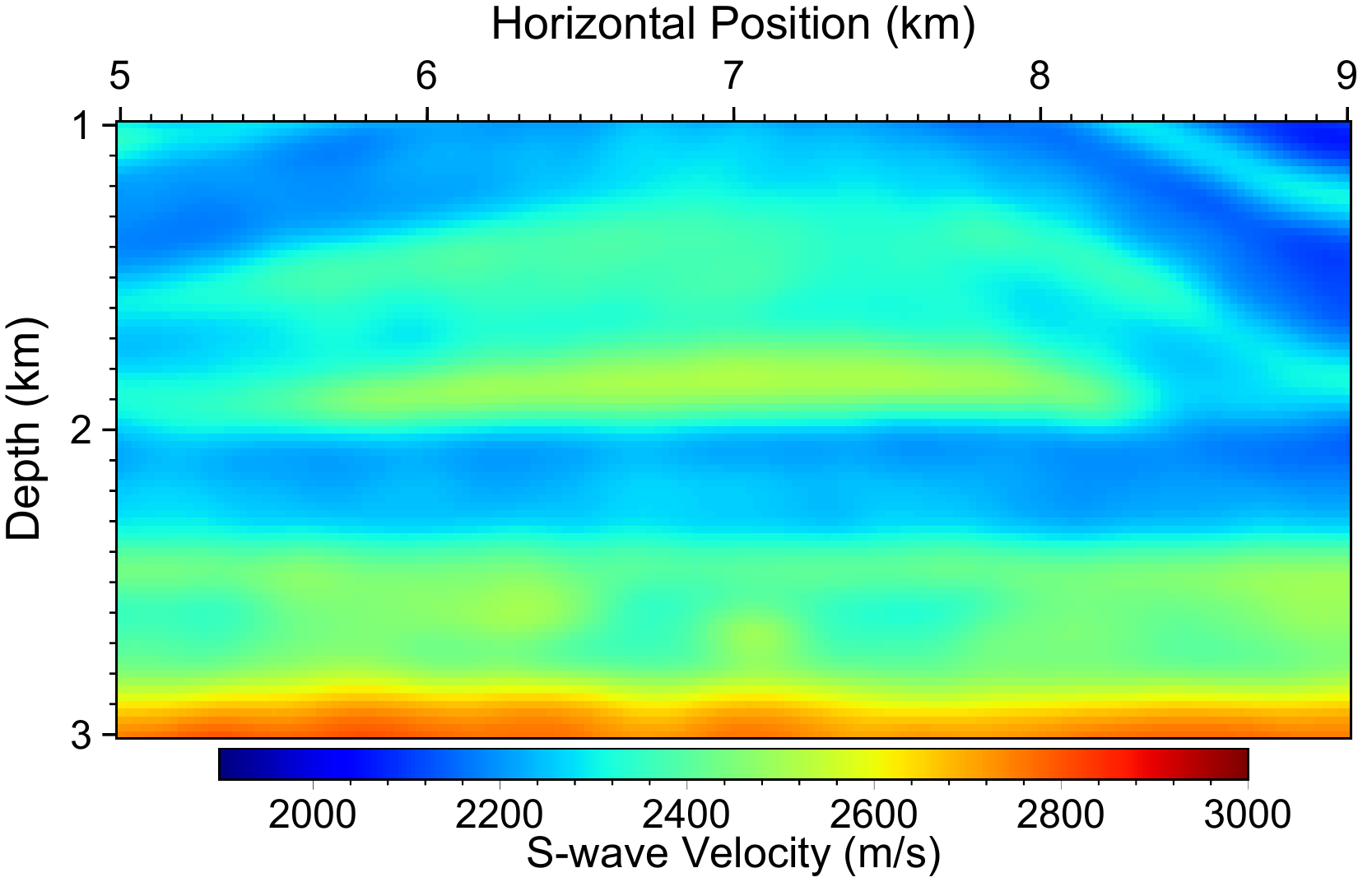}} \\
\subfloat[]{\includegraphics[width=0.485\textwidth]{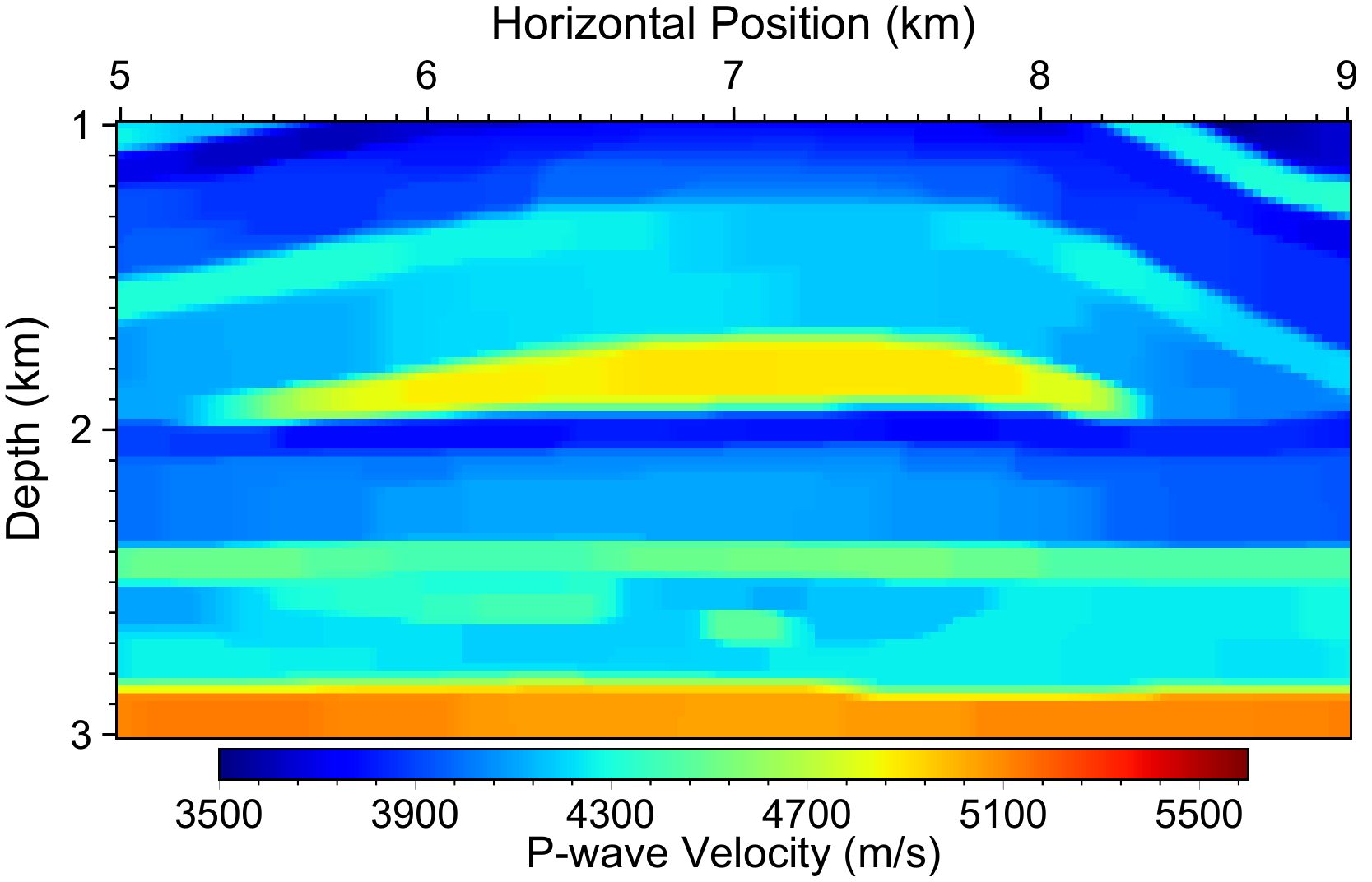}}
\subfloat[]{\includegraphics[width=0.485\textwidth]{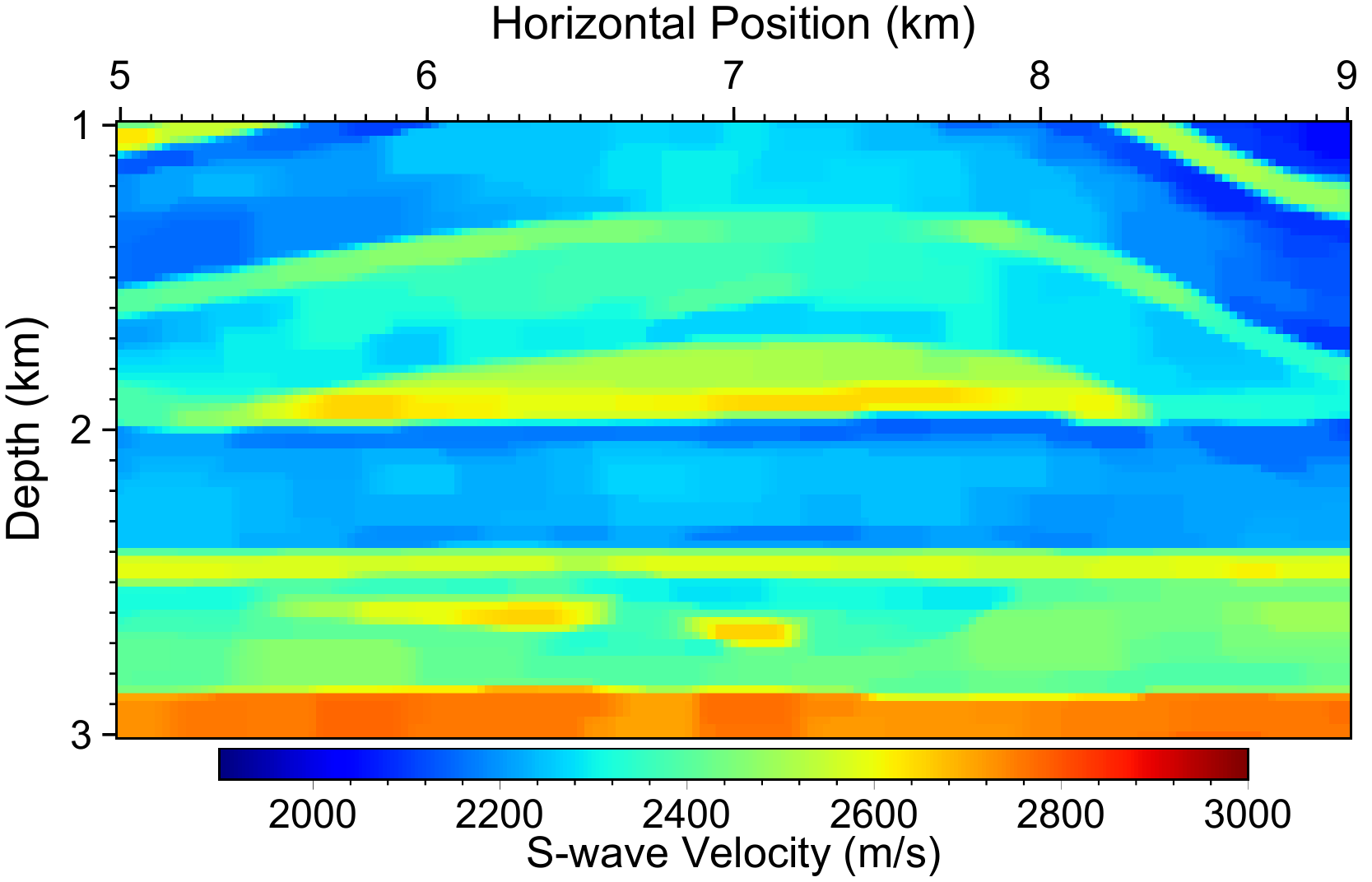}} \\
\subfloat[]{\includegraphics[width=0.485\textwidth]{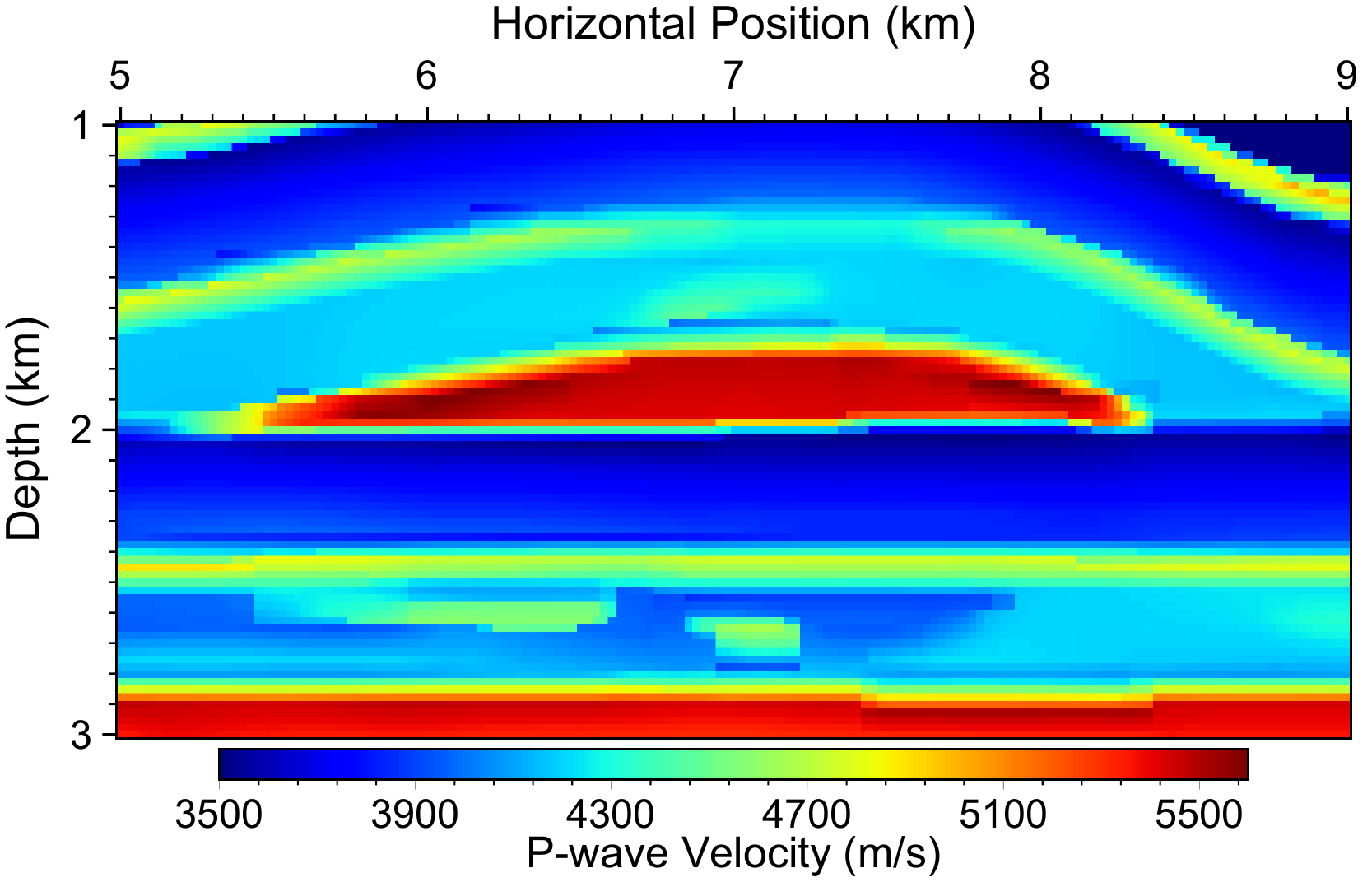}}
\subfloat[]{\includegraphics[width=0.485\textwidth]{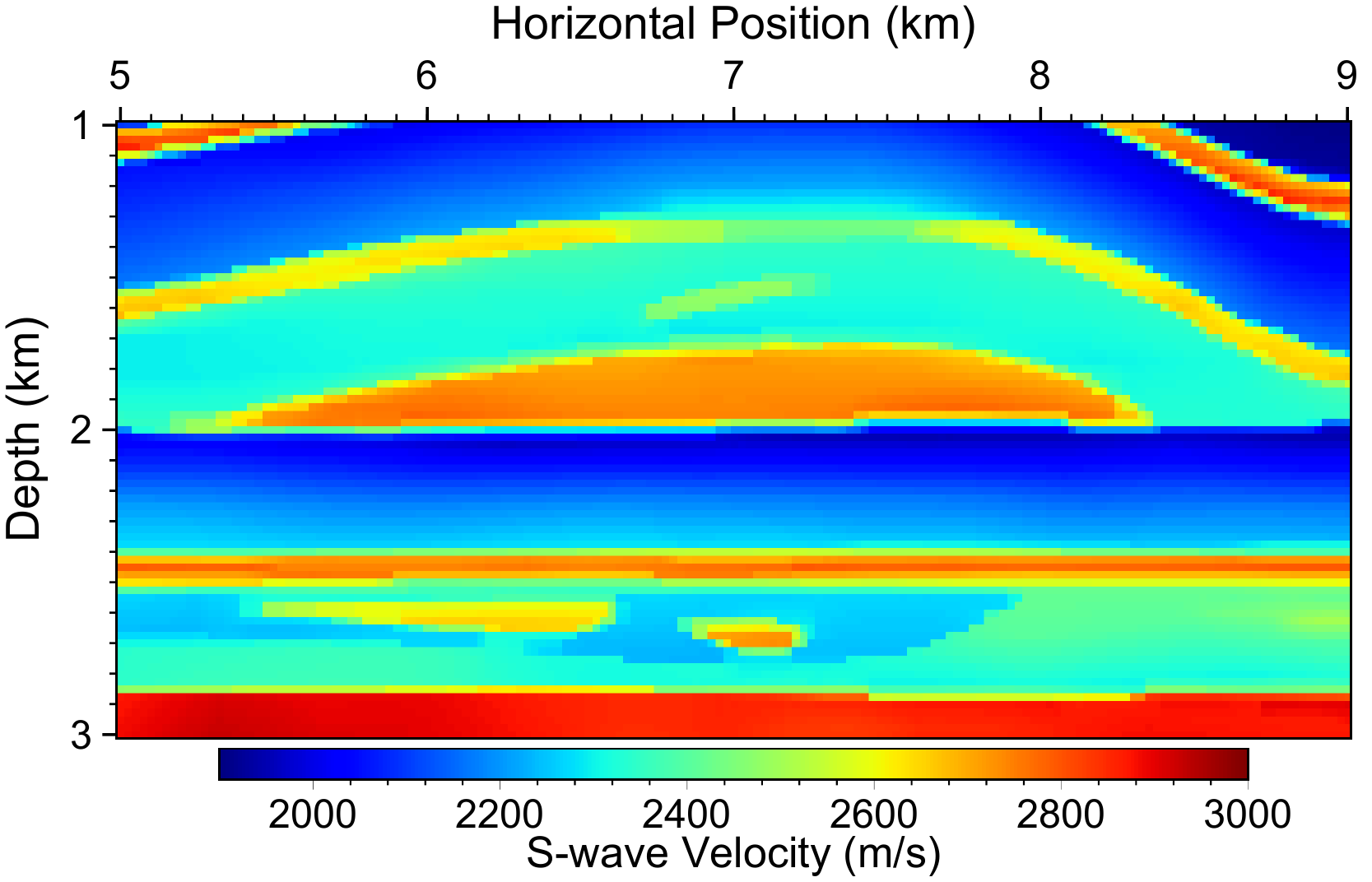}}
\caption{Zoom-in view of the inverted velocity models shown in 
   Fig.~\ref{fig:over_fwi_sparse} using (a) Tikhonov-FWI (b) TV-FWI and (c) TGPV-FWI.  (a), (c) and (e) are inverted P-wave velocities and 
   (b), (d) and (f) are inverted S-wave velocities.}
\label{fig:over_fwi_zoom_sparse}
\end{figure}

In the second numerical test, we verify the efficacy of our TGPV-FWI for 
sparse seismic data acquired using inadequate numbers of sources and 
receivers.
We use seismic data for one third of the sources (27 out of the 80 
sources) and one third of the receivers (133 out of the 399 receivers) to 
conduct inversions. That is, we use only approximately 11 percent of the full 
dataset. The inversion results obtained using the three aforementioned 
inversion methods are shown in Fig.~\ref{fig:over_fwi_sparse}, and the 
zoom-in view of the inversion results in the region of interest are shown 
in Fig.~\ref{fig:over_fwi_zoom_sparse}. Both results of the Tikhonov-FWI and TV-FWI in 
Figs.~\ref{fig:over_fwi_sparse}a-d contain more inversion artifacts than 
those in Figs.~\ref{fig:over_fwi}a-d. By contrast, our TGPV-FWI results in 
Figs.~\ref{fig:over_fwi_sparse}e and f are almost identical to those in 
Figs.~\ref{fig:over_fwi}e and f. These inversion results demonstrate the 
capability of our TGPV-FWI to accurately reconstruct seismic velocities 
in complex models using sparse seismic data. 

\begin{figure}
\centering
\subfloat[]{\includegraphics[width=0.485\textwidth]{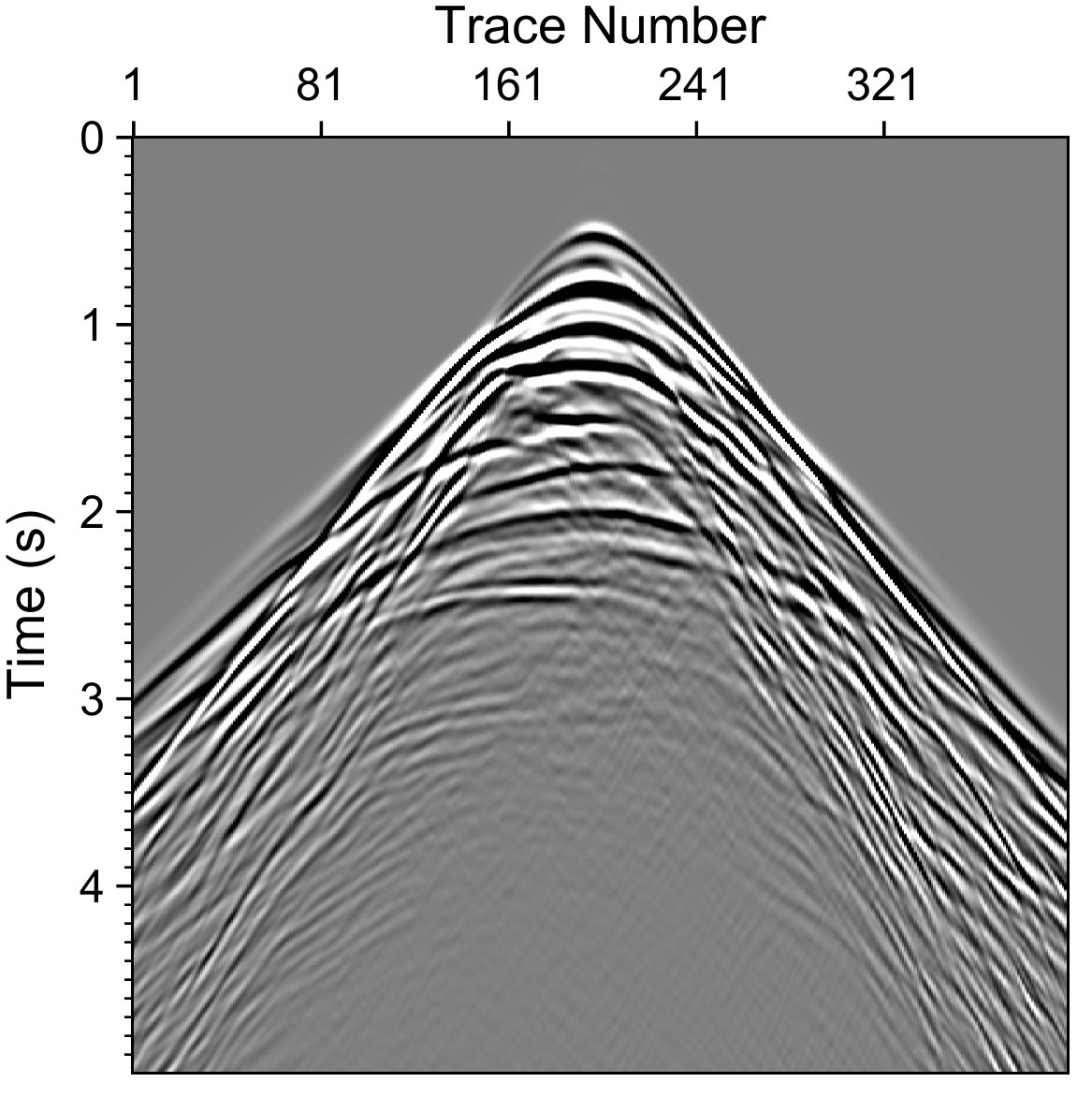}} 
\subfloat[]{\includegraphics[width=0.485\textwidth]{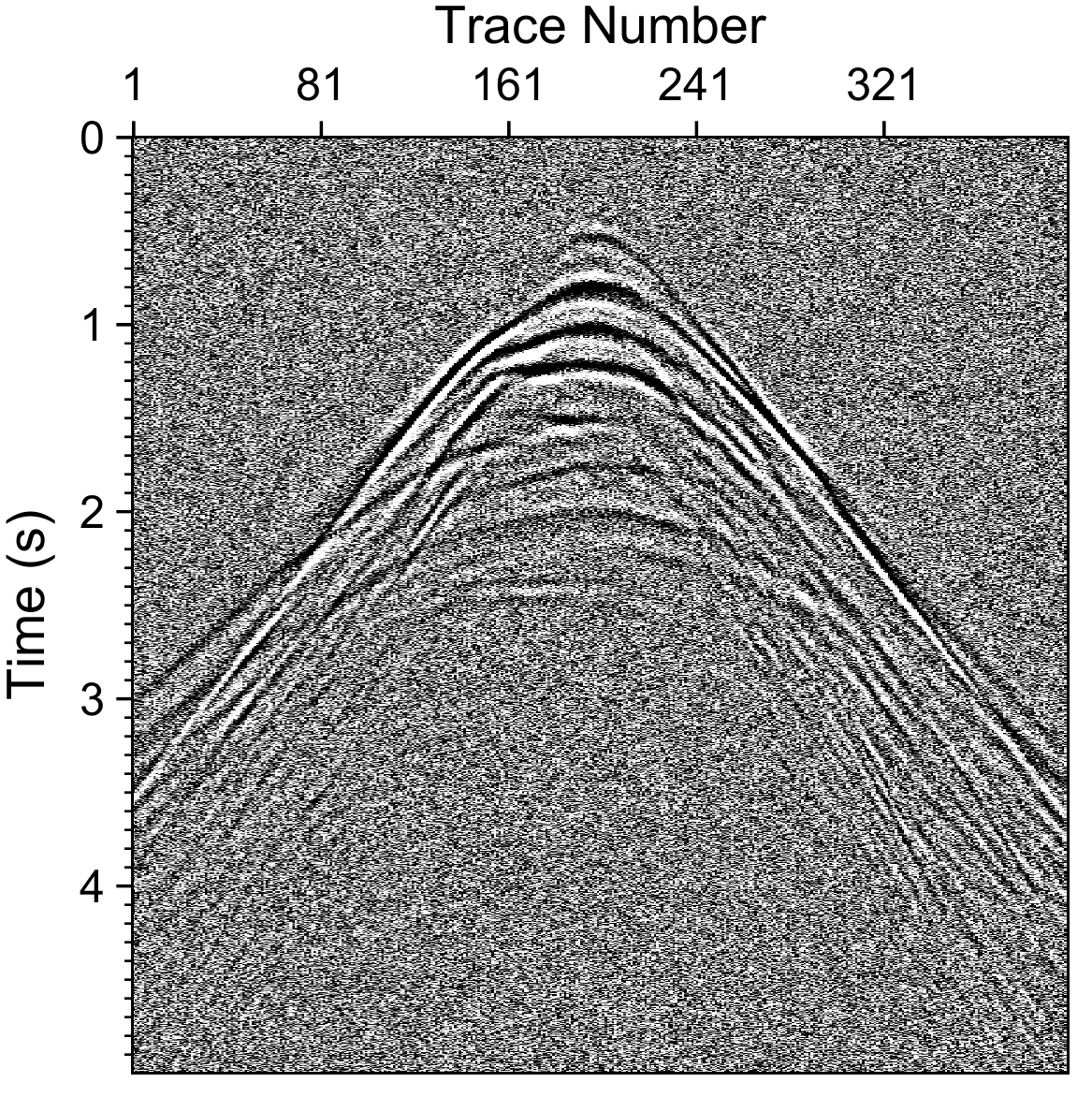}}
   \caption{Noise-free (a) and noisy (b) vertical component data of the 
   40th common-shot gather for the modified elastic SEG/EAGE overthrust 
   model.}
\label{fig:over_data}
\end{figure}

We conduct the third numerical test using noisy data.  
Fig.~\ref{fig:over_data}a shows the vertical component data of the 40th 
common-shot gather for the modified elastic SEG/EAGE overthrust model.  
Fig.~\ref{fig:over_data}b depicts the same vertical component data with 
random noise. The random noise obviously deteriorates the quality of the 
synthetic data, making many reflections indiscernible from the random 
noises. We also add the same level of random noise to the horizontal 
component.  

\begin{figure}
\centering
\subfloat[]{\includegraphics[width=0.485\textwidth]{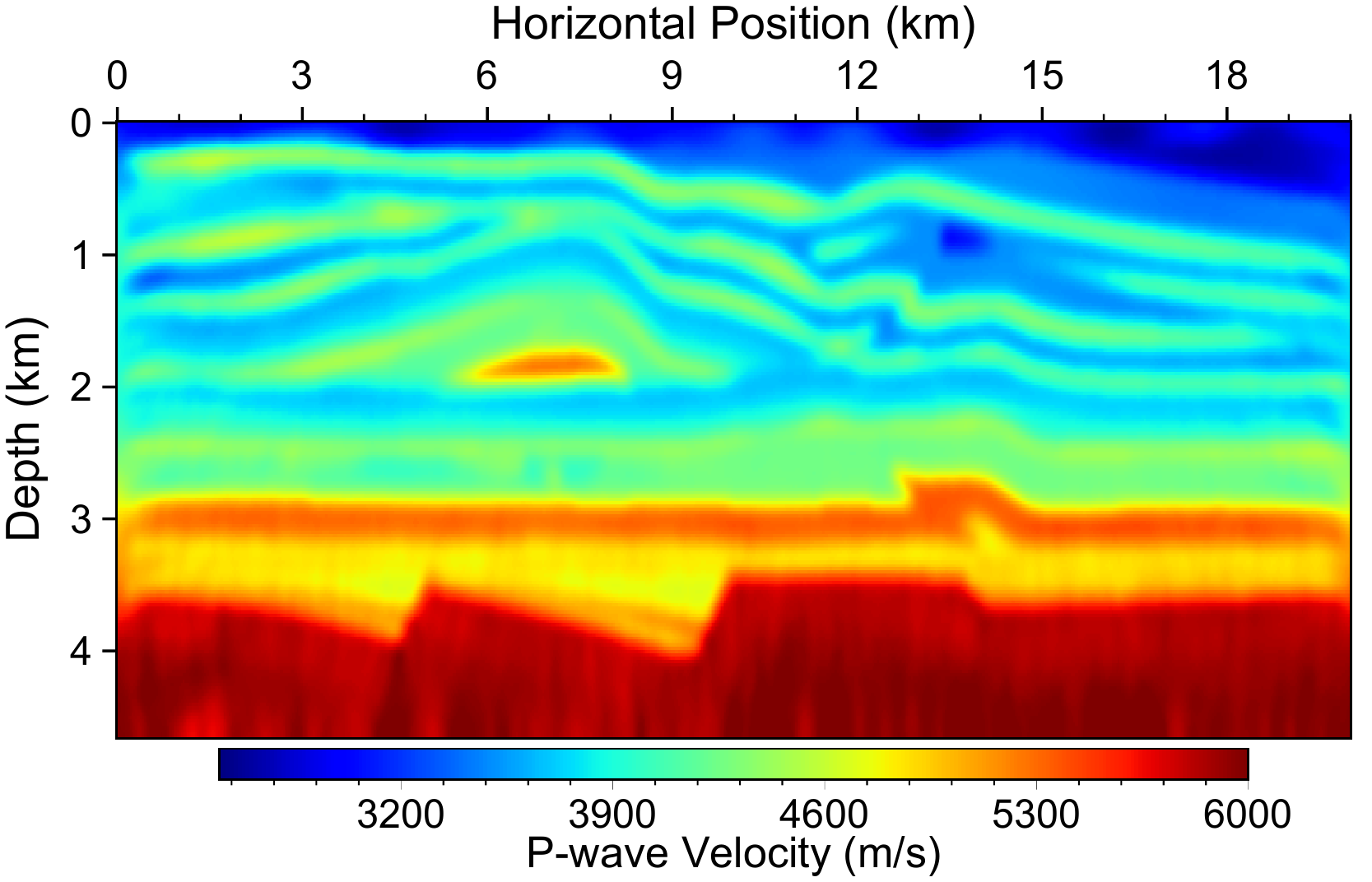}} 
\subfloat[]{\includegraphics[width=0.485\textwidth]{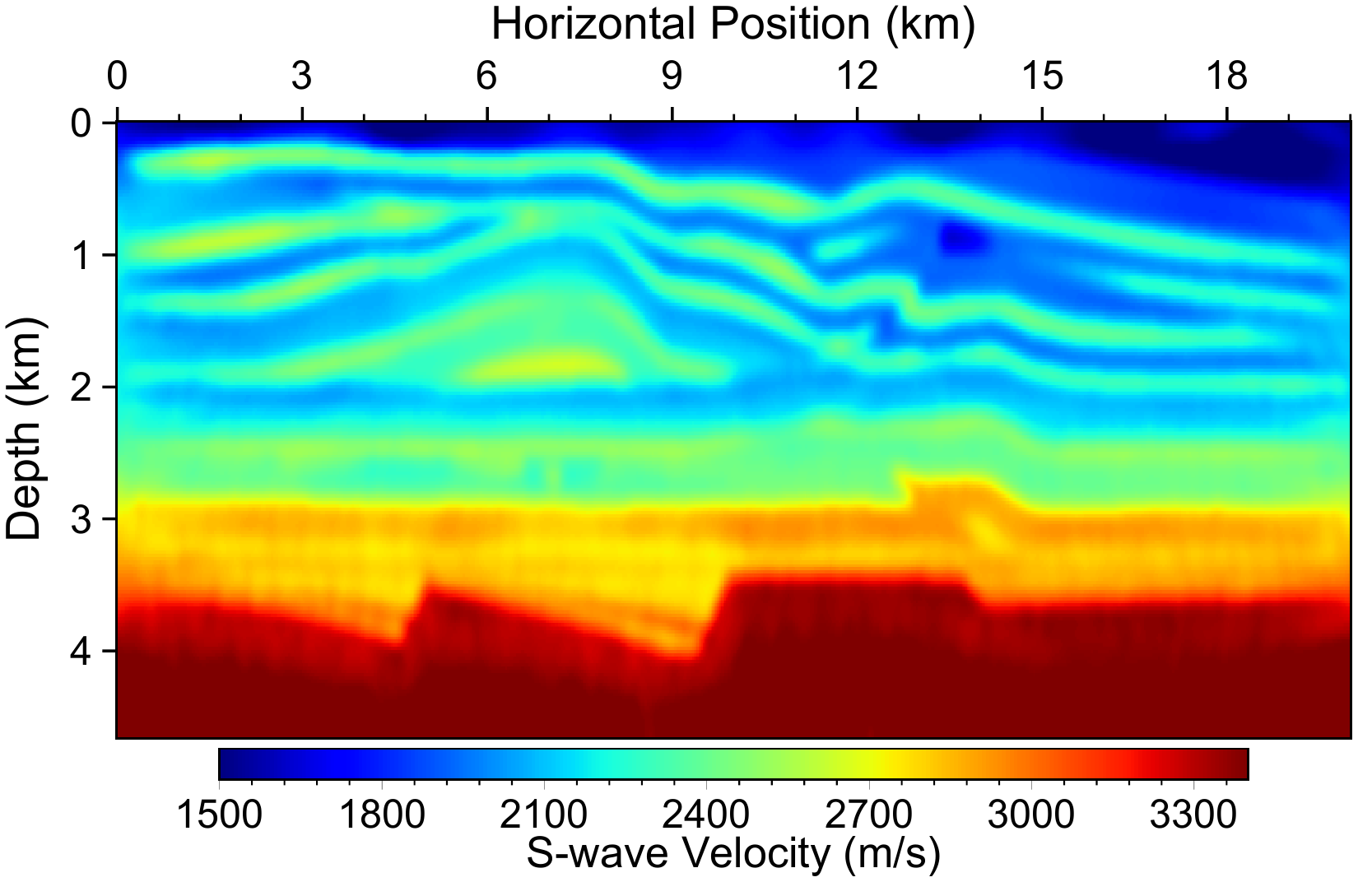}}  \\
\subfloat[]{\includegraphics[width=0.485\textwidth]{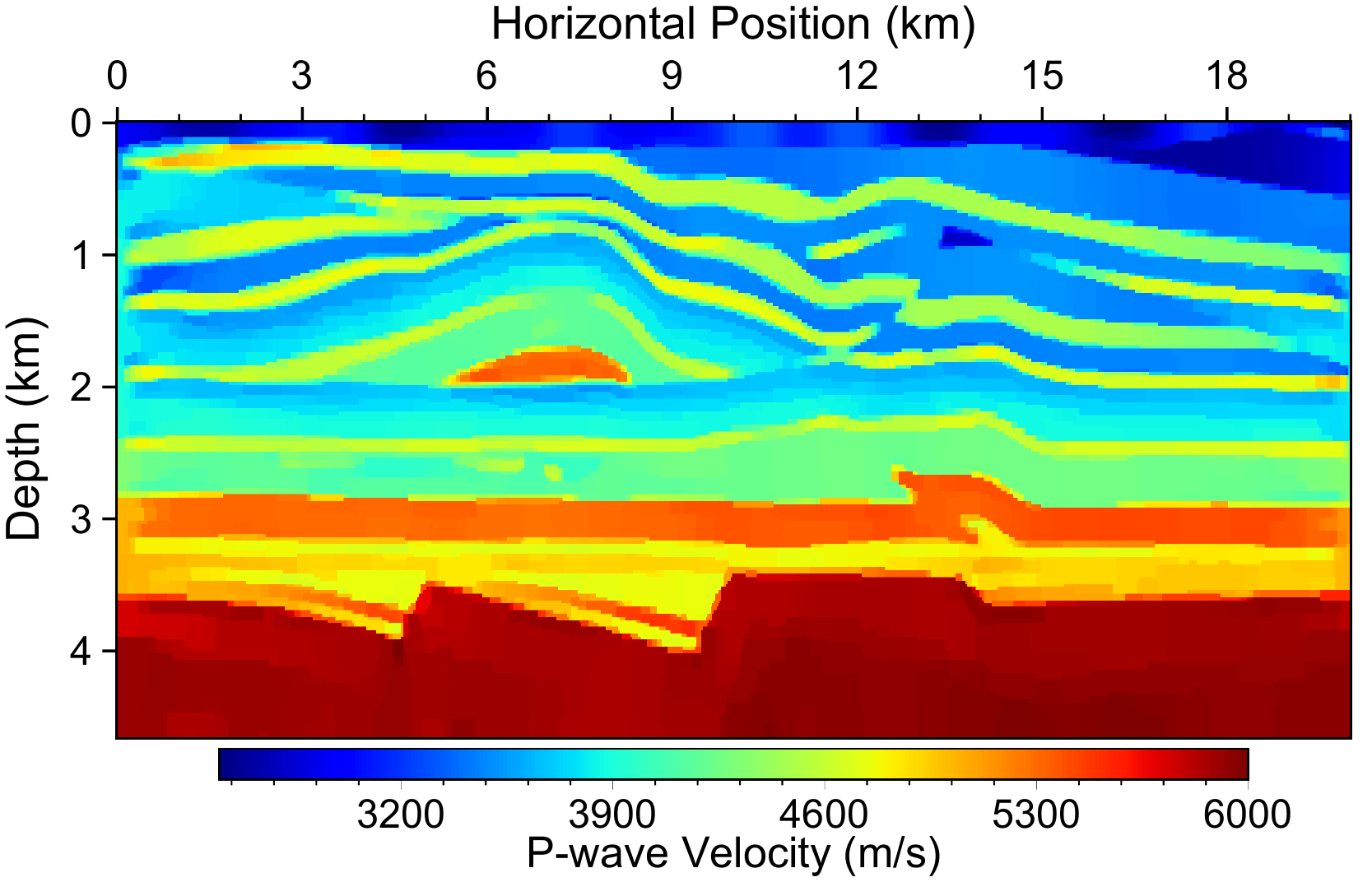}}
\subfloat[]{\includegraphics[width=0.485\textwidth]{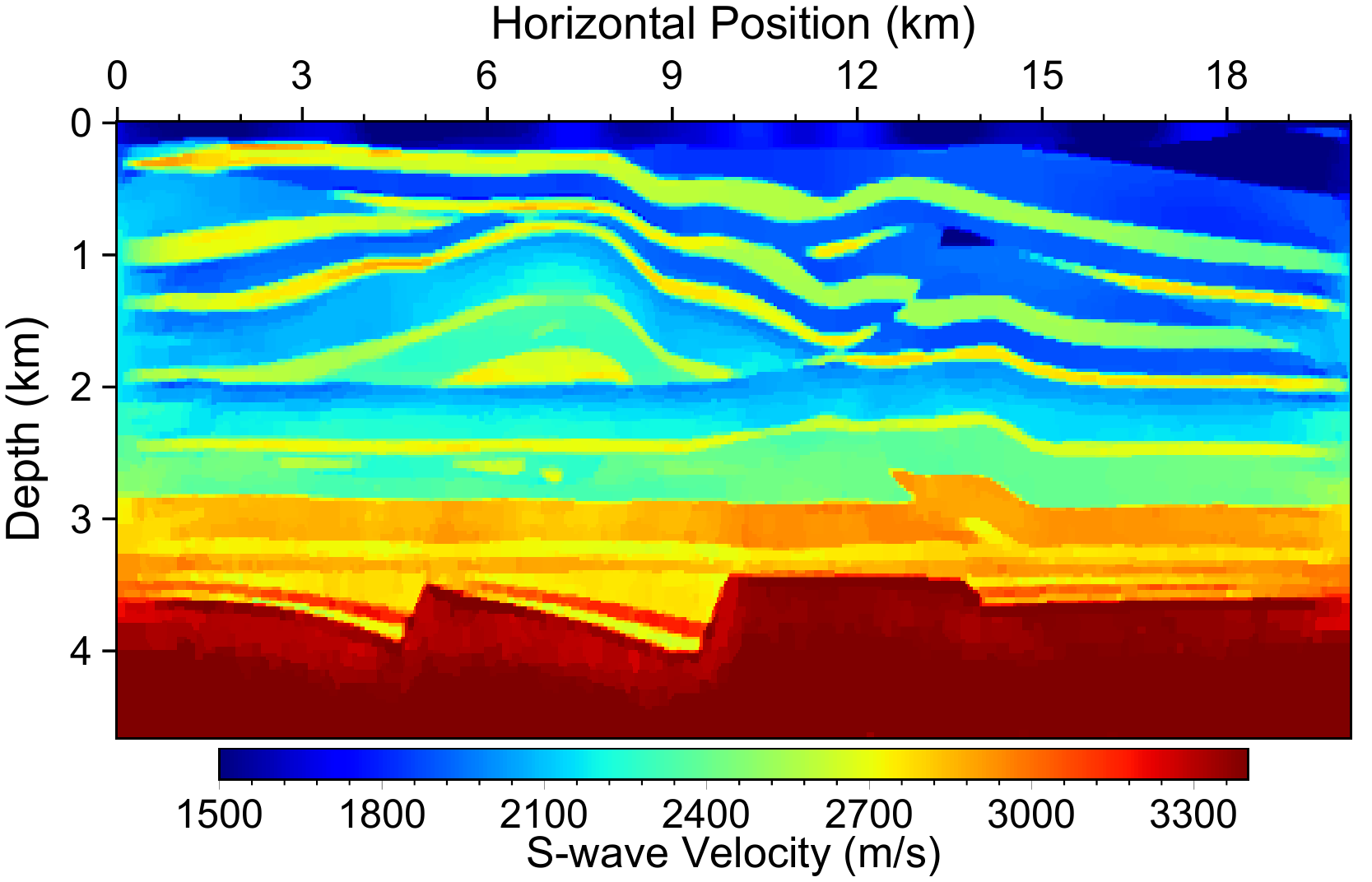}} \\
\subfloat[]{\includegraphics[width=0.485\textwidth]{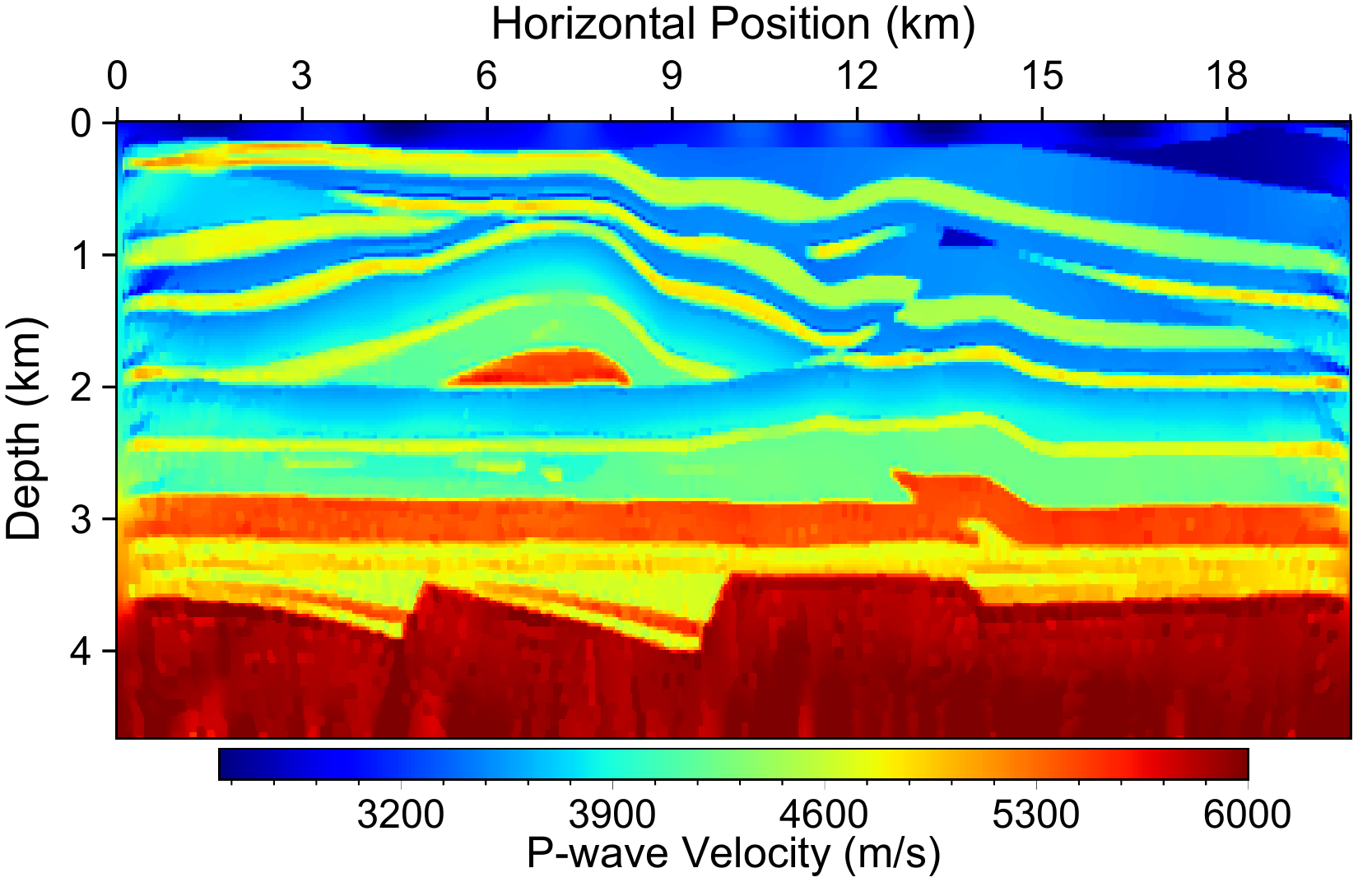}}
\subfloat[]{\includegraphics[width=0.485\textwidth]{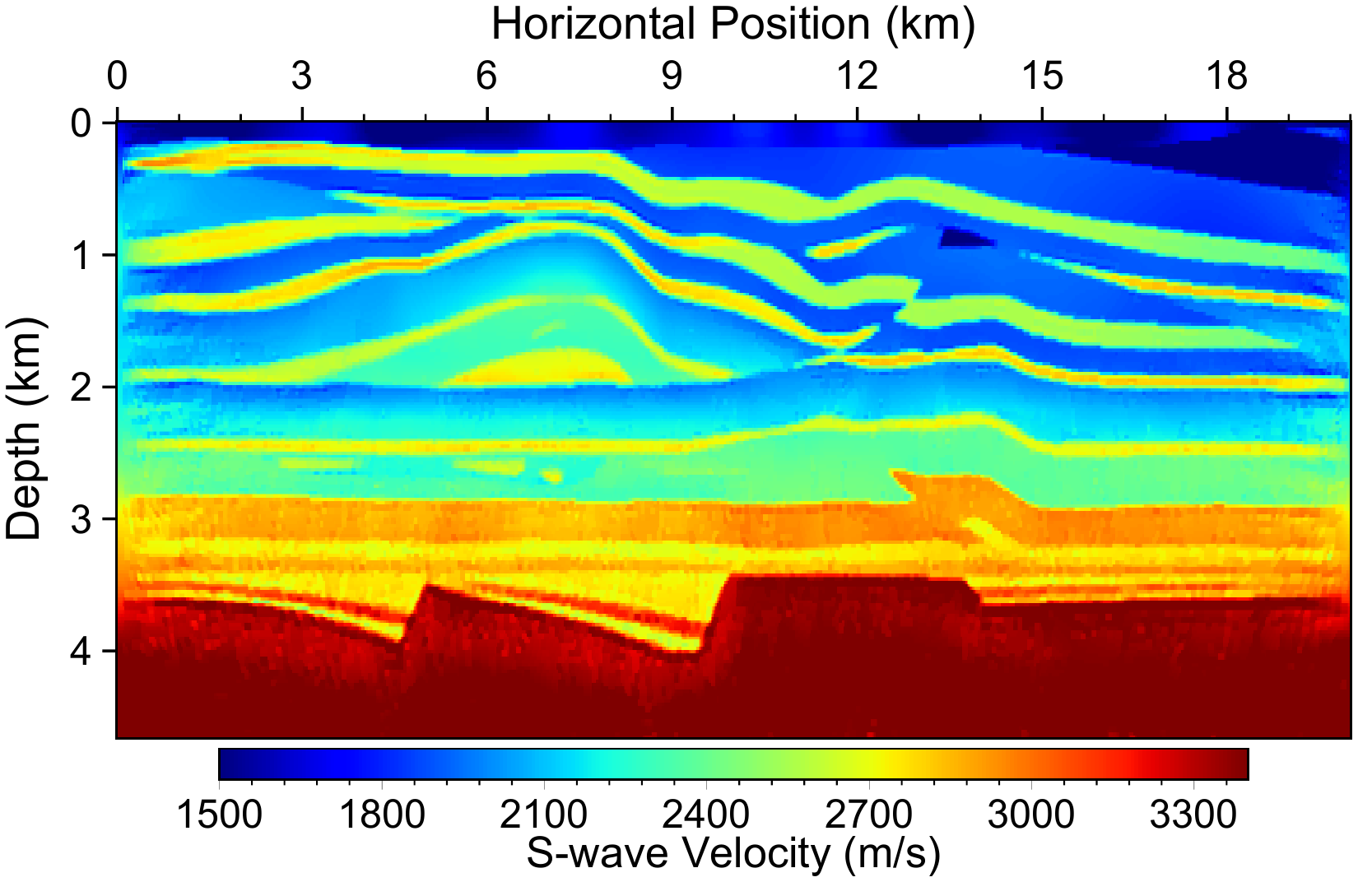}}
\caption{Inverted velocity models produced using (a) Tikhonov-FWI (b) TV-FWI and (c) TGPV-FWI. Panels in (a), (c) and (e) are inverted P-wave 
   velocities and those in (b), (d) and (f) are inverted S-wave 
   velocities. All three inversions use noisy data.}
\label{fig:over_fwi_noisy}
\end{figure}

\begin{figure}
\centering
\subfloat[]{\includegraphics[width=0.485\textwidth]{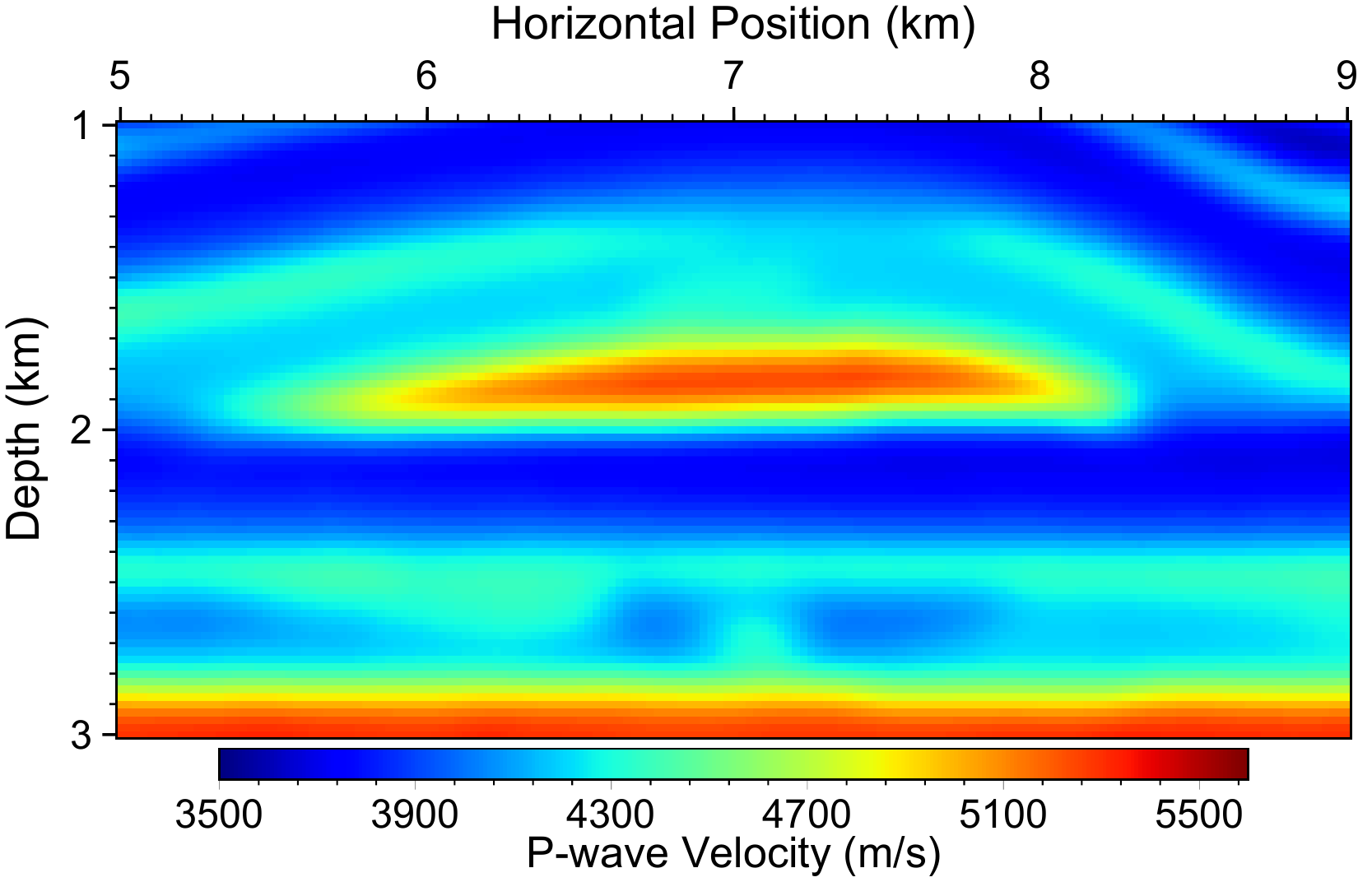}} 
\subfloat[]{\includegraphics[width=0.485\textwidth]{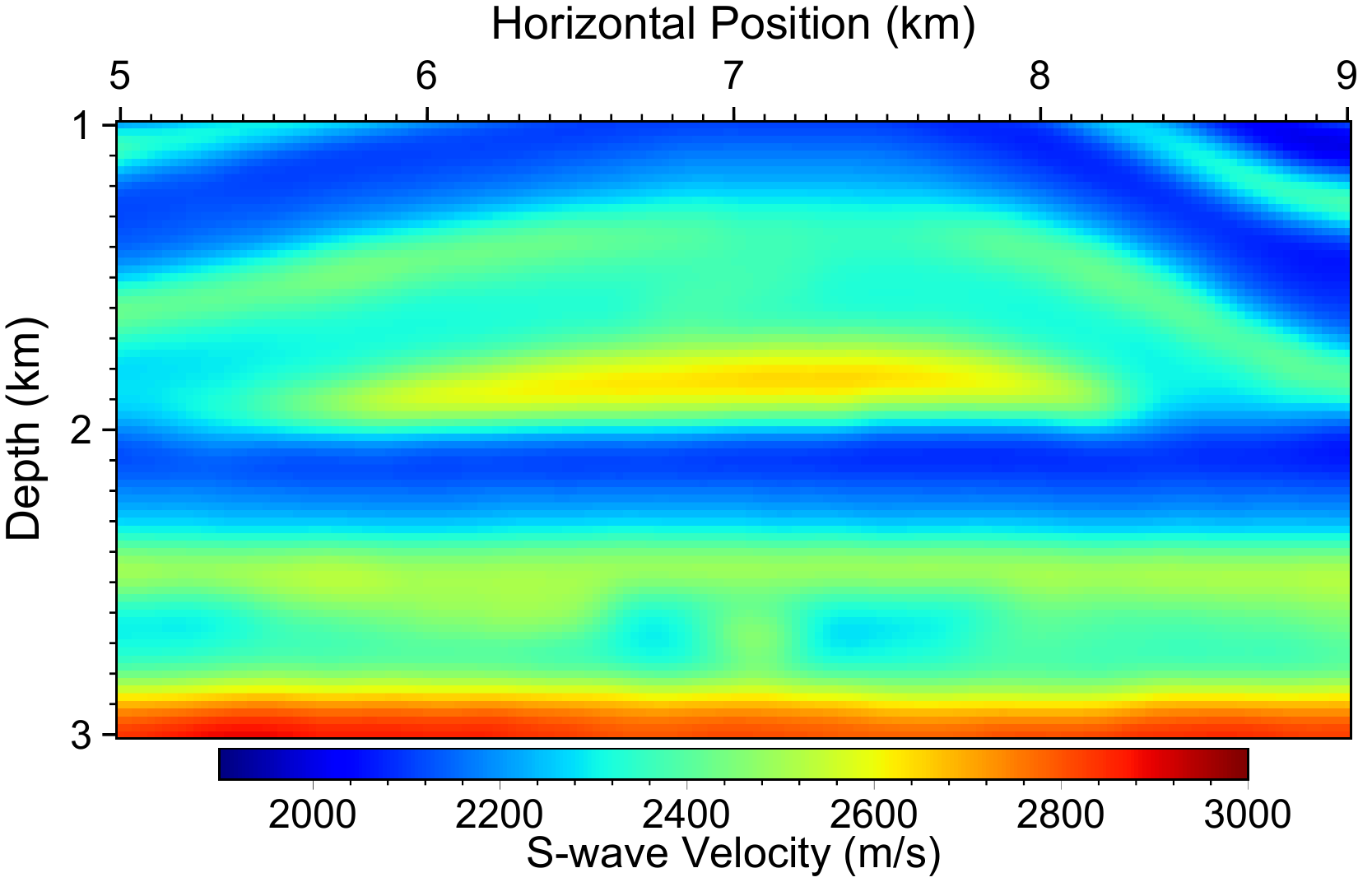}} \\
\subfloat[]{\includegraphics[width=0.485\textwidth]{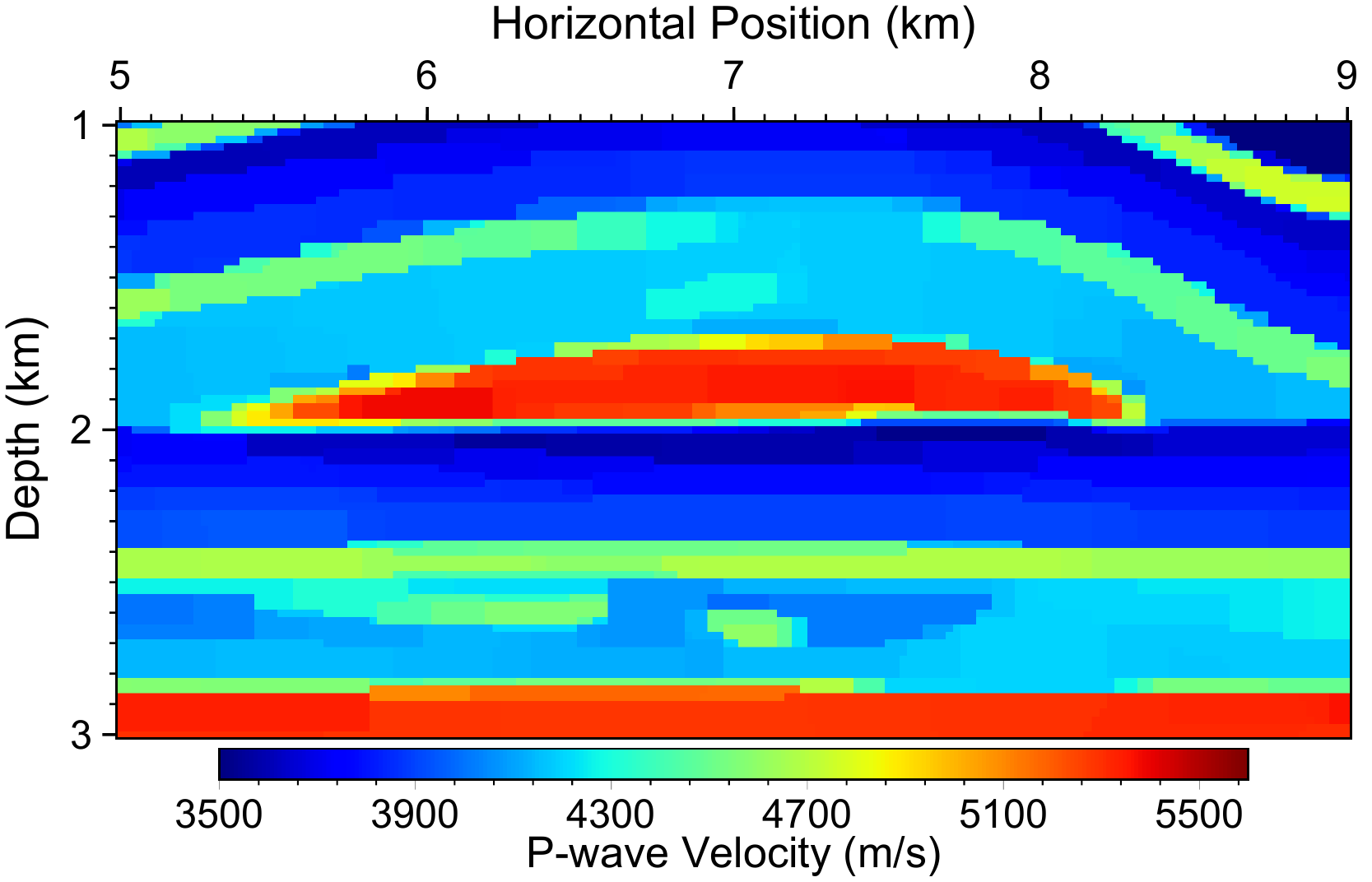}}
\subfloat[]{\includegraphics[width=0.485\textwidth]{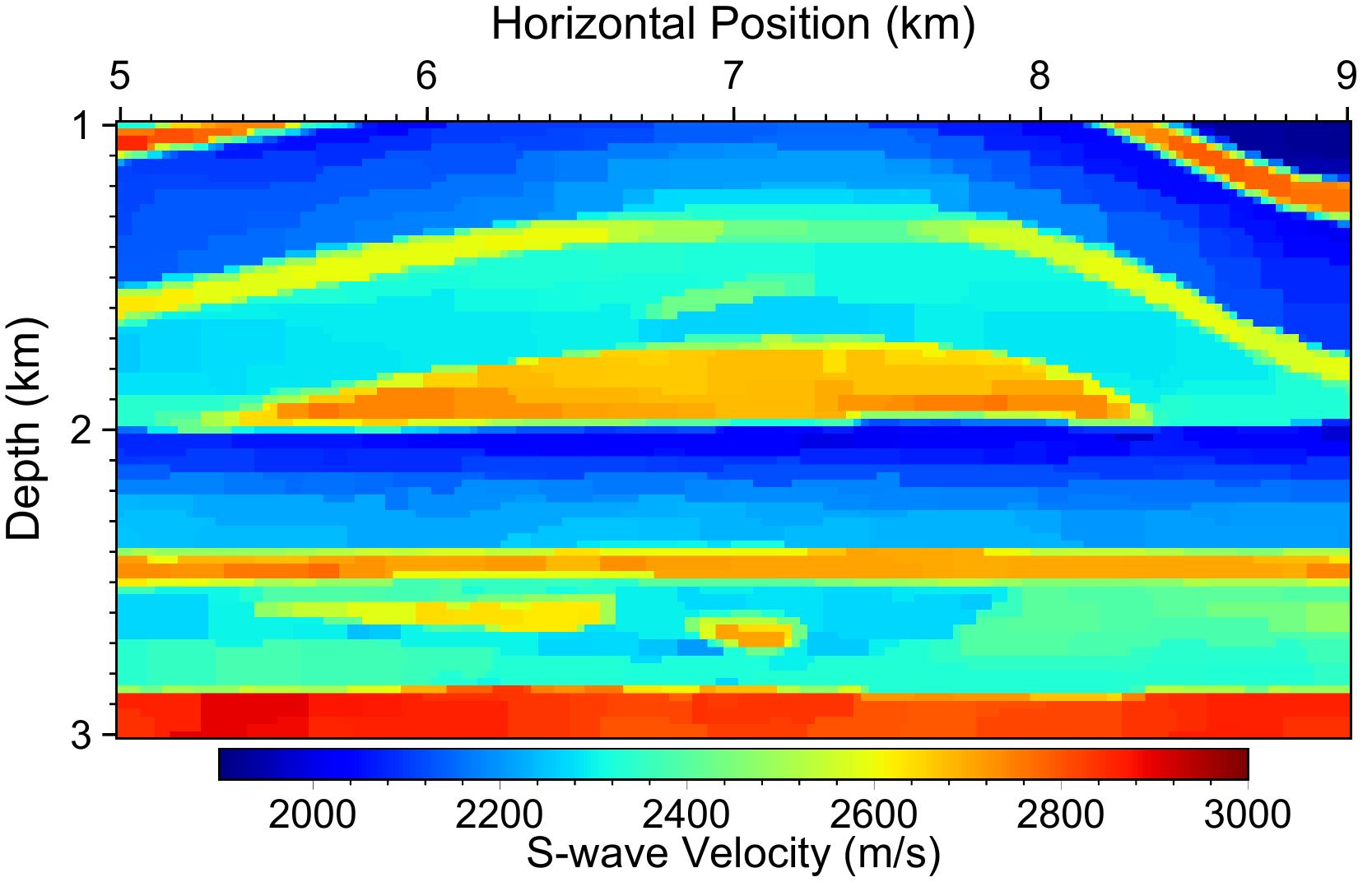}} \\
\subfloat[]{\includegraphics[width=0.485\textwidth]{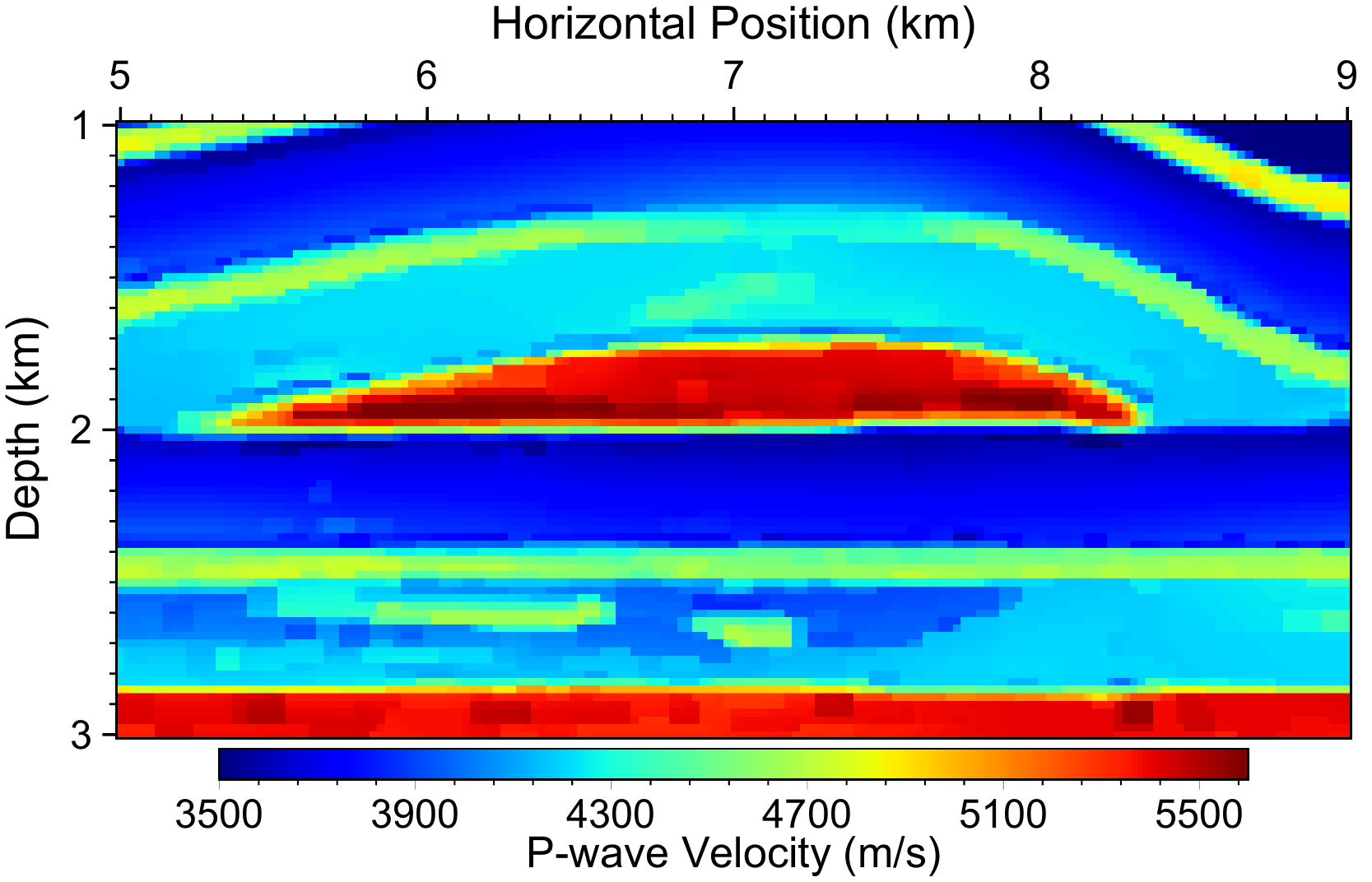}}
\subfloat[]{\includegraphics[width=0.485\textwidth]{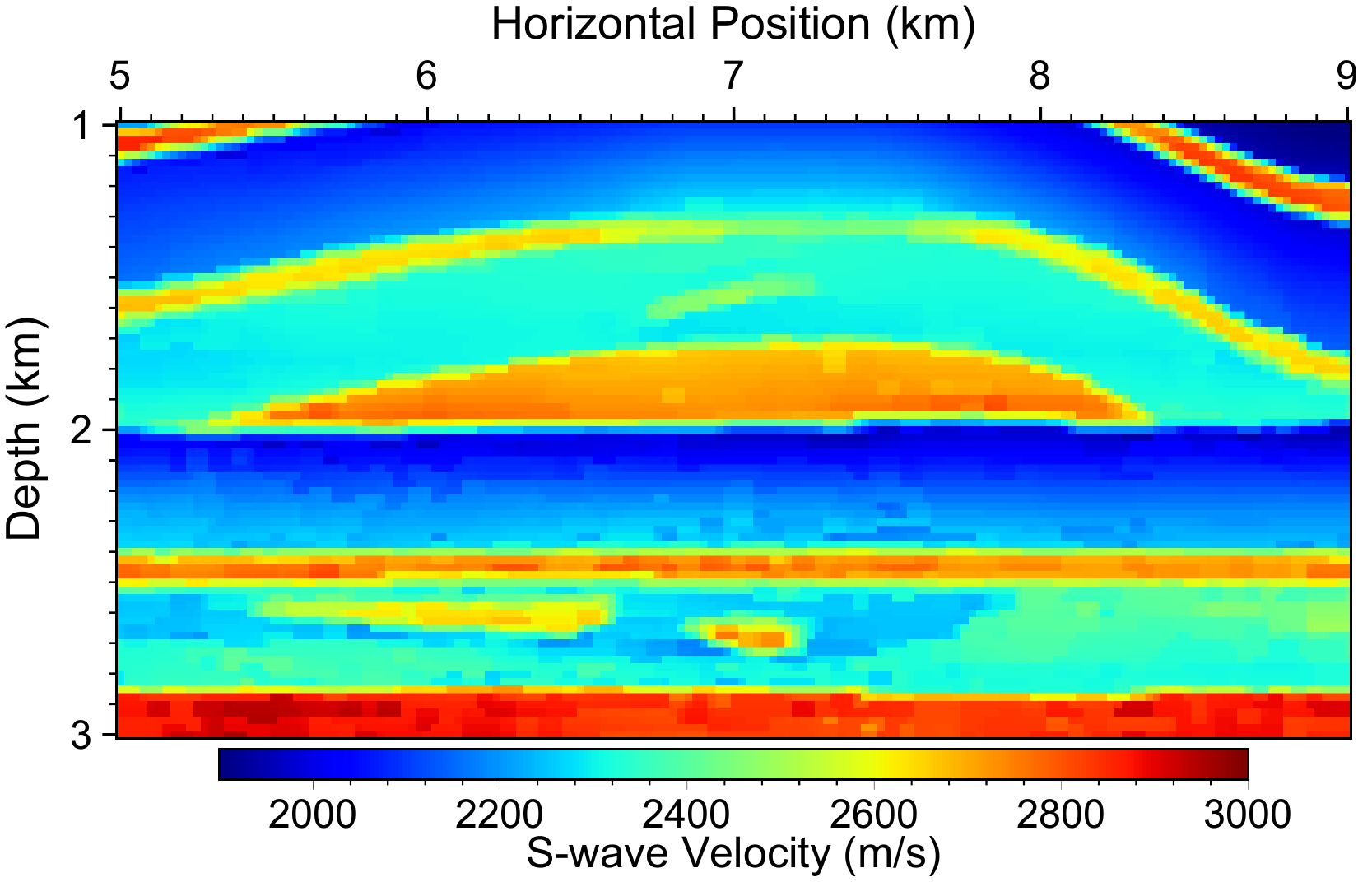}}
\caption{Zoom-in views of the region of interest in the inverted velocity 
   models shown in Fig.~\ref{fig:over_fwi_noisy} yielded using (a) Tikhonov-FWI (b) TV-FWI and (c) TGPV-FWI. Panels in  (a), (c) and (e) are inverted P-wave 
   velocities and those in (b), (d) and (f) are inverted S-wave 
   velocities.}
\label{fig:over_fwi_zoom_noisy}
\end{figure}

We show the inversion results obtained using the Tikhonov-FWI, the TV-FWI and the TGPV-FWI in Fig.~\ref{fig:over_fwi_noisy} and 
Fig.~\ref{fig:over_fwi_zoom_noisy}. In this case, the noises in the data 
leads to degraded inversion results compared with those in the previous 
two numerical tests. Nevertheless, our TGPV-FWI results shown in 
Figs.~\ref{fig:over_fwi_zoom_noisy}e and f are better than those produced 
with Tikhonov-FWI 
(Figs.~\ref{fig:over_fwi_zoom_noisy}a and b) and TV-FWI (Figs.~\ref{fig:over_fwi_zoom_noisy}c and d). 

\begin{figure}
\centering
\subfloat[]{\includegraphics[width=0.485\textwidth]{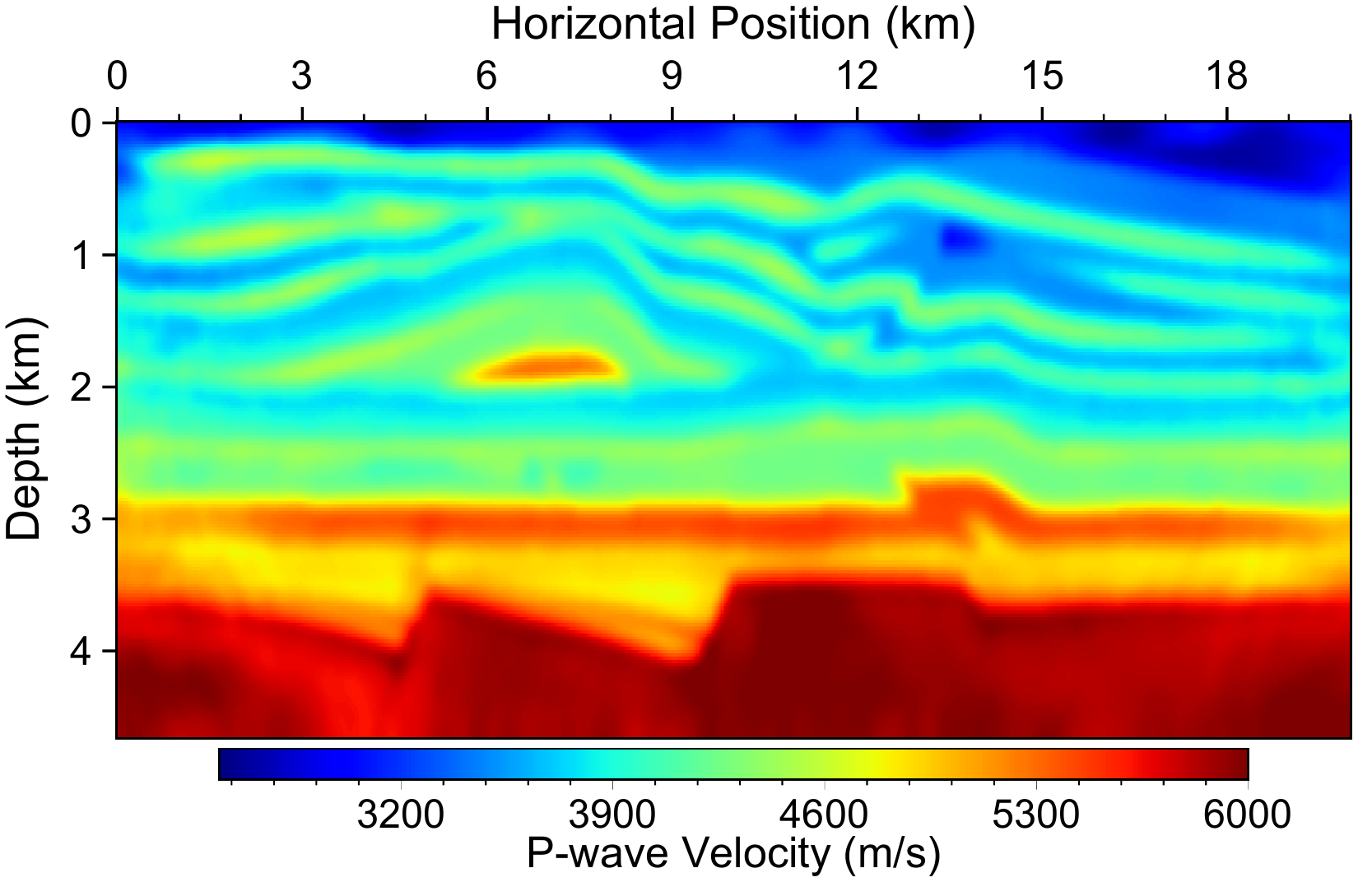}} 
\subfloat[]{\includegraphics[width=0.485\textwidth]{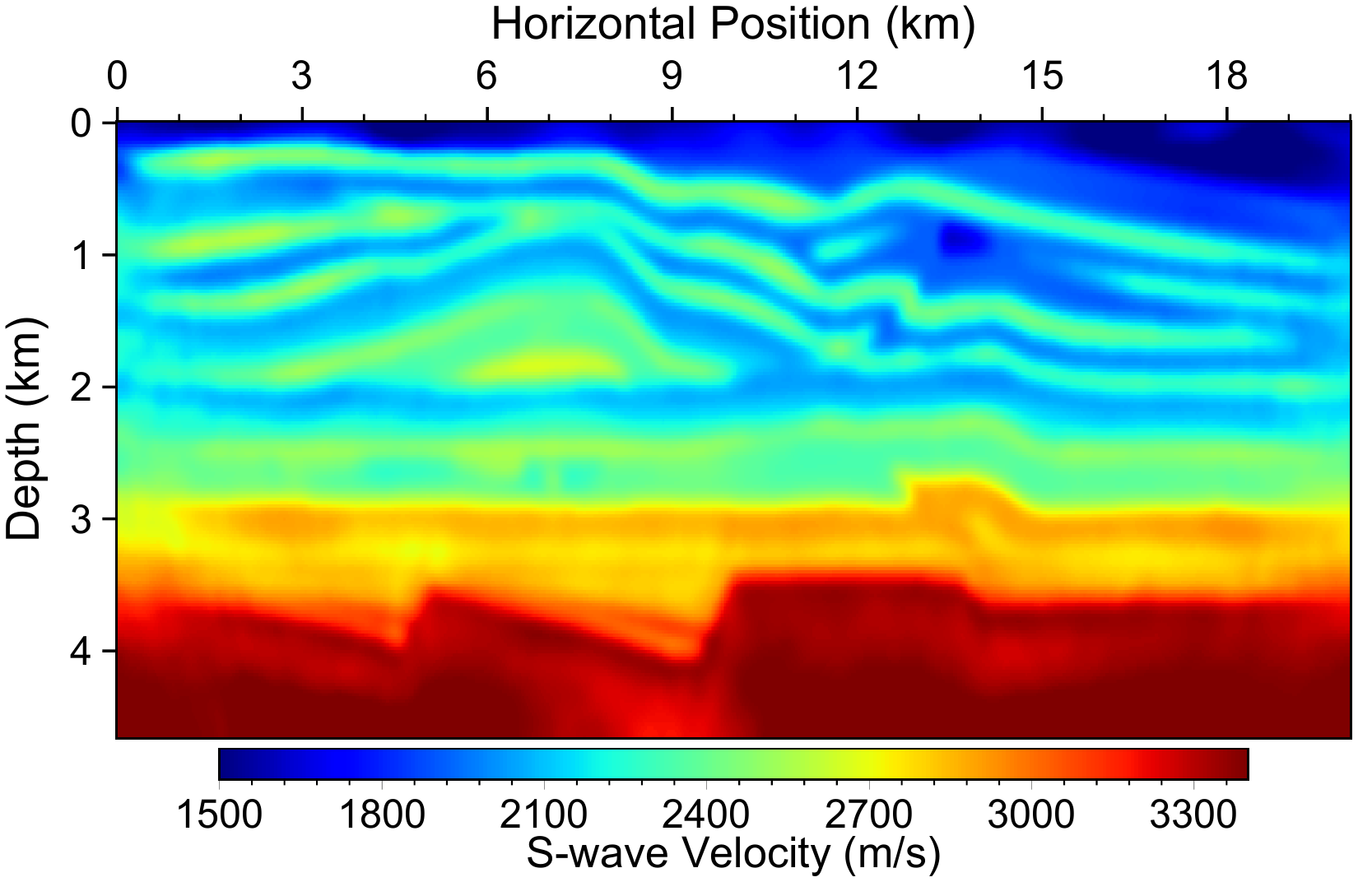}} \\
\subfloat[]{\includegraphics[width=0.485\textwidth]{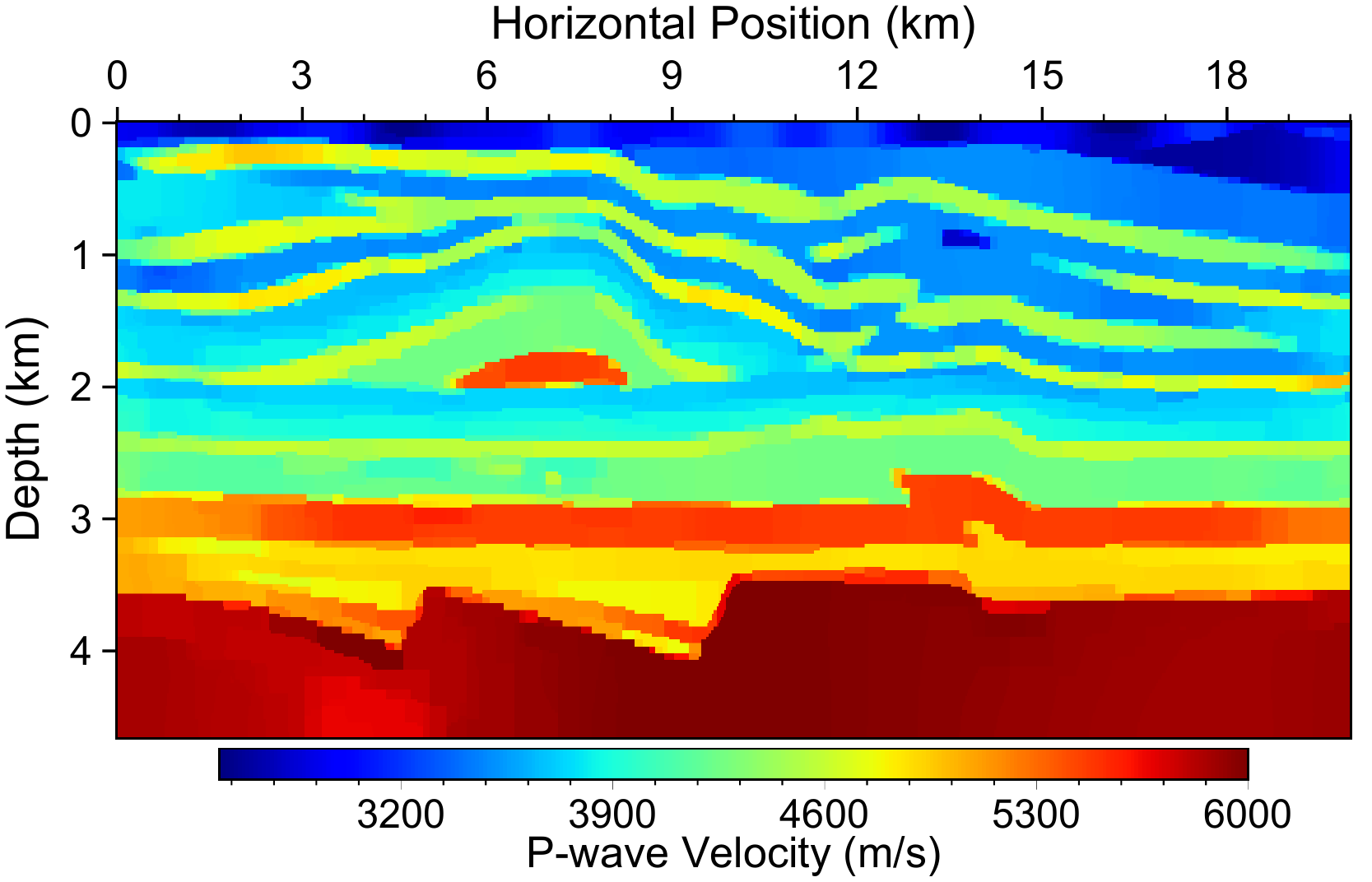}}
\subfloat[]{\includegraphics[width=0.485\textwidth]{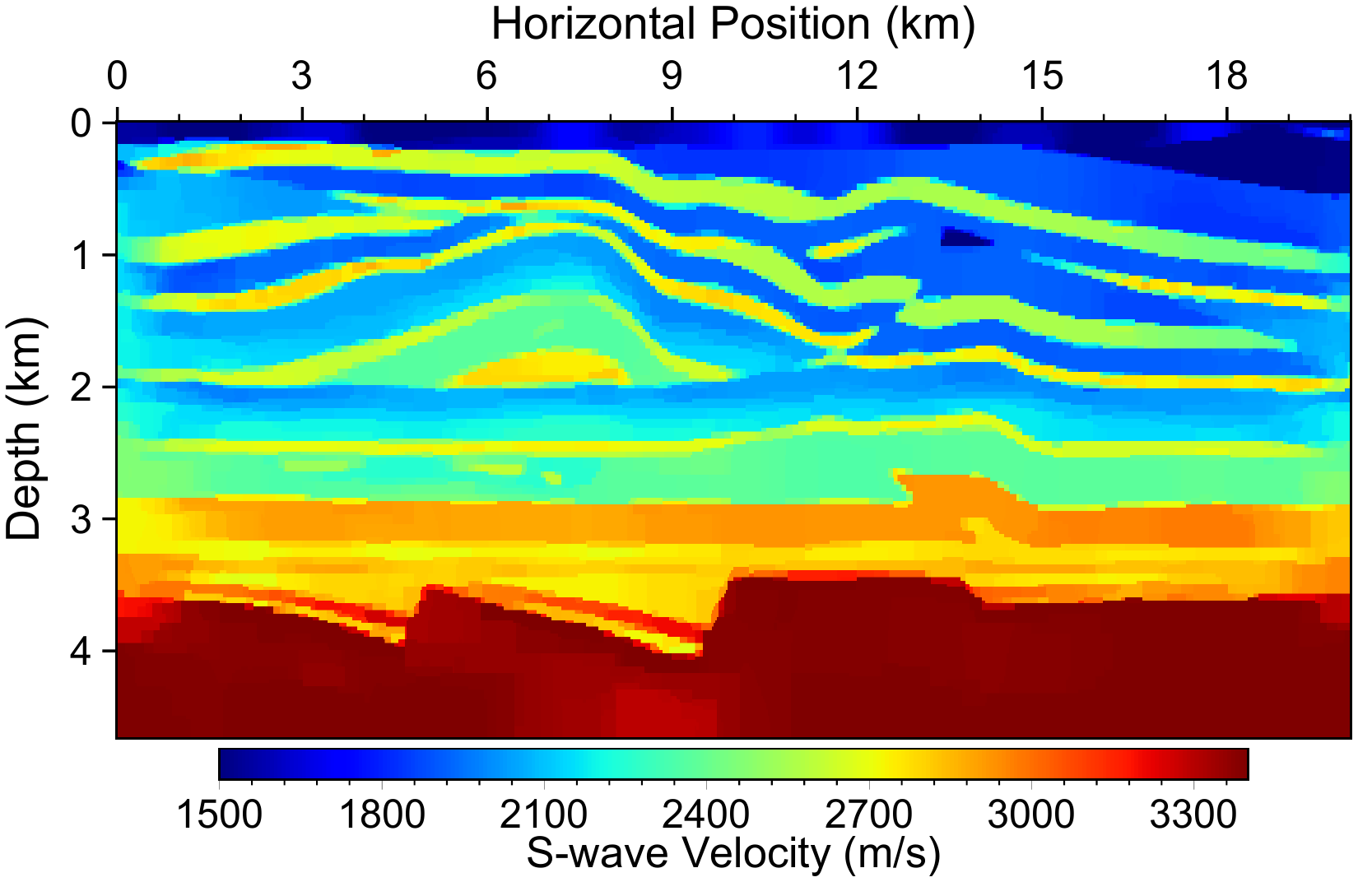}} \\
\subfloat[]{\includegraphics[width=0.485\textwidth]{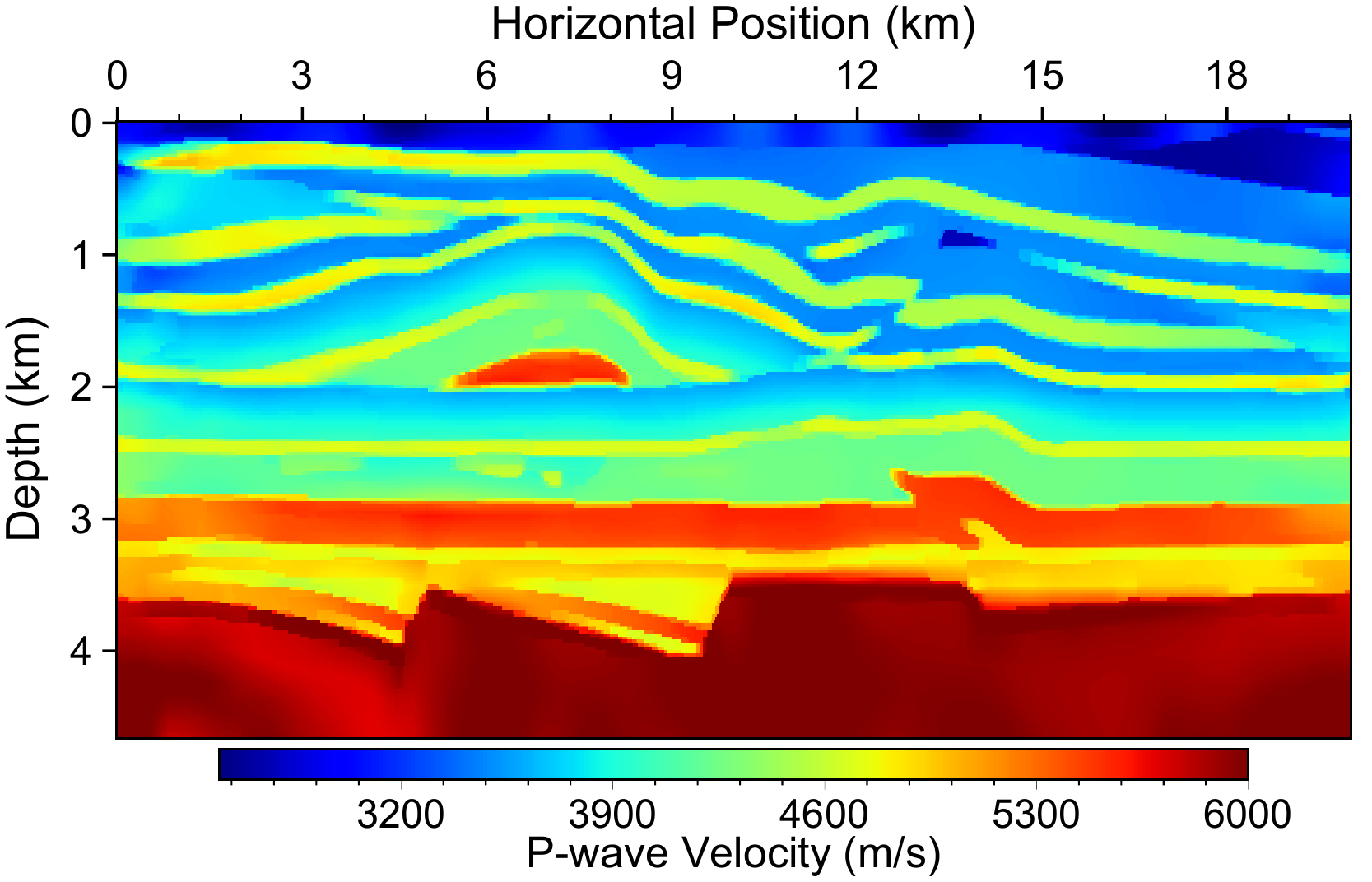}}
\subfloat[]{\includegraphics[width=0.485\textwidth]{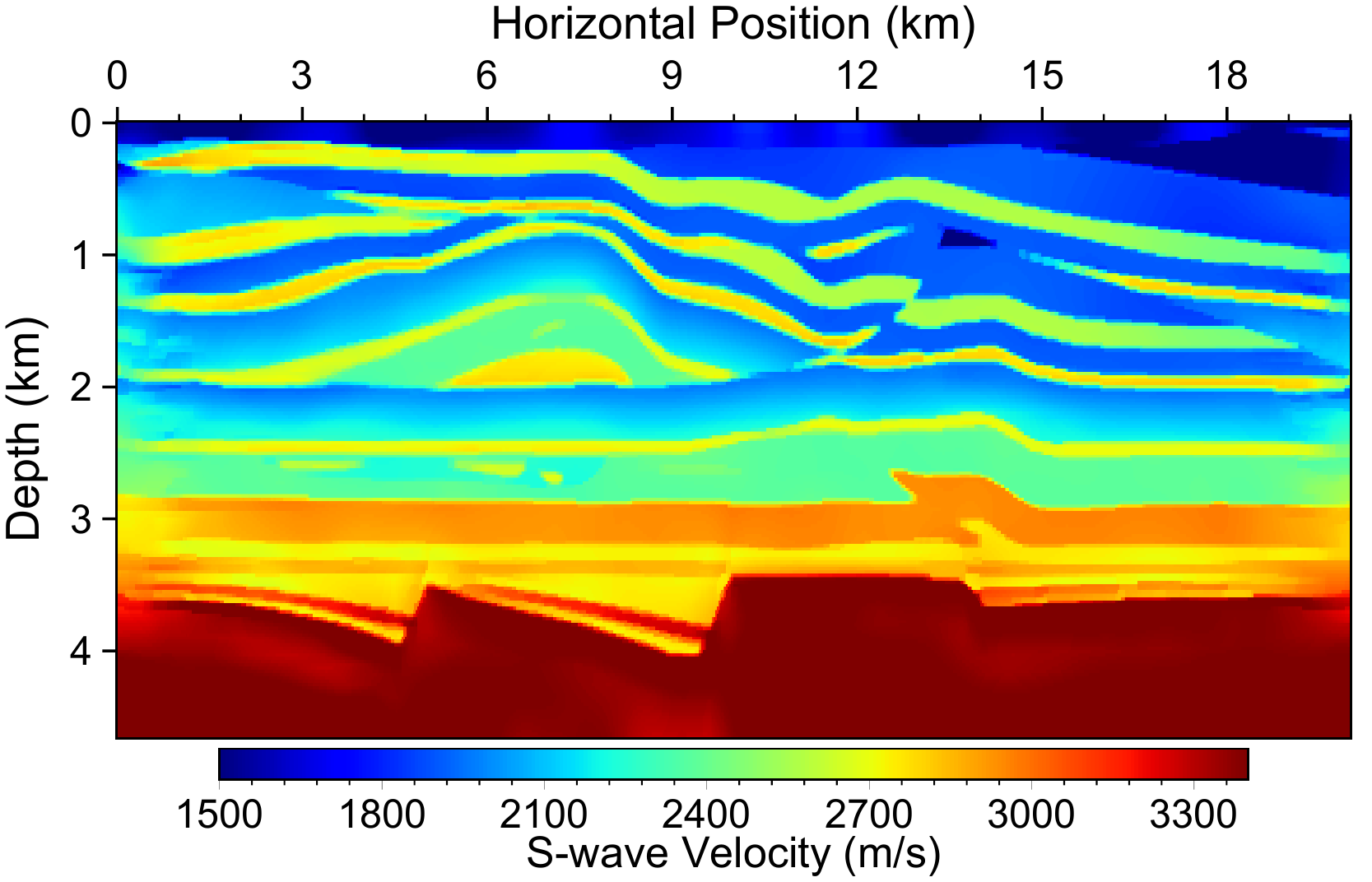}}
\caption{Inverted velocity models obtained using (a) Tikhonov-FWI (b) TV-FWI and (c) TGPV-FWI. Panels in (a), (c) and (e) are inverted P-wave 
   velocities and those in (b), (d) and (f) are inverted S-wave 
   velocities.  All three inversions use dynamic random source encoding.}
\label{fig:over_fwi_encoding}
\end{figure}

\begin{figure}
\centering
\subfloat[]{\includegraphics[width=0.485\textwidth]{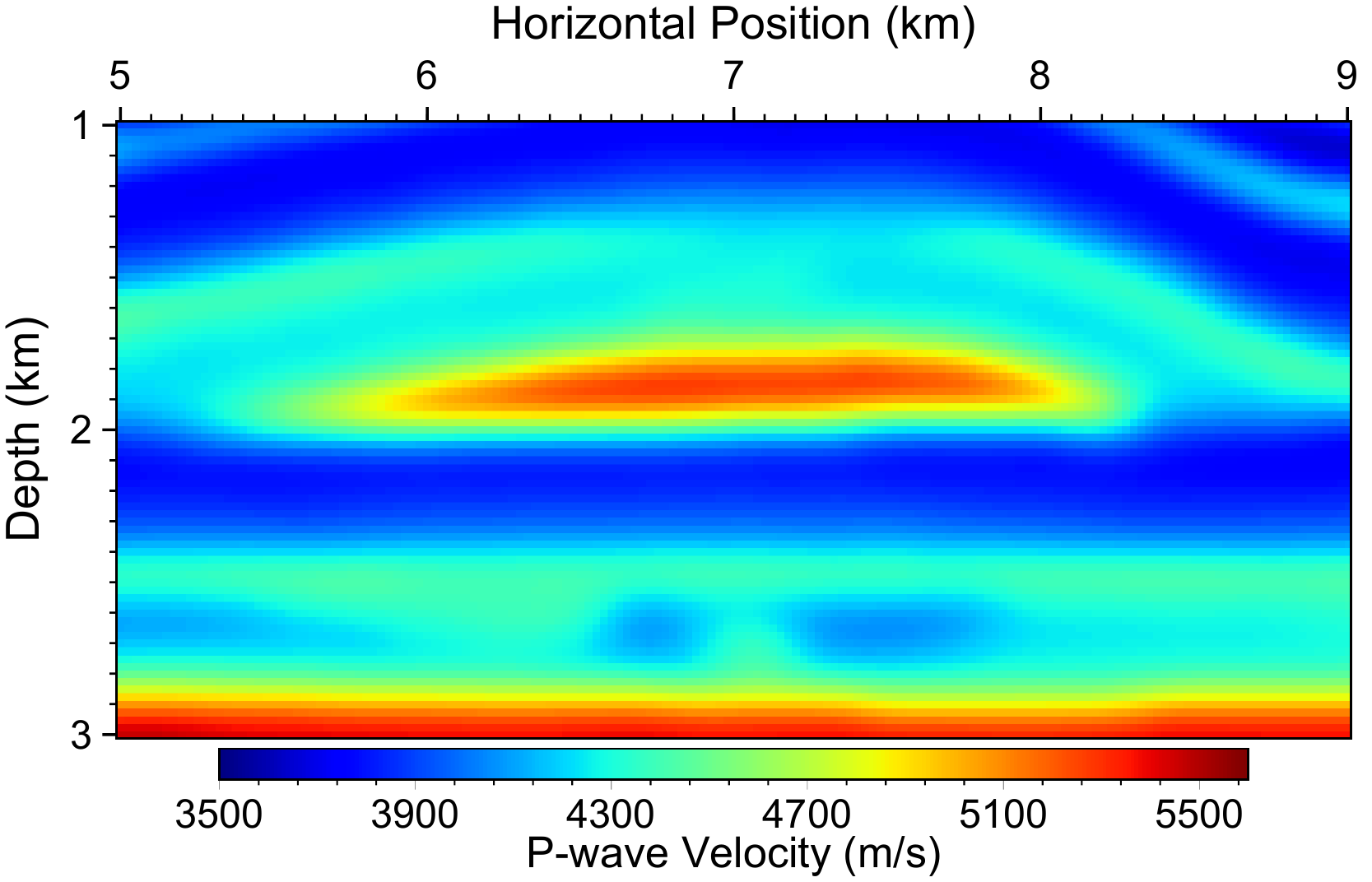}} 
\subfloat[]{\includegraphics[width=0.485\textwidth]{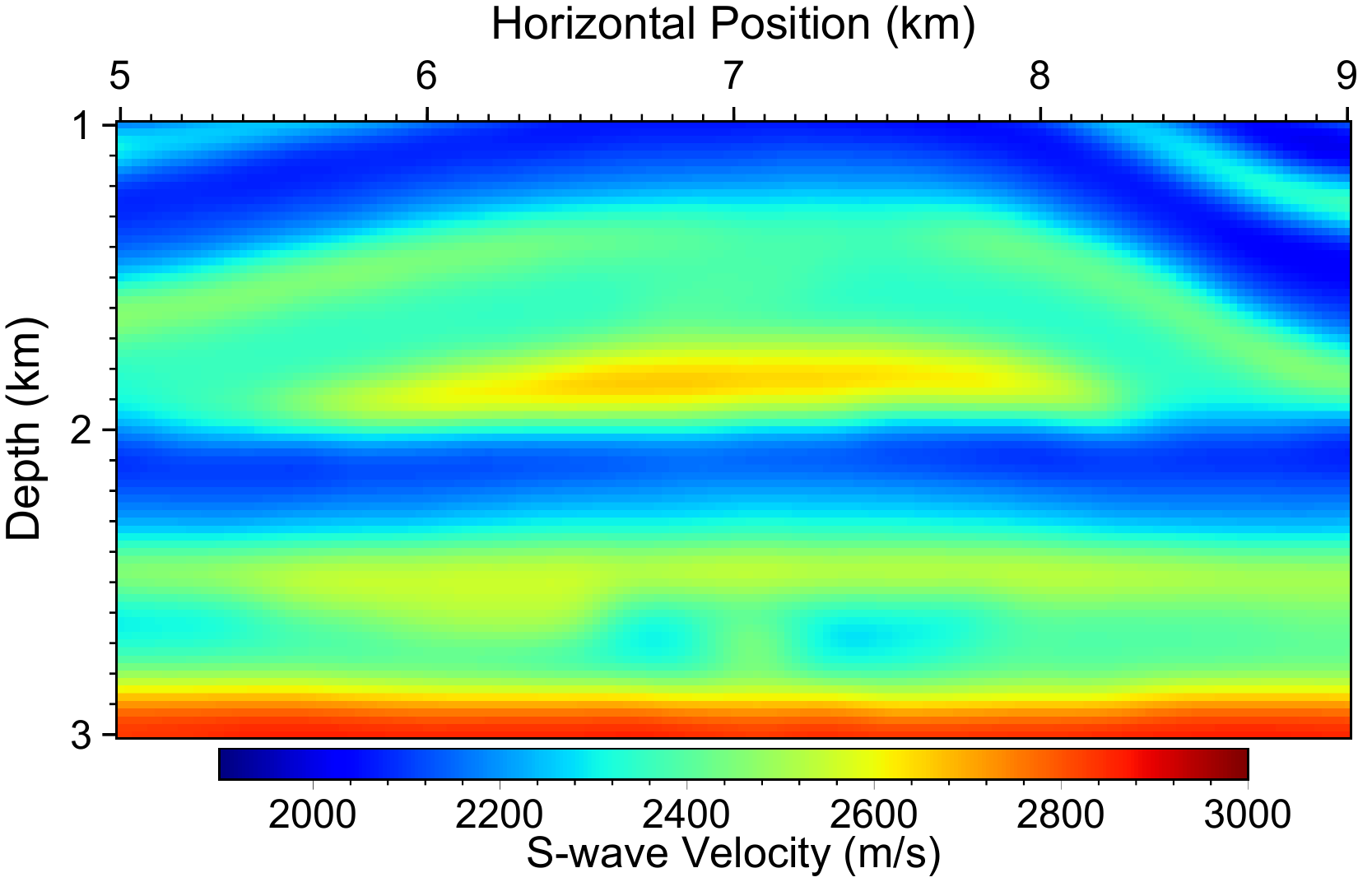}} \\
\subfloat[]{\includegraphics[width=0.485\textwidth]{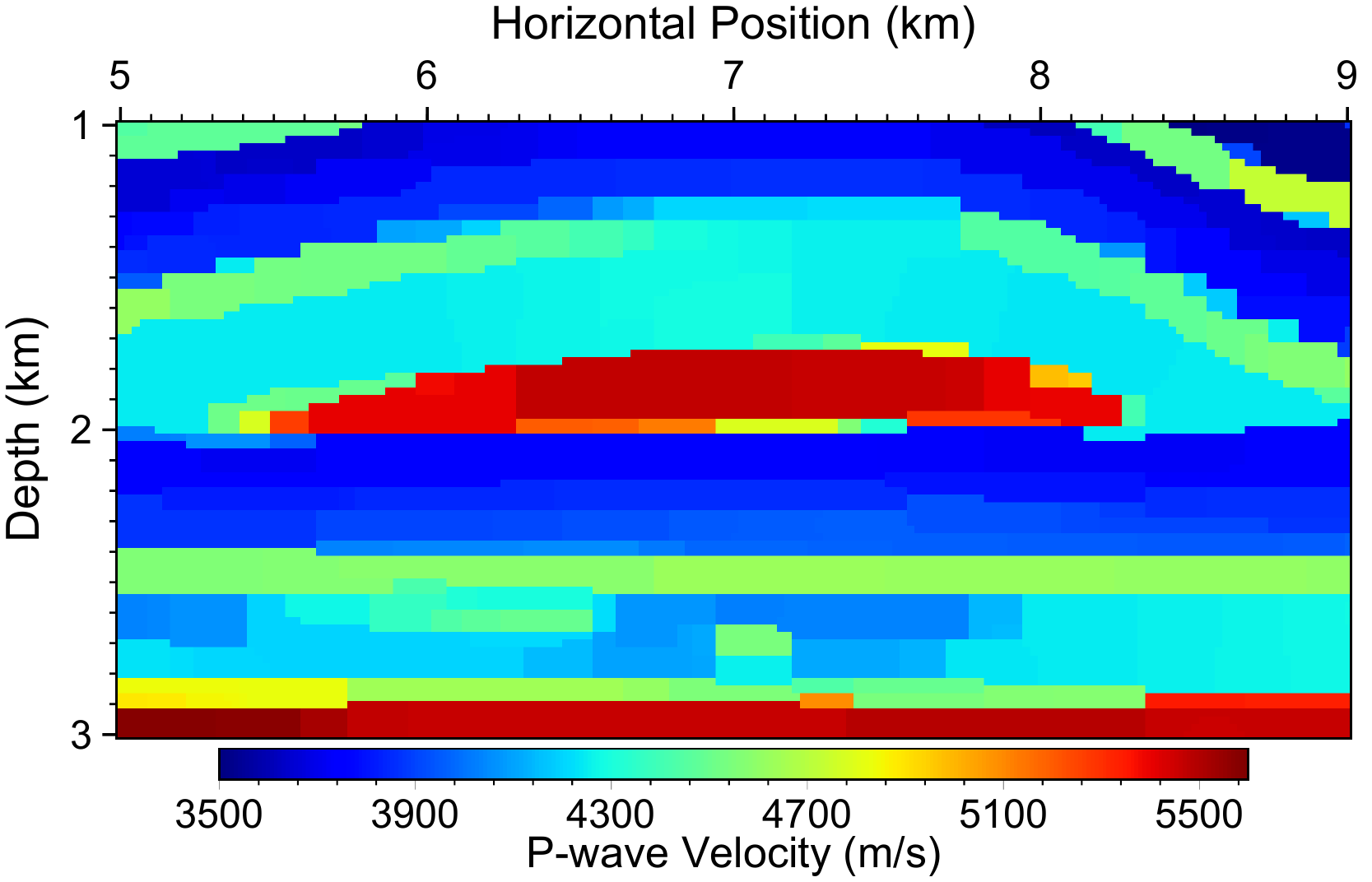}}
\subfloat[]{\includegraphics[width=0.485\textwidth]{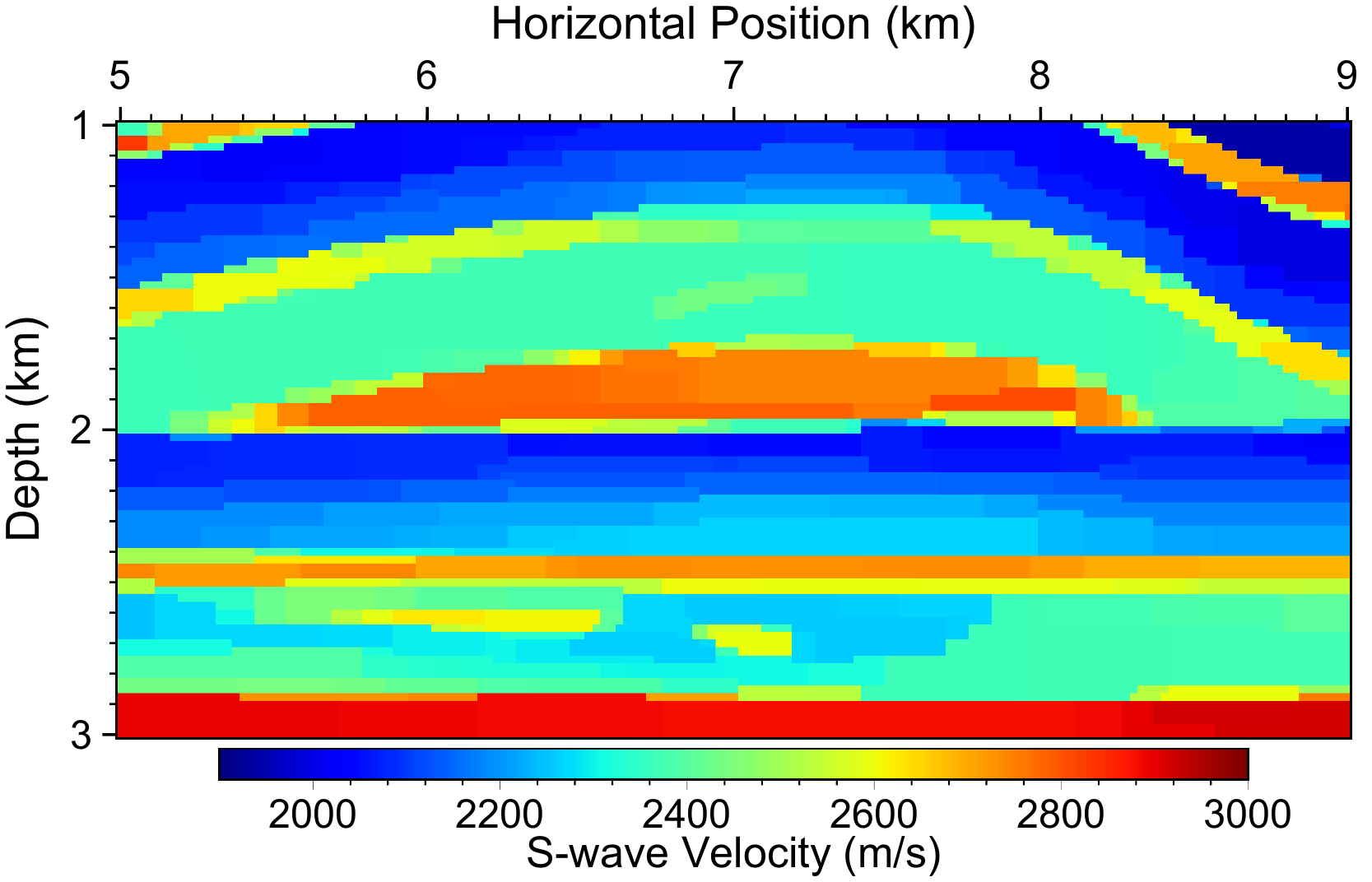}} \\
\subfloat[]{\includegraphics[width=0.485\textwidth]{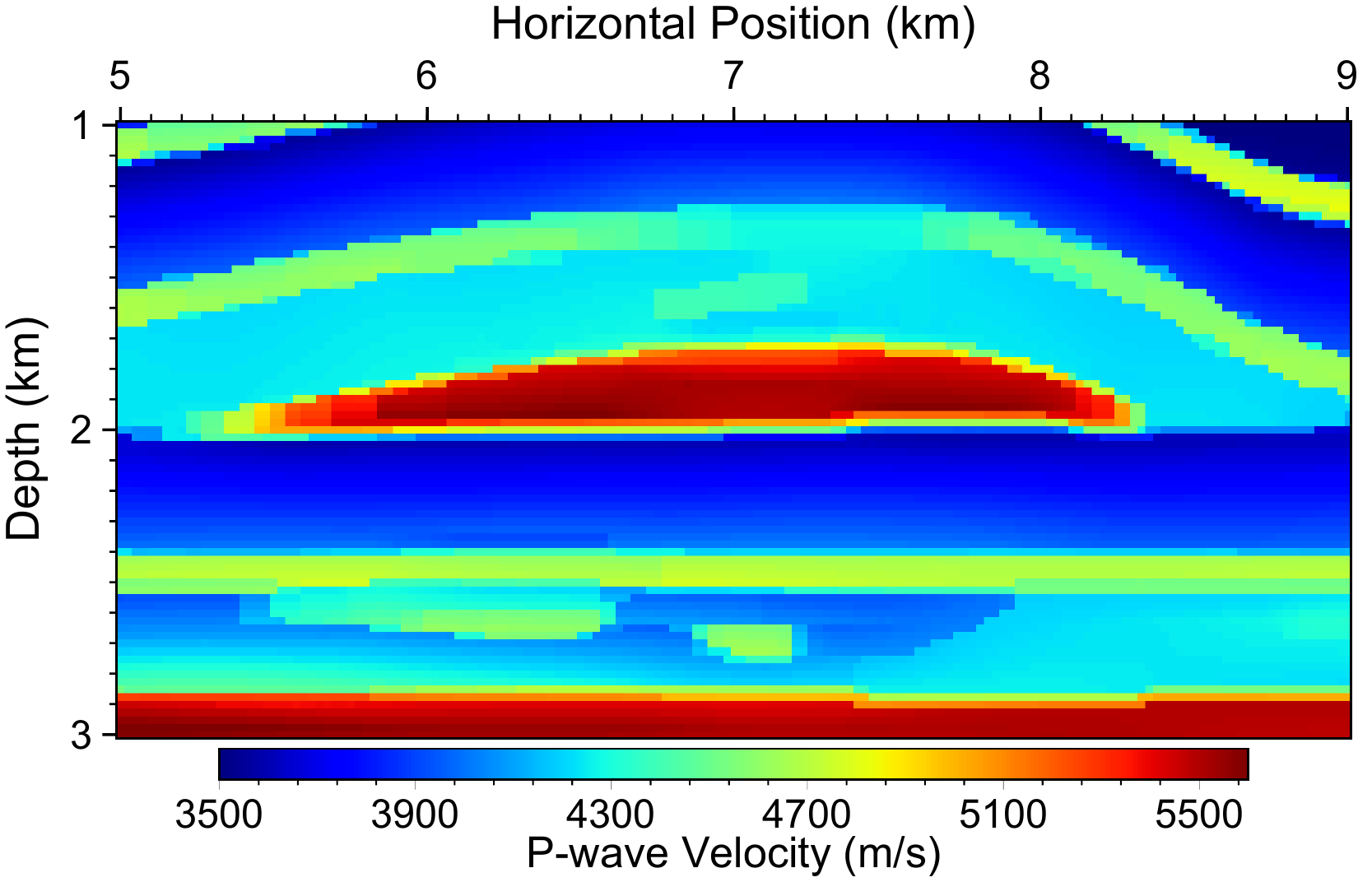}}
\subfloat[]{\includegraphics[width=0.485\textwidth]{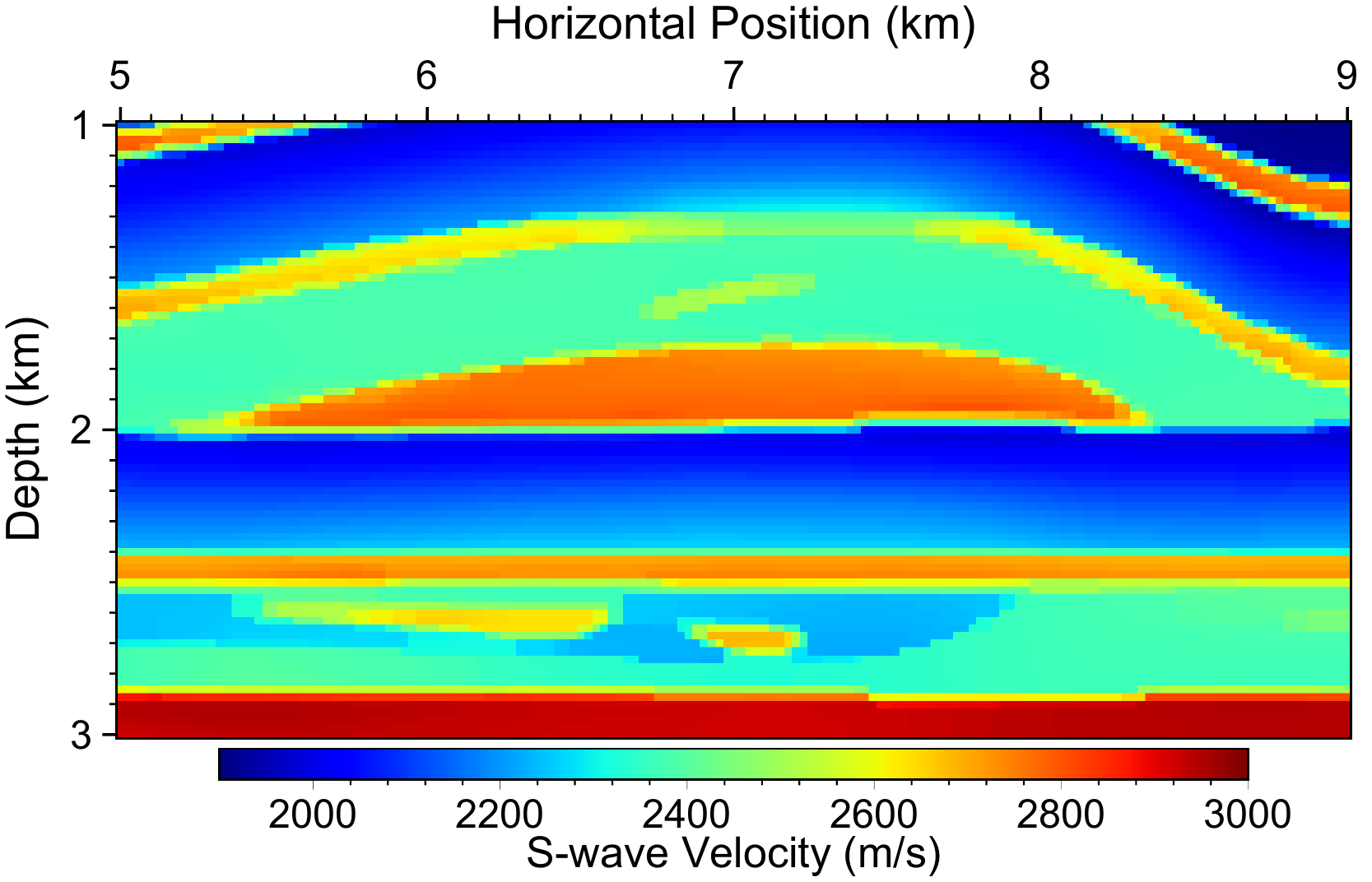}}
\caption{Zoom-in views of the region of interest in the inverted velocity 
   models shown in Fig.~\ref{fig:over_fwi_encoding} produced using (a) Tikhonov-FWI (b) TV-FWI and (c) TGPV-FWI. Panels in  (a), (c) and (e) are inverted 
   P-wave velocities and those in (b), (d) and (f) are inverted S-wave 
   velocities.}
\label{fig:over_fwi_zoom_encoding}
\end{figure}

In the fourth numerical test, we adopt dynamic source encoding to verify 
the capability of our TGPV-FWI method. We form six encoded super gathers 
using the data for 80 sources, and dynamically encode the phase and 
amplitude of the super gathers with random time delays and random 
polarity reversal, respectively, over iterations.  
Fig.~\ref{fig:over_fwi_encoding} and 
Fig.~\ref{fig:over_fwi_zoom_encoding} show the inversion results for the 
aforementioned three FWI methods. The results show that our TGPV-FWI 
produces the most accurate inversion results among the three FWI methods.

\begin{figure}
\centering
\subfloat[]{\includegraphics[width=0.485\textwidth]{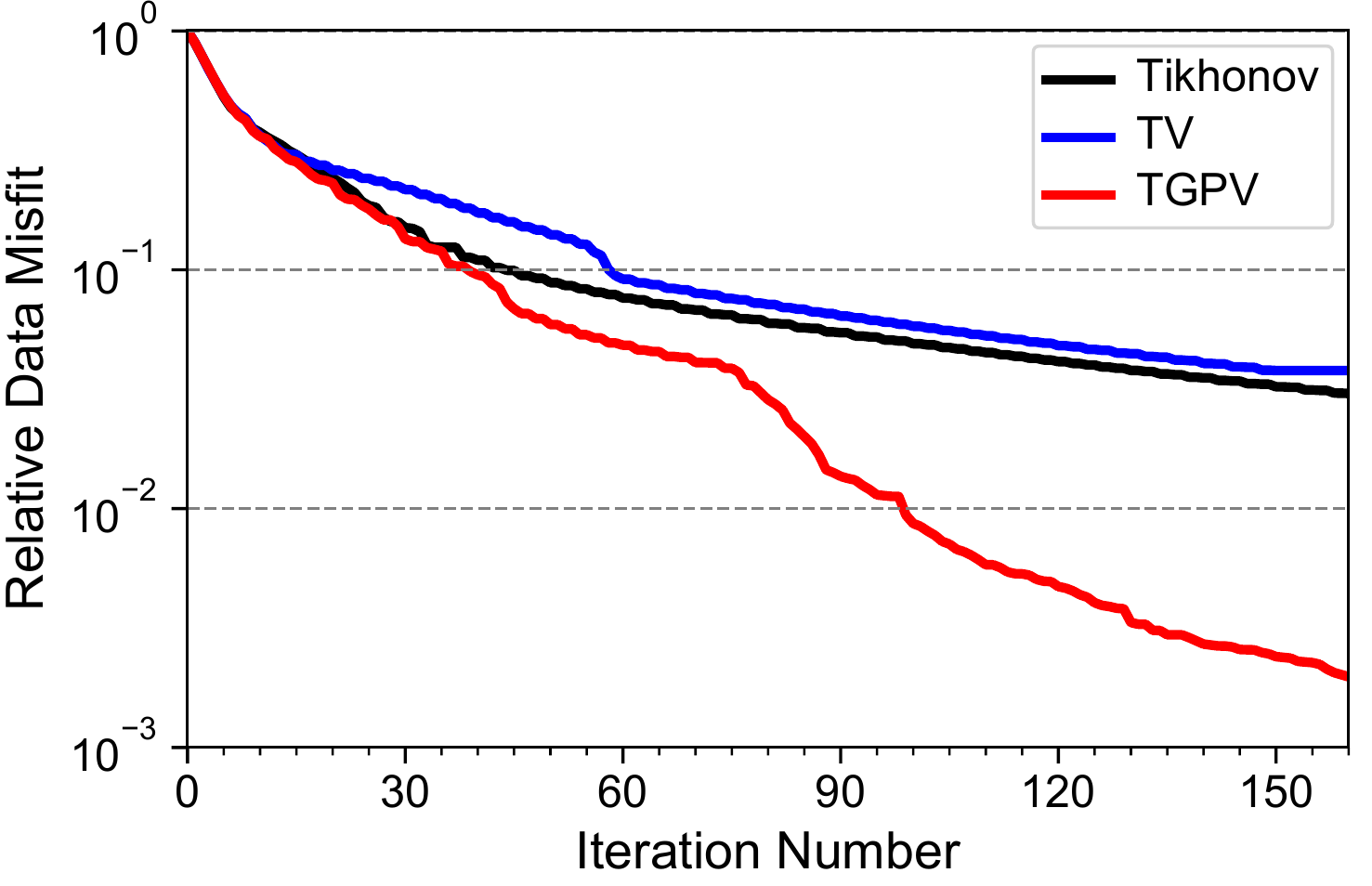}}
\subfloat[]{\includegraphics[width=0.485\textwidth]{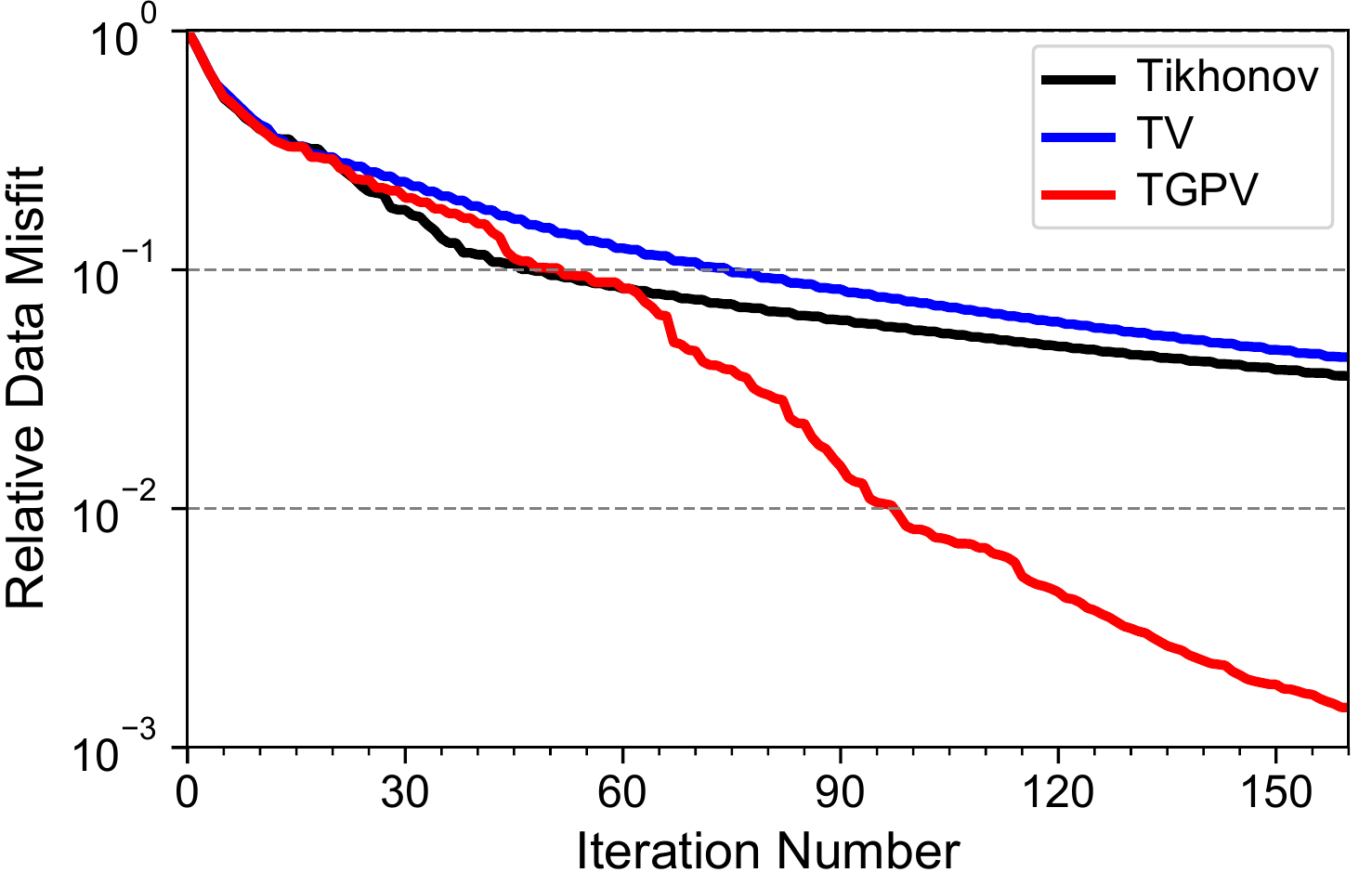}}\\
\subfloat[]{\includegraphics[width=0.485\textwidth]{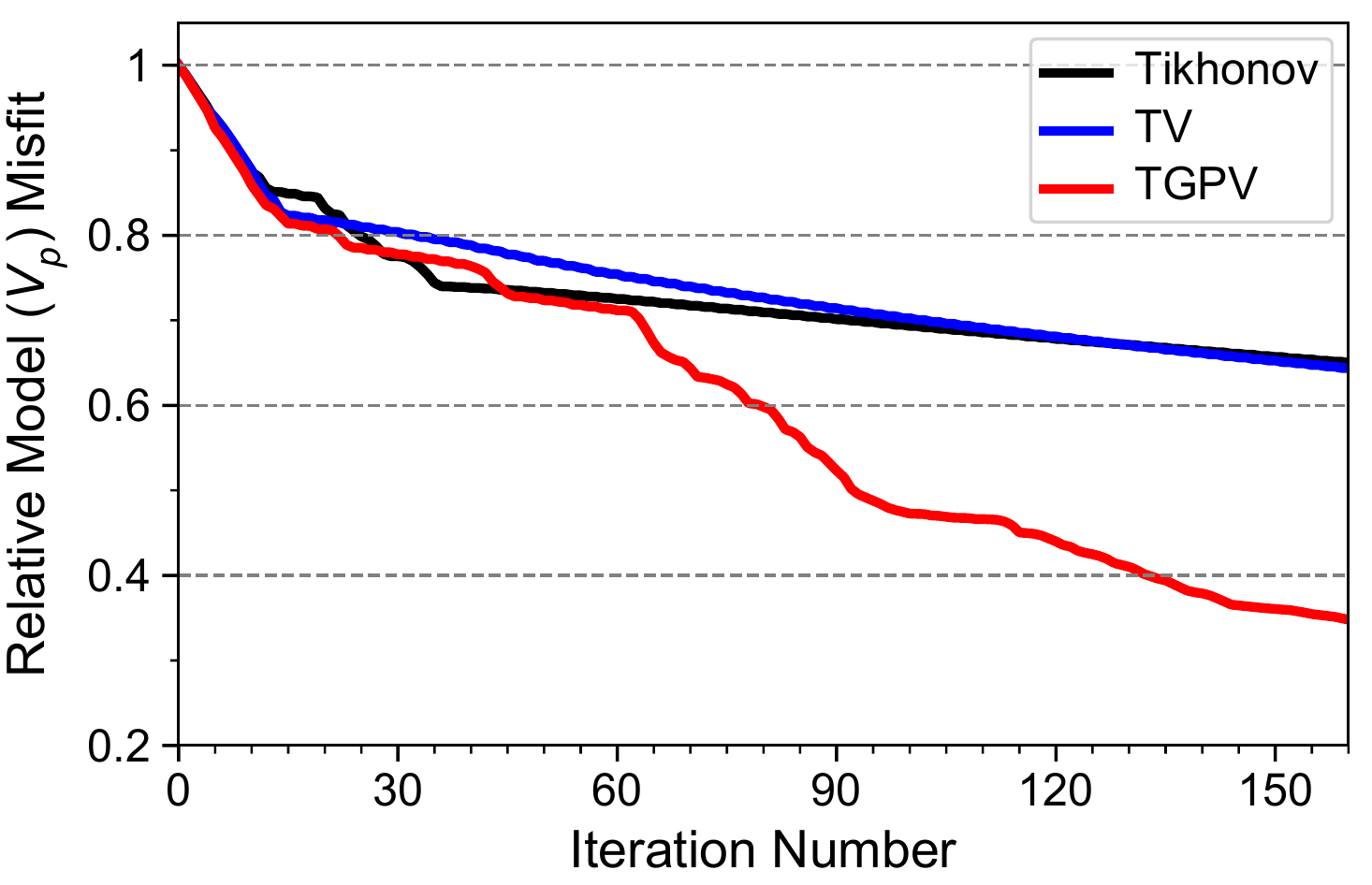}}
\subfloat[]{\includegraphics[width=0.485\textwidth]{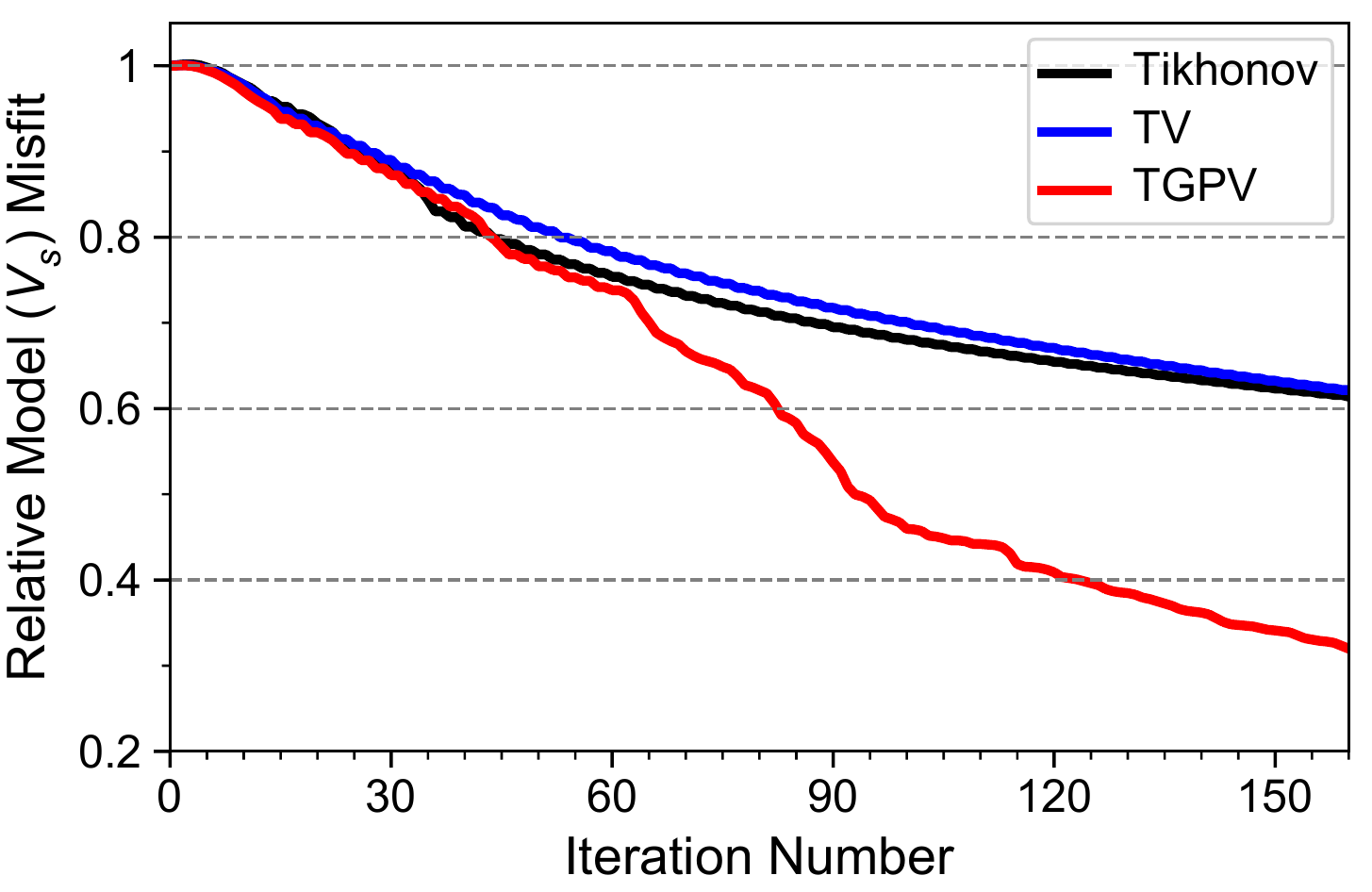}}\\
\subfloat[]{\includegraphics[width=0.485\textwidth]{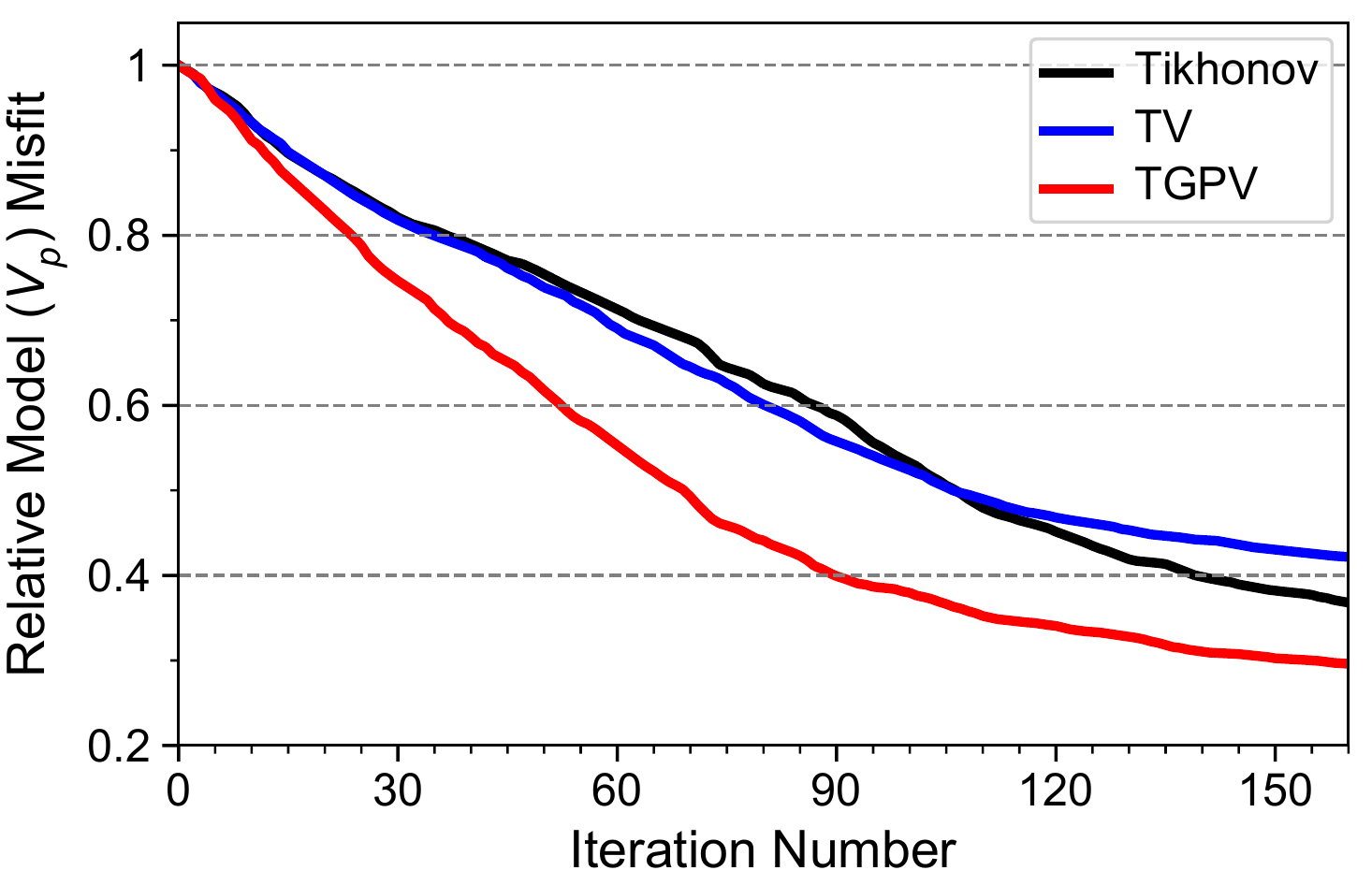}}
\subfloat[]{\includegraphics[width=0.485\textwidth]{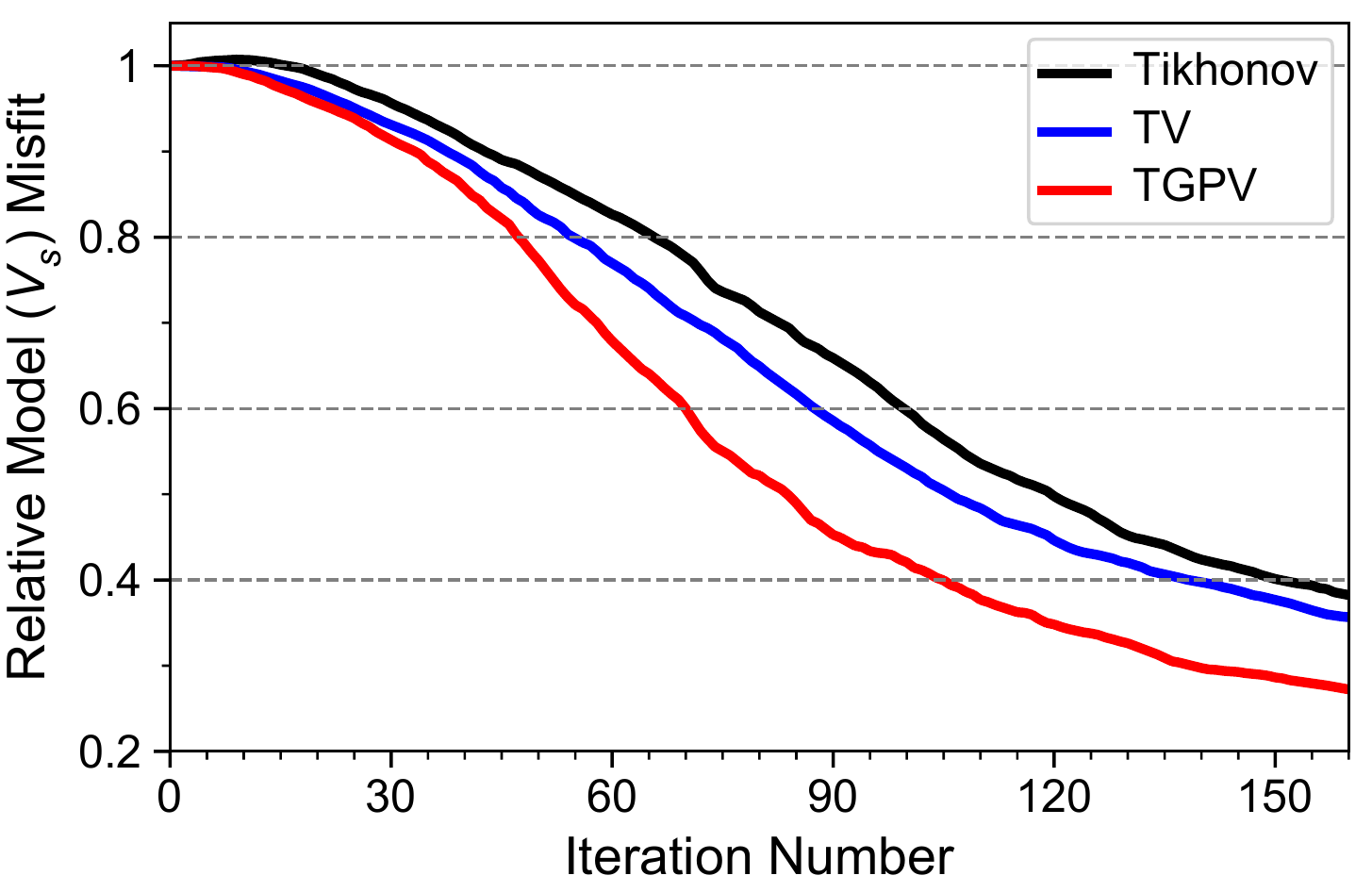}}
\caption{Data misfits of Tikhonov-FWI, TV-FWI and TGPV-FWI for (a) the first numerical test using all data and 
   for (b) the second numerical test using sparse data.
   Panels in (c) and (d) show the P- and S-wave velocity model misfits in 
   the second numerical test using sparse data. Panels in (e) and (f) 
   depict the P- and S-wave velocity model misfits in the fourth 
   numerical test using source-encoding inversions.}
   \label{fig:over_misfit}
\end{figure}

Finally, we compare the convergences of the data  and model misfits for  
the three FWI methods in Fig.~\ref{fig:over_misfit}. 
Fig.~\ref{fig:over_misfit}a shows the relative data misfits in the first 
numerical test using noise-free data with a regular source/receiver 
geometry. The data misfits of the Tikhonov-FWI and TV-FWI differ from each other insignificantly. By contrast, the 
data misfit for our TGPV-FWI decreases to a much smaller value than those 
of the other two methods. The corresponding data misfits in 
Fig.~\ref{fig:over_misfit}b for the noise-free, sparse data resemble the 
results in Fig.~\ref{fig:over_misfit}a.

Model misfit is another important indicator for FWI convergence. The 
model misfits in Fig.~\ref{fig:over_misfit}c and d for FWI with the 
Tikhonov, TV and TGPV regularizations and the sparse seismic data
demonstrate again that our TGPV regularization is the most effective  
among the three regularization schemes. The P- and S-wave velocity model 
misfits for the source-encoding FWI test displayed in 
Fig.~\ref{fig:over_misfit}c and d, respectively, indicate that the model 
misfits for the source-encoding TGPV-FWI are always smaller than those 
for the other two inversion methods.

\subsection{Field data example: Soda Lake geothermal field}

We apply our TGPV-FWI method to surface seismic data acquired at the Soda 
Lake geothermal field in Nevada, USA, and compare the result with those 
obtained using the Tikhonov-FWI and TV-FWI.

\begin{figure}
\centering
\subfloat[]{\includegraphics[width=0.6\textwidth]{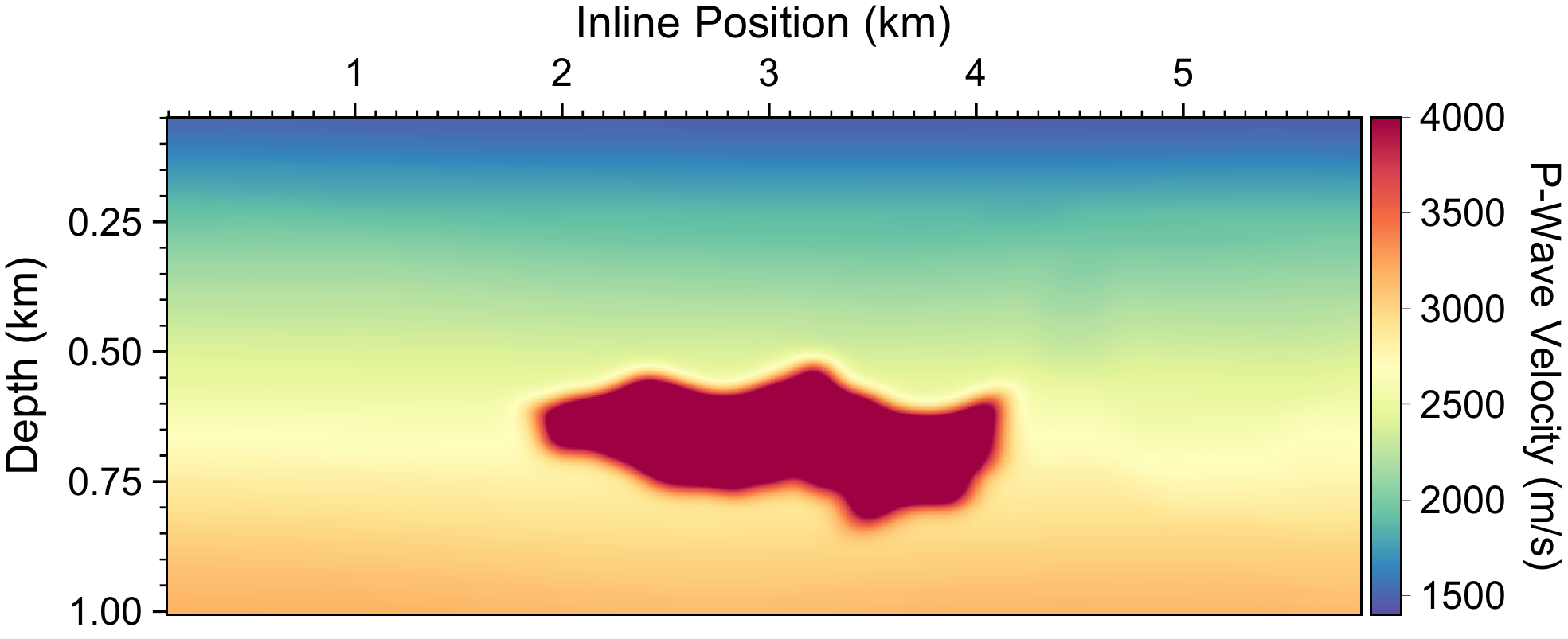}}  \\
\subfloat[]{\includegraphics[width=0.6\textwidth]{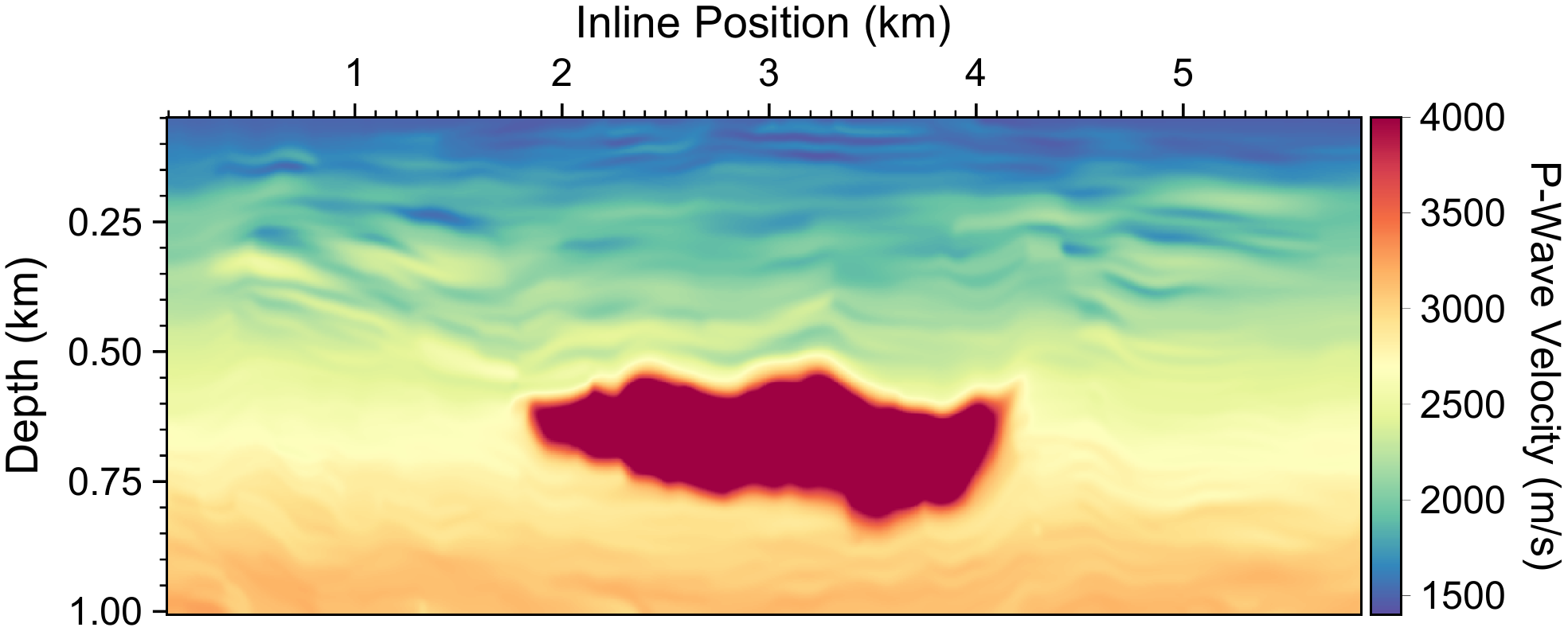}} \\
\subfloat[]{\includegraphics[width=0.6\textwidth]{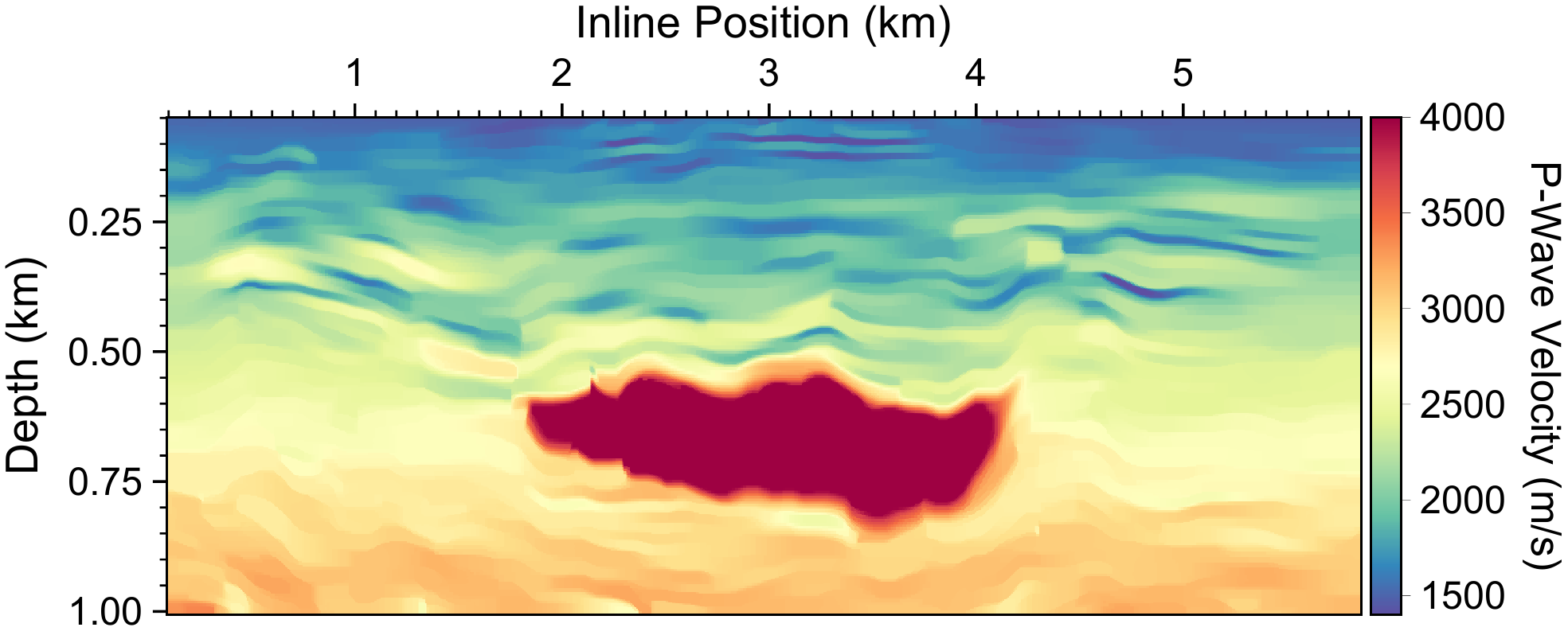}} \\
\subfloat[]{\includegraphics[width=0.6\textwidth]{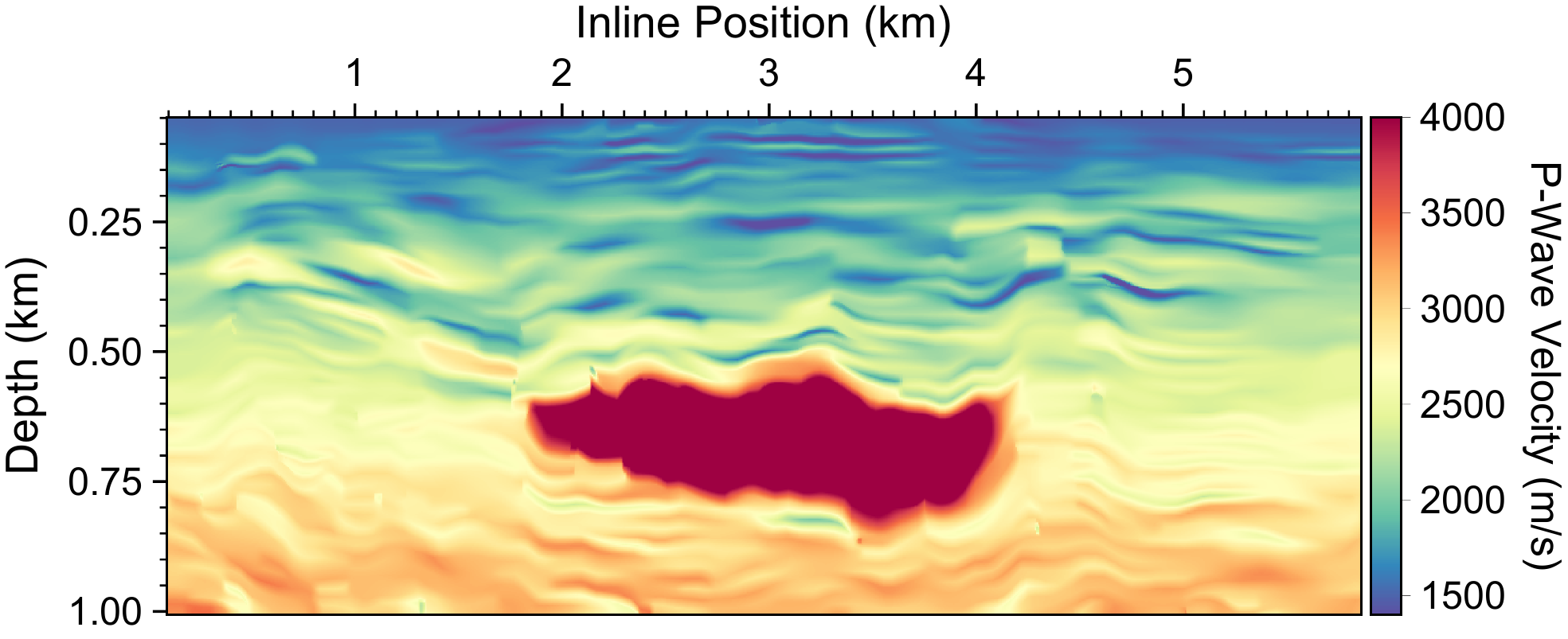}}
\caption{The P-wave migration velocity model (a) used as the initial 
   model for FWI of surface seismic data acquired at the Soda Lake 
   geothermal field, together with inversion results obtained using (a) 
   Tikhonov-FWI, (b) TV-FWI, and (c) TGPV-FWI. All three inversions 
   terminate after 100 iterations.}
\label{fig:soda_vp}
\end{figure}

Fig.~\ref{fig:soda_vp}a is a 2D P-wave velocity model for the Soda Lake 
geothermal field built using migration velocity analysis. The 
high-velocity body at the center of the model is a basalt body. The near 
surface velocity is approximately 1455~m/s. The center frequency of the 
surface seismic data is approximately 35~Hz. There are a total of 62 
shots along this 2D survey line. The source interval varies from 
approximately 30~m to 200~m. The receiver interval is approximately 66~m.  
Each common-shot gather contains 15 receivers to 46 receivers. Therefore, 
this dataset is considered to be very sparse in terms of data coverage. 

Fig.~\ref{fig:soda_vp}b, c and d are the inverted velocity models 
produced using Tikhonov-FWI, TV-FWI, and TGPV-FWI.  Comparing these three 
results, we find that Tikhonov-FWI produces only a low-resolution 
velocity model without many structural details, particularly in the 
region beneath the basalt body.

The TV-FWI result in Fig.~\ref{fig:soda_vp}b provides more details 
compared with that of Tikhonov-FWI in Fig.~\ref{fig:soda_vp}a.  The 
TGPV-FWI result in Fig.~\ref{fig:soda_vp}c has the highest resolution 
among the three results.

\begin{figure}
\centering
\includegraphics[width=0.6\textwidth]{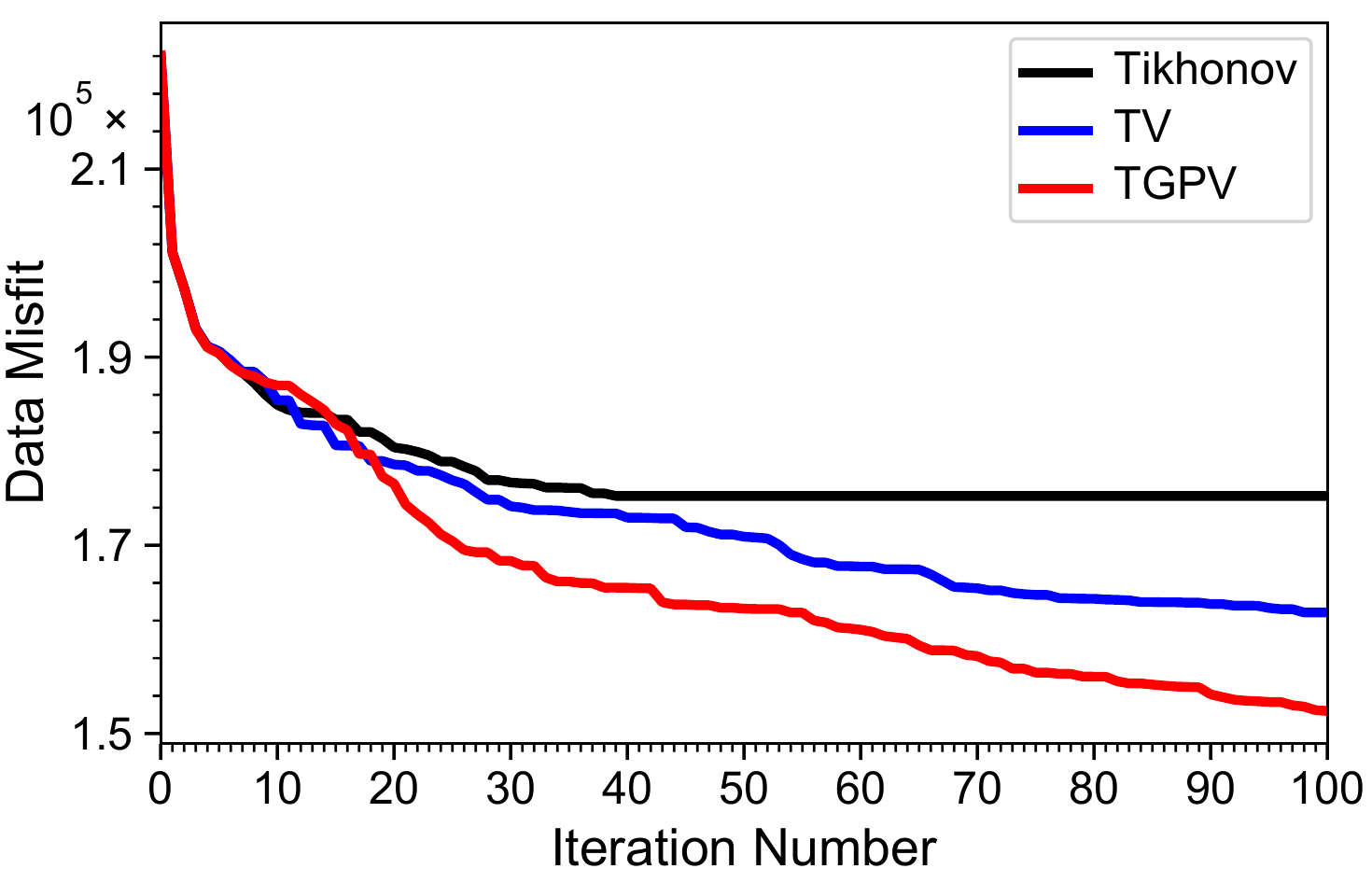}
\caption{Comparison of the convergence curves of data misfits for 
Tikhonov-FWI, TV-FWI and TGPV-FWI of surface seismic data from the Soda 
   Lake geothermal field.}
   \label{fig:soda_misfit}
\end{figure}

The accuracy of FWI with field seismic data can be validated using the 
convergence curve of the data misfit. We plot the convergence curves of  
the data misfits for the Tikhonov-FWI, TV-FWI, and our TGPV-FWI of the 
field seismic data in Fig.~\ref{fig:soda_misfit}, showing that our 
TGPV-FWI converges fastest and to the smallest value after 100 iterations 
among the three methods.

\section{Conclusions}

We have developed a novel full-waveform inversion method for acoustic  
and elastic waves using a total generalized p-variation regularization 
scheme. We decompose the regularized full-waveform inversion problem into 
two interlacing minimization problems. The first minimization problem is 
a conventional full-waveform inversion with a Tikhonov regularization 
term, and the second minimization problem is a denoising problem using 
the $\ell_p$-norm total generalized variation. We have developed an 
efficient algorithm to solve the second minimization problem based on the 
split-Bregman iteration method. We have used two synthetic data examples 
and one field data example to verify the improved accuracy of our new 
method for both acoustic and elastic full-waveform inversion. Our 
numerical results demonstrate that our new acoustic and elastic 
full-waveform inversion with the total generalized p-variation 
regularization produces accurate inversion results that preserve both 
sharp interfaces and smooth background velocity variations in the model.  
In addition, our new inversion method produces reliable inversion results 
when using sparse seismic data, noisy data, and dynamic source encoding.  
Our results of synthetic and field seismic data demonstrate that our new 
inversion method can be used as a robust and accurate tool for subsurface 
velocity model building, subsurface reservoir characterization, and 
large-scale tomographic reconstruction using Earthquake data.

\section{Acknowledgments}

This work was supported by U.S.\ Department of Energy through contract 
DE-AC52-06NA25396 to Los Alamos National Laboratory (LANL). The 
computation was performed using the super-computers of LANL's 
Institutional Computing Program. We thank Benxin Chi of LANL and Rick 
Chartrand of Descartes Labs for helpful discussions. Surface seismic data 
from the Soda Lake geothermal field was provided by Magma Energy (U.S.) 
Corp.

\bibliographystyle{seg}
\bibliography{refs}

\newpage 

\renewcommand{\theequation}{A-\arabic{equation}}

\section*{Appendix A: Solution to the second minimization problem in the 2D case}
Eq.~\eqref{eq:fwi_u} is equivalent to the following problem:
\begin{align}
\{\bfu^*,\bfw^*,\bfh^*,\bfs^*\} &= \argmin\limits_{\bfu,\bfw,\bfh,\bfs} \left\{  \frac{\mu}{2} 
\|\bfm^{(l+1)}-\bfu\|_2^2 + \alpha_0 \|\bfh\|_p^p + \alpha_1 
\|\bfs\|_p^p\right\}, \\ \text{s.t.} \quad \bfh & = \begin{bmatrix}
\bfh_x \\
\bfh_y
\end{bmatrix} =\nabla \bfu- \bfw, \\
\bfs &= \begin{bmatrix}
\bfs_{xx} & \bfs_{xy} \\
\bfs_{xy} & \bfs_{yy} 
\end{bmatrix} = \varepsilon(\bfw),
\label{eq:fwi2sub}
\end{align}
where $\bfh$ and $\bfs$ are dual variables, $\mu=1/\lambda_2$ is a parameter chosen for convenience, and $\bfw = [\bfw_x, \bfw_y]^{\mathrm{T}}$.

Using the split-Bregman iteration technique \cite[]{Goldstein_Osher_2009,Chartrand_2009}, we reformulate the constrained minimization problem as the following optimization system:
\begin{align}
\left\{\bfu^{(k+1)},\bfw^{(k+1)},\bfh^{(k+1)},\bfs^{(k+1)} \right\} &= \argmin\limits_{\bfu,\bfw,\bfh,\bfs} \left\{\alpha_0 \|\bfh\|_p + \frac{\eta_0}{2} \|\bfh-\tilde{\bfh}^{(k)}-(\nabla\bfu-\bfw)\|_2^2 \right. \nonumber \\
& \quad \left.+ \alpha_1 \|\bfs\|^p_p + \frac{\eta_1}{2} \|\bfs-\tilde{\bfs}^{(k)}-\varepsilon(\bfw)\|_2^2 +\frac{\mu}{2} \|\bfu- \bfm^{(l+1)}\|_2^2\right\}, \label{eq:uwhs}\\
\tilde{\bfh}^{(k+1)} & = \tilde{\bfh}^{(k)} + \left[(\nabla\bfu^{(k+1)}-\bfw^{(k+1)})-\bfh^{(k+1)}\right], \\
\tilde{\bfs}^{(k+1)} & = \tilde{\bfs}^{(k)} + \left[\varepsilon(\bfw^{(k+1)})-\bfs^{(k+1)}\right],
\end{align}
where $\tilde{\bfh}$ and $\tilde{\bfs}$ are auxiliary split-Bregman 
variables. Note that in the above formulation, $l$ is the index of FWI 
inversion iteration, while $k$ is the index of the split-Bregman 
iteration in the second minimization problem. In addition, $\eta_0$ and 
$\eta_1$ are two regularization parameters for the constrained 
minimization problem in eq.~\eqref{eq:uwhs}. 

We formulate the following four subproblems associated with the variables 
$\bfh$, $\bfs$, $\bfu$ and $\bfw$ from eq.~\eqref{eq:uwhs}. 

The first subproblem is the minimization of $\bfu$:
\begin{equation} 
\argmin\limits_{\bfu} \left\{\frac{\mu}{2} \|\bfu-\bfm^{(l+1)}\|_2^2 + \frac{\eta_0}{2} \|\bfh-\tilde{\bfh}- (\nabla \bfu -\bfw) \|_2^2 \right\}. 
\label{eq:subu}
\end{equation}
The first-order optimality condition of eq.~\eqref{eq:subu} is
\begin{equation}\label{eq:sub_u}
(\mu\mathbf{I}-\eta_0 \nabla^{\mathrm{T}}\nabla) \bfu^{(k+1)} = \mu \bfm^{(l+1)} 
+\eta_0 \nabla_x^{\mathrm{T}} (\bfh_x + \bfw_x-\tilde{\bfh}_x) 
+\eta_0 \nabla_y^{\mathrm{T}} (\bfh_y + \bfw_y-\tilde{\bfh}_y). 
\end{equation}
Because the linear system in eq.~\eqref{eq:sub_u} is strictly diagonally dominant, it can be efficiently solved with the Gauss-Seidel iteration:
\begin{align}
\label{eq:discrete_sub_u}
\bfu_{i,j}^{(k+1)} & = \frac{\eta_0}{\mu+4\eta_0} \left[\bfu^{(k)}_{i+1,j}+\bfu^{(k)}_{i-1,j}+\bfu^{(k)}_{i,j+1}+\bfu^{(k)}_{i,j-1}\right] \nonumber \\
&+\frac{\eta_0}{\mu+4\eta_0} \left[\bfh^{(k)}_{x|i-1,j}-\bfh^{(k)}_{x|i,j}+\bfw^{(k)}_{x|i-1,j}-\bfw^{(k)}_{x|i,j}-\left(\tilde{\bfh}^{(k)}_{x|i-1,j}-\tilde{\bfh}^{(k)}_{x|i,j}\right)\right] \nonumber \\
&+\frac{\eta_0}{\mu+4\eta_0} \left[\bfh^{(k)}_{y|i,j-1}-\bfh^{(k)}_{y|i,j}+\bfw^{(k)}_{y|i,j-1}-\bfw^{(k)}_{y|i,j}-\left(\tilde{\bfh}^{(k)}_{y|i,j-1}-\tilde{\bfh}^{(k)}_{y|i,j}\right)\right] \nonumber \\
&+ \frac{\mu}{\mu+4\eta_0} \bfm^{(l+1)}_{i,j} 
\end{align}
\citet{Goldstein_Osher_2009} developed a similar solution for the 
total-variation image denoising problem.

The second subproblem is the minimization of $\bfw$:
\begin{equation}
\argmin\limits_{\bfw} \left\{ \frac{\eta_0}{2} \|\bfh-\tilde{\bfh}- (\nabla \bfu -\bfw) \|_2^2 + \frac{\eta_1}{2} \|\bfs -\tilde{\bfs} - \varepsilon(\bfw) \|_2^2\right\}. 
\label{eq:subw}
\end{equation}
The first-order optimality condition of eq.~\eqref{eq:subw} leads to two linear systems:
\begin{align}
(\eta_0 \mathbf{I} - \eta_1 \nabla_x^{\mathrm{T}} \nabla_x - \frac{1}{2} \eta_1 \nabla_y^{\mathrm{T}} \nabla_y) \bfw_x & = 
-\eta_0 (\bfh_x-\tilde{\bfh}_x -\nabla_x\bfu)
+\eta_1 \nabla_x^{\mathrm{T}}(\bfs_{xx}-\tilde{\bfs}_{xx}) \nonumber \\
&+\eta_1 \nabla_y^{\mathrm{T}}(\bfs_{xy}-\tilde{\bfs}_{xy}-\frac{1}{2}\nabla_x \bfw_y), \label{eq:sub_w_1} \\
(\eta_0 \mathbf{I} - \frac{1}{2} \eta_1 \nabla_x^{\mathrm{T}} \nabla_x - \eta_1 \nabla_y^{\mathrm{T}} \nabla_y) \bfw_y & =-\eta_0 (\bfh_y-\tilde{\bfh}_y -\nabla_y\bfu) 
+\eta_1 \nabla_y^{\mathrm{T}}(\bfs_{yy}-\tilde{\bfs}_{yy}) \nonumber \\
&+\eta_1 \nabla_x^{\mathrm{T}}(\bfs_{xy}-\tilde{\bfs}_{xy}-\frac{1}{2}\nabla_y \bfw_x),  \label{eq:sub_w_2} 
\end{align}
which can also be solved via the Gauss-Seidel iteration because of their 
strict diagonal dominance. We use eq.~\eqref{eq:sub_w_1} to explain the 
procedure, and the solution to eq.~\eqref{eq:sub_w_2} can be obtained 
analogously. 

For convenience, we first define and compute an intermediate variable $\mathbf{v}_{y,x}$ for $\nabla_x \bfw_y$ as
\begin{equation}
\mathbf{v}^{(k)}_{y,x|i,j} = \bfw^{(k)}_{y|i+1,j}-\bfw^{(k)}_{y|i,j}.
\end{equation}
We then obtain
\begin{align}
\label{eq:discrete_sub_wx}
\bfw_{x|i,j}^{(k+1)} & = \frac{\eta_1}{\eta_0+3\eta_1} \left[\bfw^{(k)}_{x|i+1,j}+\bfw^{(k)}_{x|i-1,j}+\frac{1}{2}\left(\bfw^{(k)}_{x|i,j+1}+\bfw^{(k)}_{x|i,j-1}\right)\right] \nonumber \\
&-\frac{\eta_0}{\eta_0+3\eta_1} \left[\bfh^{(k)}_{x|i,j}-\tilde{\bfh}^{(k)}_{x|i,j}-\left(\bfu^{(k+1)}_{i+1,j}-\bfu^{(k+1)}_{i,j}\right)\right] \nonumber \\
&+\frac{\eta_1}{\eta_0+3\eta_1} \left[\bfs^{(k)}_{xx|i-1,j}-\bfs^{(k)}_{xx|i,j}-\left(\tilde{\bfs}^{(k)}_{xx|i-1,j}-\tilde{\bfs}^{(k)}_{xx|i,j}\right)\right] \nonumber \\
&+\frac{\eta_1}{\eta_0+3\eta_1} \left[\bfs^{(k)}_{xy|i,j-1}-\bfs^{(k)}_{xy|i,j}-\left(\tilde{\bfs}^{(k)}_{xy|i,j-1}-\tilde{\bfs}^{(k)}_{xy|i,j}\right)-\frac{1}{2}\left(\mathbf{v}^{(k)}_{y,x|i,j-1}-\mathbf{v}^{(k)}_{y,x|i,j}\right)\right].
\end{align}

The third subproblem is the minimization of $\bfh$: 
\begin{equation}
\argmin\limits_{\bfh} \left\{ \alpha_0 \|\bfh\|^p_p + \frac{\eta_0}{2} \|\bfh-\tilde{\bfh} - (\nabla \bfu -\bfw) \|_2^2\right\}. 
\label{eq:subh}
\end{equation}
This problem can be efficiently solved with
\begin{equation}
\bfh^{(k+1)} = \mathcal{S}_p \left((\nabla \bfu^{(k+1)}-\bfw^{(k+1)})+\tilde{\bfh}^{(k)},\frac{\alpha_0}{\eta_0}\right), 
\end{equation}
where the generalized $p$-shrinkage $\mathcal{S}_p$ reads \cite[]{Chartrand_2013}
\begin{equation}
\mathcal{S}_p\left(\xi,\frac{1}{\beta}\right) = \max(|\xi|-\beta^{p-2}|\xi|^{p-1},0) \frac{\xi}{|\xi|}. 
\end{equation}

The shrinkage operation is an element-wise operation and therefore can be solved with fairly small computational cost. 

The fourth subproblem is the minimization of $\bfs$: 
\begin{equation}
\argmin\limits_{\bfs} \left\{\alpha_1 \|\bfs\|^p_p + \frac{\eta_1}{2} \|\bfs -\tilde{\bfs} - \varepsilon(\bfw) \|_2^2\right\}. 
\label{eq:subs}
\end{equation}
which can also be solved with the generalized $p$-shrinkage:
\begin{equation}
\bfs^{(k+1)} = \mathcal{S}_p \left(\varepsilon(\bfw^{(k+1)}) + \bfs^{(k)}, \frac{\alpha_1}{\eta_1}\right)
\end{equation}

Lastly, the update of the auxiliary split-Bregman variables is trivial. For example, for $\tilde{\bfh}$, we have
\begin{align}
\tilde{\bfh}^{(k+1)}_{x|i,j} & =\tilde{\bfh}^{(k)}_{x|i,j} + \left(\bfu^{(k+1)}_{i+1,j}-\bfu^{(k+1)}_{i,j}-\bfw^{(k+1)}_{x|i,j}-\bfh^{(k+1)}_{x|i,j}\right), \\
\tilde{\bfh}^{(k+1)}_{y|i,j} & =\tilde{\bfh}^{(k)}_{y|i,j} + \left(\bfu^{(k+1)}_{i,j+1}-\bfu^{(k+1)}_{i,j}-\bfw^{(k+1)}_{y|i,j}-\bfh^{(k+1)}_{y|i,j}\right).
\end{align}
And the update of the auxiliary $\tilde{\bfs}$ is similar. 

\renewcommand{\theequation}{B-\arabic{equation}}
\section*{Appendix B: Solution to the second minimization problem in the 3D case}
\setcounter{equation}{0}
\renewcommand{\thesubsection}{(\Alph{subsection})}

In the 3D case, the TGPV regularization term in eq.~\eqref{eq:fwilp} reads:
\begin{equation}\label{eq:reg_tgpv_3d}
\mathcal{T}_p(\bfm) = \argmin\limits_{\bfw} \alpha_0 \|\nabla \bfm - \bfw \|^p_p + \alpha_1 \|\varepsilon(\bfw)\|^p_p, \qquad (0< p < 1)
\end{equation}
where the gradient is
\begin{equation}
\nabla \bfm = \begin{bmatrix}
\nabla_x \bfm \\
\nabla_y \bfm \\
\nabla_z \bfm
\end{bmatrix},
\end{equation}
and the symmetric gradient $\varepsilon(\cdot)$  is
\begin{equation}
\varepsilon(\bfw) = \begin{bmatrix}
\nabla_x \bfw_x & \frac{1}{2}(\nabla_x \bfw_y + \nabla_y \bfw_x) &\frac{1}{2}(\nabla_x \bfw_z + \nabla_z \bfw_x) \\
\frac{1}{2}(\nabla_x \bfw_y + \nabla_y \bfw_x) & \nabla_y \bfw_y & \frac{1}{2}(\nabla_y \bfw_z + \nabla_z \bfw_y) \\
\frac{1}{2}(\nabla_x \bfw_z + \nabla_z \bfw_x) & \frac{1}{2}(\nabla_y \bfw_z + \nabla_z \bfw_y) & \nabla_z \bfw_z 
\end{bmatrix}.
\end{equation}

We still use a dual-variable alternating-direction minimization strategy to solve the regularized FWI. We transform the second minimization problem to 
\begin{align}
\argmin\limits_{\bfu,\bfw,\bfh,\bfs} & \frac{\mu}{2} \|\bfu-\bfm^{(k+1)}\|_2^2 + \alpha_0 \|\bfh\|^p_p + \alpha_1 \|\bfs\|^p_p, \\ 
\text{s.t.} \quad & \bfh = \begin{bmatrix}
\bfh_x \\
\bfh_y \\
\bfh_z 
\end{bmatrix} =\nabla \bfu- \bfw, \\
& \bfs = \begin{bmatrix}
\bfs_{xx} & \bfs_{xy} & \bfs_{xz} \\
\bfs_{xy} & \bfs_{yy} & \bfs_{yz} \\
\bfs_{xz} & \bfs_{yz} & \bfs_{zz} \\
\end{bmatrix} = \varepsilon(\bfw), 
\end{align}
which also contains four subproblems as is in the 2D case. 

The first-order optimality condition for the minimization of variable $\bfu$ is 
\begin{equation}\label{eq:sub_u_3d}
(\mu\mathbf{I}-\eta_0 \nabla^{\mathrm{T}}\nabla) \bfu^{(k+1)} = \mu \bfm^{(k+1)}
+\eta_0 \nabla_x^{\mathrm{T}} (\bfh_x + \bfw_x-\tilde{\bfh}_x) 
+\eta_0 \nabla_y^{\mathrm{T}} (\bfh_y + \bfw_y-\tilde{\bfh}_y) 
+\eta_0 \nabla_z^{\mathrm{T}} (\bfh_z + \bfw_z-\tilde{\bfh}_z). 
\end{equation}

The Gauss-Seidel solution to this linear system is
\begin{align}
\label{eq:discrete_sub_u_3d}
\bfu_{i,j,k}^{(k+1)} & = \frac{\eta_0}{\mu+6\eta_0} \left[\bfu^{(k)}_{i+1,j,k}+\bfu^{(k)}_{i-1,j,k}+\bfu^{(k)}_{i,j+1,k}+\bfu^{(k)}_{i,j-1,k}+\bfu^{(k)}_{i,j,k+1}+\bfu^{(k)}_{i,j,k-1}\right] \nonumber \\
&+\frac{\eta_0}{\mu+6\eta_0} \left[\bfh^{(k)}_{x|i-1,j,k}-\bfh^{(k)}_{x|i,j,k}+\bfw^{(k)}_{x|i-1,j,k}-\bfw^{(k)}_{x|i,j,k}-(\tilde{\bfh}^{(k)}_{x|i-1,j,k}-\tilde{\bfh}^{(k)}_{x|i,j,k})\right] \nonumber \\
&+\frac{\eta_0}{\mu+6\eta_0} \left[\bfh^{(k)}_{y|i,j-1,k}-\bfh^{(k)}_{y|i,j,k}+\bfw^{(k)}_{y|i,j-1,k}-\bfw^{(k)}_{y|i,j,k}-(\tilde{\bfh}^{(k)}_{y|i,j-1,k}-\tilde{\bfh}^{(k)}_{y|i,j,k})\right] \nonumber \\
&+\frac{\eta_0}{\mu+6\eta_0} \left[\bfh^{(k)}_{z|i,j,k-1}-\bfh^{(k)}_{z|i,j,k}+\bfw^{(k)}_{z|i,j,k-1}-\bfw^{(k)}_{z|i,j,k}-(\tilde{\bfh}^{(k)}_{z|i,j,k-1}-\tilde{\bfh}^{(k)}_{z|i,j,k})\right] \nonumber \\
&+ \frac{\mu}{\mu+6\eta_0} \bfm^{(k+1)}_{i,j,k}.
\end{align}

The first-order optimality condition for the minimization of variable $\bfw$ is a system composed of three equations:
\begin{align}
(\eta_0 \mathbf{I} - \eta_1 \nabla_x^{\mathrm{T}} \nabla_x - \frac{1}{2} \eta_1 \nabla_y^{\mathrm{T}} \nabla_y -\frac{1}{2} \eta_1 \nabla_z^{\mathrm{T}} \nabla_z) \bfw_x & = 
-\eta_0 (\bfh_x-\tilde{\bfh}_x -\nabla_x\bfu) \nonumber \\
&+\eta_1 \nabla_x^{\mathrm{T}}(\bfs_{xx}-\tilde{\bfs}_{xx}) \nonumber \\
&+\eta_1 \nabla_y^{\mathrm{T}}(\bfs_{xy}-\tilde{\bfs}_{xy}-\frac{1}{2}\nabla_x \bfw_y) \nonumber \\
&+\eta_1 \nabla_z^{\mathrm{T}}(\bfs_{xz}-\tilde{\bfs}_{xz}-\frac{1}{2}\nabla_x \bfw_z), \label{eq:sub_w3_1} \\
(\eta_0 \mathbf{I} - \frac{1}{2} \eta_1 \nabla_x^{\mathrm{T}} \nabla_x - \eta_1 \nabla_y^{\mathrm{T}} \nabla_y - \frac{1}{2} \eta_1 \nabla_z^{\mathrm{T}} \nabla_z) \bfw_y & =-\eta_0 (\bfh_y-\tilde{\bfh}_y -\nabla_y\bfu) \nonumber \\
&+\eta_1 \nabla_x^{\mathrm{T}}(\bfs_{xy}-\tilde{\bfs}_{xy}-\frac{1}{2}\nabla_y \bfw_x) \nonumber\\
&+\eta_1 \nabla_y^{\mathrm{T}}(\bfs_{yy}-\tilde{\bfs}_{yy}) \nonumber \\
&+\eta_1 \nabla_z^{\mathrm{T}}(\bfs_{yz}-\tilde{\bfs}_{yz}-\frac{1}{2}\nabla_y \bfw_z),  \label{eq:sub_w3_2} \\
(\eta_0 \mathbf{I} - \frac{1}{2} \eta_1 \nabla_x^{\mathrm{T}} \nabla_x - \frac{1}{2}\eta_1 \nabla_y^{\mathrm{T}} \nabla_y - \eta_1 \nabla_z^{\mathrm{T}} \nabla_z) \bfw_z & =-\eta_0 (\bfh_z-\tilde{\bfh}_z -\nabla_z\bfu) \nonumber \\
&+\eta_1 \nabla_x^{\mathrm{T}}(\bfs_{xz}-\tilde{\bfs}_{xz}-\frac{1}{2}\nabla_z \bfw_x) \nonumber\\
&+\eta_1 \nabla_y^{\mathrm{T}}(\bfs_{yz}-\tilde{\bfs}_{yz}-\frac{1}{2}\nabla_z \bfw_y) \nonumber \\
&+\eta_1 \nabla_z^{\mathrm{T}}(\bfs_{zz}-\tilde{\bfs}_{zz}).  \label{eq:sub_w3_3} 
\end{align}

Solutions to the above three minimization equations can be obtained with the Gauss-Seidel iteration analogously to those for the 2D case. For instance, for the first equation \eqref{eq:sub_w3_1}, we first define and compute two intermediate variables $\mathbf{v}_{y,x}$ for $\nabla_x \bfw_y$ and $\mathbf{v}_{z,x}$ for $\nabla_x \bfw_z$ as
\begin{align}
\mathbf{v}^{(k)}_{y,x|i,j,k} &= \bfw^{(k)}_{y|i+1,j,k}-\bfw^{(k)}_{y|i,j,k}, \\
\mathbf{v}^{(k)}_{z,x|i,j,k} &= \bfw^{(k)}_{z|i+1,j,k}-\bfw^{(k)}_{z|i,j,k}.
\end{align}

We then obtain
\begin{align}
\label{eq:discrete_sub_w3x}
\bfw_{x|i,j,k}^{(k+1)} & = \frac{\eta_1}{\eta_0+4\eta_1} \left[
\bfw^{(k)}_{x|i+1,j,k}+\bfw^{(k)}_{x|i-1,j,k}
+\frac{1}{2}\left(\bfw^{(k)}_{x|i,j+1,k}+\bfw^{(k)}_{x|i,j-1,k}\right)
+\frac{1}{2}\left(\bfw^{(k)}_{x|i,j,k+1}+\bfw^{(k)}_{x|i,j,k-1}\right)
\right] \nonumber \\
&-\frac{\eta_0}{\eta_0+4\eta_1} \left[\bfh^{(k)}_{x|i,j,k}-\tilde{\bfh}^{(k)}_{x|i,j,k}-\left(\bfu^{(k+1)}_{i+1,j,k}-\bfu^{(k+1)}_{i,j,k}\right)\right] \nonumber \\
&+\frac{\eta_1}{\eta_0+4\eta_1} \left[\bfs^{(k)}_{xx|i-1,j,k}-\bfs^{(k)}_{xx|i,j,k}-\left(\tilde{\bfs}^{(k)}_{xx|i-1,j,k}-\tilde{\bfs}^{(k)}_{xx|i,j,k}\right)\right] \nonumber \\
&+\frac{\eta_1}{\eta_0+4\eta_1} \left[\bfs^{(k)}_{xy|i,j-1,k}-\bfs^{(k)}_{xy|i,j,k}-\left(\tilde{\bfs}^{(k)}_{xy|i,j-1,k}-\tilde{\bfs}^{(k)}_{xy|i,j,k}\right)-\frac{1}{2}\left(\mathbf{v}^{(k)}_{y,x|i,j-1,k}-\mathbf{v}^{(k)}_{y,x|i,j,k}\right)\right] \nonumber \\
&+\frac{\eta_1}{\eta_0+4\eta_1} \left[\bfs^{(k)}_{xz|i,j,k-1}-\bfs^{(k)}_{xz|i,j,k}-\left(\tilde{\bfs}^{(k)}_{xz|i,j,k-1}-\tilde{\bfs}^{(k)}_{xz|i,j,k}\right)-\frac{1}{2}\left(\mathbf{v}^{(k)}_{z,x|i,j,k-1}-\mathbf{v}^{(k)}_{z,x|i,j,k}\right)\right]. 
\end{align}
Solutions to the other two equations can be obtained similarly. 

The minimization problems for the other variables, including the split-Bregman variables, are trivial to solve. 

\renewcommand{\theequation}{C-\arabic{equation}}
\section*{Appendix C: Algorithm implementation of the TGPV-FWI}
\setcounter{equation}{0}
We summarize the algorithm of our TGPV-FWI in the following Algorithm~\ref{alg:tgpv_fwi}. 

\vskip\baselineskip
\begin{algorithm}[H]
	\DontPrintSemicolon
	\SetKwInOut{Input}{input}
	\SetKwInOut{Output}{output}
	
	\Input{Observed data $\bfd$, initial model $\bfm^{(0)}$, initial misfit $\epsilon_0=\mathtt{HUGE\_VALUE}$}
	
	\While{$\epsilon > \epsilon_{\min}$}{
		
		1) Compute the adjoint source based on certain type of misfit function\;
		2) Compute the gradient $\mathbf{g}_{k}$ using the adjoint-state method\;
		3) Solve the second subproblem for model $\bfm^{(k)}$: \\
		\Input{$\bfm^{(k)}$, maximum number of split-Bregman iteration $n_{\max}$ and $m_{\max}$, norm $p$, regularization parameter $\lambda_2$ (or equivalently $1/\mu$)}
		\While{$n < n_{\max}$}{
			\While{$m < m_{\max}$}{
				3.1.1) Minimization of $\bfu$ using eq.~\eqref{eq:subu}\;
				3.1.2) Minimization of $\bfw$ using eq.~\eqref{eq:subw}\;
				3.1.3) Minimization of $\bfh$ using eq.~\eqref{eq:subh}\;
				3.1.4) Minimization of $\bfs$ using eq.~\eqref{eq:subs} \;
			}
			3.2.1) Update of the split-Bregman variable $\tilde{\bfh}$\;
			3.2.2) Update of the split-Bregman variable $\tilde{\bfs}$\;
		}
		\Output{$\mathbf{u}$}
                4) The Tikhonov regularization using \begin{equation*}
		\mathbf{g}^{(k+1)} \leftarrow \mathbf{g}^{(k+1)} + \lambda_1(\bfm^{(k+1)}-\mathbf{u}^{(k)})
		\end{equation*}
		
		5) Compute the search direction, say $\boldsymbol{\phi}^{(k+1)}$, based on CG or L-BFGS framework\;
		6) Compute the optimal step length $\beta$ using the perturbation method or the line search method\;
		7) Update the model using the computed search direction $\boldsymbol{\phi}^{(k+1)}$ and the optimal step length $\beta$\;
	}
	
	\Output{The updated model $\bfm_N$ (or equivalently $\bfu_N$)}
	
	\label{alg:tgpv_fwi}
	\caption{TGPV-FWI.}
\end{algorithm}

\vskip\baselineskip

In Step (3) of this algorithm, $n_{\max}$ is the maximum number of the 
outer split-Bregman iterations, which can also be replaced by a criterion 
such as $||\bfu^{(k)}-\bfu^{(k-1)}||_2<\varepsilon_0$ where 
$\varepsilon_0$ is a small number; $m_{\max}$ is the maximum number of 
inner split-Bregman iterations, and we find that $m_{\max}=2$ suffices to 
produce an accurate result for the second minimization problem in various 
FWI problems. The computation time of the second minimization problem is 
trivially small compared to the entire FWI inversion process. 

\end{document}